\documentclass[epj]{svjour}
\usepackage{graphics}
\newcommand{\zaa}{{\rm Astron.~Astrophys.}}
\newcommand{\zaas}{{\rm Astron.~Astrophys.~Supl.}}
\newcommand{\zapj}{{\rm Astrophys.~J.}}
\newcommand{\zapjl}{{\rm Astrophys.~J.~Lett.}}
\newcommand{\zapjs}{{\rm Astrophys.~J.~S.}}

\newcommand{\zepj}{{\rm European Physical Journal}}
\newcommand{\znp}{{\rm Nucl.~Phys.}}
\newcommand{\znpa}{{\rm Nucl.~Phys. A}}
\newcommand{\zpl}{{\rm Phys.~Lett.}}
\newcommand{\zpr}{{\rm Phys.~Rev.}}
\newcommand{\zprd}{{\rm Phys.~Rev. D}}
\newcommand{\zprc}{{\rm Phys.~Rev. C}}
\newcommand{\zprl}{{\rm Phys.~Rev.~Lett.}}

\newcommand{\znim}{{\rm Nucl.~Inst.~and~Meth.}}
\newcommand{\zadndt}{{\rm At. Data Nucl. Data Tables}}
\newcommand{\zmnras}{{\rm Mon. Not. R. Astron. Soc.}}

\newcommand{\zrmp}{{\rm Rev. Mod. Phys.}}
\newcommand{\zjcap}{{\rm J. Cosmology Astropart. Phys.}}
\newcommand{\zjpg}{{\rm J. Phys. G}} 

\newcommand{\cago}{$^{12}$C($\alpha,\gamma)^{16}$O}
\newcommand{\cano}{$^{13}$C($\alpha$,n)$^{16}$O}
\newcommand{\nean}{$^{22}$Ne($\alpha$,n)$^{25}$Mg}
\newcommand{\aaag}{$^4$He($\alpha\alpha,\gamma)^{12}$C}
\newcommand{\aabe}{$^4$He($\alpha,\gamma)^{8}$Be}
\newcommand{\beac}{$^8$Be($\alpha,\gamma)^{12}$C}
\newcommand{\np}{$^{26}$Al(n,p)$^{26}$Mg}
\newcommand{\na}{$^{26}$Al(n,$\alpha$)$^{23}$Na}
\newcommand{\sfac}{$S$--factor}

\newcommand{\deu}{D}
\newcommand{\tro}{$^3$He}
\newcommand{\qua}{$^4$He}
\newcommand{\six}{$^{6}$Li}
\newcommand{\sep}{$^{7}$Li}
\newcommand{\neu}{$^{9}$Be}

\newcommand{\onz}{$^{11}$B}
\newcommand{\hli}{$^4$He, D, $^3$He and $^{7}$Li}
\newcommand{\zdag}{D$(\alpha,\gamma)^6$Li}
\newcommand{\bedp}{$^7$Be(d,p)2$\alpha$}
\newcommand{\carb}{$^{12}$C}
\newcommand{\hui}{$^{8}$Be}
\newcommand{\bery}{$^{7}$Be}
\newcommand{\alr}{\ensuremath{^{26}\mathrm{Al}}}
\newcommand{\nas}{\ensuremath{^{23}\mathrm{Na}}}
\newcommand{\als}{\ensuremath{^{27}\mathrm{Al}}}

\newcommand{\lap}{\mathrel{ \rlap{\raise.5ex\hbox{$<$}}
	            {\lower.5ex\hbox{$\sim$}}  } }
\newcommand{\gap}{\mathrel{ \rlap{\raise.5ex\hbox{$>$}}
                    {\lower.5ex\hbox{$\sim$}}  } }

\begin{document}
\title{Recent results in nuclear astrophysics}
\author{Alain Coc\inst{1}, Fa\"{\i}rouz Hammache\inst{2} \and J\"urgen Kiener\inst{1}
}                     
%
%
\institute{Centre de Sciences Nucl\'eaires et de Sciences de la
Mati\`ere  (CSNSM), CNRS/IN2P3 et  Universit\'e~Paris~Sud~11, UMR~8609,
B\^atiment 104, 91405 Orsay Campus, France \and 
Institut de Physique Nucl\'eaire d'Orsay (IPNO),  CNRS/IN2P3 et  
Universit\'e~Paris~Sud~11, UMR8608, 91406 Orsay  Campus, France}

\date{Received: \today / Revised version: date}
%
\abstract{
In this review, we emphasize the interplay between astrophysical observations, modeling,  and
nuclear physics laboratory experiments. Several important nuclear cross sections 
for astrophysics have long been identified e.g. \cago\ for stellar evolution, or \cano\
and \nean\ as neutron sources for the s--process.  
More recently, observations  of lithium abundances in the oldest
stars, or of nuclear gamma--ray lines from space, have required new laboratory experiments.
New evaluation of thermonuclear reaction rates now includes the associated rate 
uncertainties that are used in astrophysical models to $i$) estimate final uncertainties on nucleosynthesis yields and $ii$)
identify those reactions that require further experimental investigation. 
Sometimes direct cross section measurements are possible, but more generally the use of 
indirect methods is compulsory in view of the very low cross sections.
Non--thermal processes are often overlooked but are also important for nuclear astrophysics,
e.g. in gamma--ray emission from solar flares or in the interaction of cosmic rays with matter, and 
also motivate laboratory experiments.  
Finally, we show that beyond the historical motivations of nuclear astrophysics, understanding
$i$) the energy sources that drive stellar evolution and $ii$) the origin of the elements can also    
be used to give new insights into physics beyond the standard model.
\PACS{
      {PACS-key}{describing text of that key}   \and
      {PACS-key}{describing text of that key}
     } 
} 
\maketitle
%


\section{Introduction}
\label{s:intro}

\begin{figure}
\resizebox{0.45\textwidth}{!}{%
  \includegraphics{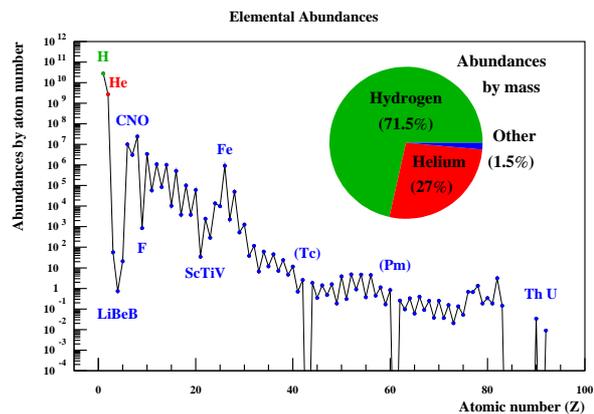}
}
\caption{Solar system elemental abundances \cite{zuni1,zuni2}. "LiBeB", "F", "ScTiV" labels
indicate underabundant elements whose nucleosynthesis is peculiar.}
\label{f:zuni}       
\end{figure}

Nuclear astrophysics was born from the quest of the energy source of stars 
and the origin of the chemical elements (Fig.~\ref{f:zuni}). 

In 1948, Alpher, Bethe and Gamow ($\alpha\beta\gamma$) \cite{ABG} proposed that the elements were
produced "during a rapid expansion and cooling of the primordial matter".
In 1957, Burbridge, Burbridge, Fowler \& Hoyle (B2FH) \cite{B2FH},  and independently, Cameron \cite{Cameron}, presented an alternative option where
elements are formed during the different phases of stellar evolution\footnote{
For extensive historical accounts of the development of nuclear astrophysics see 
Ref.~\cite{Celnikier2006,Shaviv2012} and Ref~\cite{AT99,Iliadis2007,JI11} for  comprehensive 
presentations of the present day domain. 
}.
Hence, at that epoch, the following nucleosynthetic sites ($\circ$ \cite{ABG} and $\bullet$ \cite{B2FH})
were already identified:
\begin{itemize}
\item[$\circ$] Primordial nucleosynthesis,
\item [$\bullet$] hydrogen burning and helium burning, 
\item [$\bullet$]``$e$'' process (iron peak),
\item [$\bullet$] ``$x$" process (Li, Be, B),
\item [$\bullet$] $r$ process (rapid neutron capture),
\item [$\bullet$] $s$ process (slow neutron capture),
\item [$\bullet$] $p$ process (proton rich).
\end{itemize}
Amazingly, more than 50 years later, even though considerable progress has been made
in the domain, this list has practically not changed. The ``$e$" process corresponds to nuclear statistical 
equilibrium that feeds the most tightly bound nuclei around iron, and the  ``$x$" (for unknown ) process 
is now identified with non--thermal nucleosynthesis (\S~\ref{s:non}) resulting from the interaction of cosmic rays with
interstellar matter. 
Only the subsequent burning processes (C, Ne, O, Si burning phases) and neutrino--processes
are missing. 
If changes in our overall understanding of nucleosynthesis since  B2FH are small, tremendous progress has been made in 
certain areas, like big bang nucleosynthesis, non-thermal nucleosynthesis, hydrostatic hydrogen burning and the s--process, where practically all important 
reactions are identified and their cross sections  measured.

In this review, we will present selected issues concerning these thermonuclear processes that occur in stars
or during the first minutes that followed  the big bang.  For stellar nucleosynthesis, we concentrate on reactions between charged particles, since the topic of neutron 
reactions in astrophysics has recently been reviewed \cite{Rei14}. We concentrate furthermore on (hydrostatic and explosive) stellar burning sites with relatively 
well-established conditions, where the uncertainty of a particular reaction rate can have an important impact  on nucleosynthesis yields. Of the many recent 
experiments devoted to those studies, we have selected for illustration one particular experiment for each topic that is described in detail. 
The choices naturally tended towards experiments that we know best.

Another topic presented covers nuclear reactions induced by high-energy non-thermal particles resulting from acceleration processes. 
In fact, particle acceleration occurs throughout the Universe: from inside the heliosphere, where solar flares are the most energetic phenomena, 
to supernova shock waves in our own and distant Galaxies and near supermassive black holes that power active galactic nuclei. 
We will not discuss another important and active field of nuclear astrophysics: dense matter properties (in particular the Equation of State) that are required
for the modeling of neutron stars. We refer the reader to Lattimer \& Prakash \cite{EOS} for a review.

\subsection{Hydrogen--burning reactions}
\label{s:hburn}

In the last 20 years, the most important reactions involved in the hydrogen burning phase (p-p chain and CNO cycle) have been extensively studied 
and are considered nowadays  to be well known at solar and quiescent hydrogen burning temperatures. It is sufficient to refer to the recent evaluations 
of reaction rates by Adelberger et al. \cite{Ade11}, NACRE-II \cite{NACRE2} and Iliadis et al. \cite{rateval10b,rateval10c,rateval10d}.  
Thanks, in particular, to the underground accelerator LUNA in Gran Sasso \cite{LUNA}, high--precision cross section measurements 
were achieved for $^3$He($^3$He,2p)$^4$He \cite{Bon99} and $^2$H(p,$\gamma$)$^3$He \cite{Cas02}. 
In particular, the  $^3$He($^3$He,2p)$^4$He measurement \cite{Bon99}, was the first to be performed in an energy range overlapping the Gamow window.
However, in addition to underground experiments, surface experiments contributed to 
the study of $^3$He($\alpha$,$\gamma$)$^7$Be (see references in the rate evaluation work of deBoer et al. \cite{deB14}) and 
$^{14}$N(p,$\gamma$)$^{15}$O \cite{Mar11,Run05} reactions. 
Another very important reaction where our knowledge was greatly improved  is $^7$Be(p,$\gamma$)$^8$B, the major source of the solar neutrinos detected in Homestake, Super--Kamiokande and SNO.
The main experimental difficulty here is that  $^7$Be is unstable with a 53--day half--life so that either a radioactive target or beam are required
for a direct measurement. It has been studied by many laboratories using direct and indirect methods (see Refs.~\cite{Ade11,NACRE2} for details). 
Thanks to the various experiments, the rates of all important reactions at solar energies are now known to better than 8\% \cite{Ade11}, enabling the use of 
solar hydrogen burning as a remote laboratory. This was a prerequisite for an important recent accomplishment in physics: the solution of the solar 
neutrino problem and the concurrent contribution to the establishment of neutrino oscillations.

The situation is somewhat different for {\em explosive} hydrogen burning that occurs, in particular,
in nova explosions (\S~\ref{s:novae}) and X--ray bursts \cite{Par13}. 
As the relevant energies are higher, so are the cross sections which makes them in principle more accessible to experiment.
However, many reactions occurring in explosive hydrogen burning involve radioactive species with lifetimes down to $\sim$1~s (\S~\ref{s:novae}).
In these cases it is not possible to use a radioactive target, as in most $^7$Be(p,$\gamma$)$^8$B experiments, so that radioactive beams are needed instead. 
This is a major source of difficulties because of the present day scarcity of {\em low--energy} radioactive beam facilities, limited number of 
available isotopes and the low beam intensities ($\lap10^6$~s), compared to stable beams. This can, however, be partially compensated
by the use of indirect techniques. We shall present in this review examples concerning the $^{18}$F(p,$\alpha)^{15}$O (\S~\ref{s:f18p}) and
$^{25}$Al(p,$\gamma$)$^{26}$Si (\S~\ref{s:al25pg}) reactions.

\subsection{Helium--burning reactions}
\label{s:heburn}

\begin{figure}
\resizebox{0.45\textwidth}{!}{%
  \includegraphics{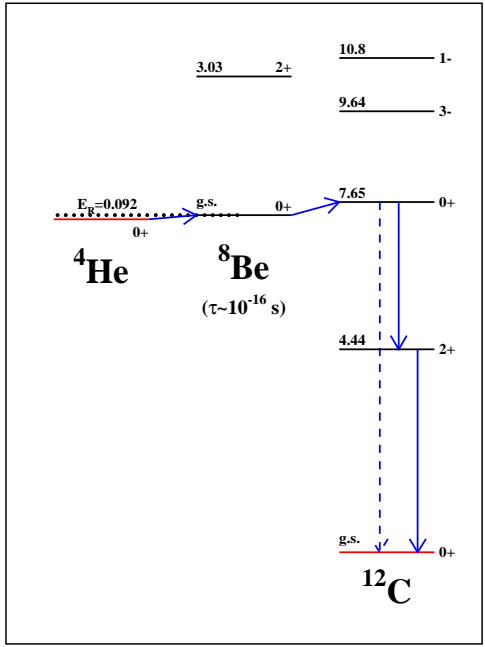}
}
\caption{The \aaag\ reaction, the first stage of helium burning proceeds in two steps: formation of $^8$Be, followed by a second alpha particle capture.}
\label{f:aaag}       
\end{figure}

The \aaag\ reaction plays a special role in the synthesis of the elements as it bridges the gap
between \qua\ and \carb. The absence of particle--bound A=5 and 8 nuclei  prevents 
\qua+p, n or $\alpha$ captures.
It proceeds in two steps, through  a resonance, as shown
in Fig.~\ref{f:aaag}.
The triple-$\alpha$ reaction begins when two alpha particles fuse to produce a $^8$Be nucleus, 
whose lifetime is only $\sim10^{-16}$~s. It is, however, sufficiently long to allow for a second alpha capture 
into the second excited level of $^{12}$C, at 7.65~MeV above the ground state. 
This excited state of $^{12}$C corresponds to an $\ell$ = 0 resonance, postulated by Hoyle \cite{hoyle} 
(see  \cite{Shaviv2012} for an historical account) to enhance the cross section during the helium burning phase. 
There has been controversy regarding the \aaag\ rate in the widely--used compilation of thermonuclear reaction rates NACRE \cite{NACRE}.
A theoretical calculation suggested a 20--order--of--magnitude enhancement  of the rate at 10~MK. This is now refuted by new calculations
that point out the very slow convergence of coupled--channel expansion as the source of the discrepancy \cite{Aka14}.
The difference, at low temperature,  is now reduced to less than an order of magnitude
(see Ref.~\cite{Bay14} for a summary of the present situation).

\begin{figure}
\resizebox{0.45\textwidth}{!}{%
  \includegraphics{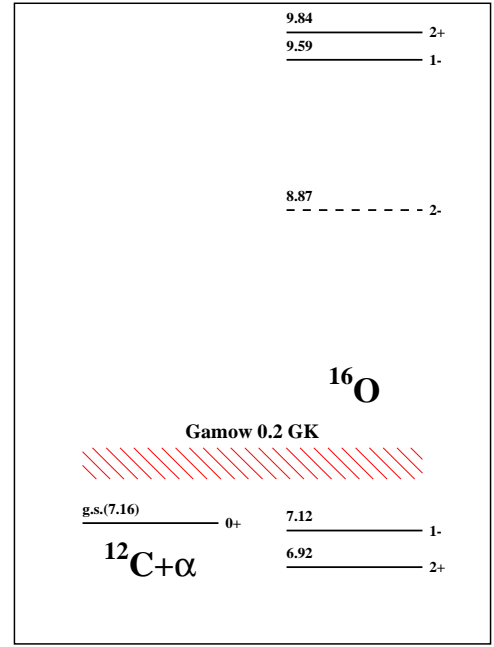}
}
\caption{The \cago\  reaction, that compete with the \aaag\ one during helium burning, proceeds through
the tails of high energy or subthreshold states. The hatched area represents the Gamow window at 2$\times10^8$~K.}
\label{f:cago}       
\end{figure}

The radiative--capture reaction $^{12}$C($\alpha,\gamma$)$^{16}$O is one of the most important reactions in astrophysics. 
The helium burning phase is essentially governed by the \aaag\ and $^{12}$C($\alpha,\gamma$)$^{16}$O reactions and their rates
 determine the ratio of $^{12}$C and $^{16}$O in the helium-burning ashes.  Consequently, the \cago\ reaction  influences strongly the subsequent nucleosynthesis 
 processes for massive stars and their final nucleosynthesis yields \cite{WW93}. At the Gamow peak energy $E_G$ = 300 keV where the 
 reaction occurs during the He burning stage, the expected cross section is extremely small (about 10$^{-17}$ barn) and therefore impossible 
 to measure directly. The extrapolation to thermonuclear burning energies is particularly difficult for this reaction, because the radiative capture 
 process has several contributions and the most important ones, E1 and E2 transitions to the ground state, are strongly influenced by  the high 
 energy tails of the 2$^+$, 6.92--MeV  and 1$^-$, 7.12--MeV states of $^{16}$O, below the 7.16~MeV $^{12}$C+$\alpha$ threshold (Fig.~\ref{f:cago}), 
 whose $\alpha$-reduced widths are not very well known.   
 There is also radiative capture to excited states of  $^{16}$O at $E_x$ = 6.05, 6.13, 6.92 and 7.12 MeV, with smaller cross sections than the 
 ground--state transitions, but needed to achieve the global accuracy of 10-15\%, required for stellar modeling. 
The determination of  $^{12}$C($\alpha,\gamma$)$^{16}$O  thermonuclear reaction rate  at helium burning temperatures has probably received 
the biggest effort of experimental 
nuclear astrophysics ever dedicated to one single reaction. It has included many direct measurements with $\alpha$-particle and $^{12}$C beams, 
as well as a variety of indirect techniques, including elastic scattering, $^{16}$N decay, $\alpha$-particle transfer reactions and Coulomb breakup. 
An overview of direct measurements  and references can be found in NACRE \cite{NACRE} 
and NACRE-II \cite{NACRE2}, as well as recent references of indirect measurements in NACRE-II.  Total astrophysical S-factors 
$S$(0.3 MeV) =  148 keVb \cite{NACRE2} and $S$(0.3 MeV) =  161 keVb \cite{Schue05}  with less than 20\% uncertainty have been extracted 
from the data, but these analyses are contested, by arguing inconsistencies between different data sets \cite{Gai13}.  
A definite answer has probably to wait for new measurements at low energies $E_{cm}$ $\leq$ 1.5 MeV with significant progress  in background suppression and 
improved detection. 

Measurements of several other important reactions occurring during the helium burning phase, such as $^{14}$C($\alpha$,$\gamma$)$^{18}$O, $^{15}$N($\alpha$,$\gamma$)$^{19}$F$,
^{18}$O($\alpha$,$\gamma$)$^{22}$Ne and $^{14}$O($\alpha$,$\gamma$)$^{18}$F, need to be experimentally improved and are among the prime scientific objectives of the underground laboratory LUNA~MV project. 

\subsection{Advanced stages of stellar evolution}
\label{s:advanced}

Following hydrogen and helium burning, and depending  on their masses, stars will successively undergo
 further burning processes: carbon, neon, oxygen and silicon burning, before, eventually exploding
as  ``core--collapse supernovae" for the most massive ones. We will not discuss these advanced stages of stellar
evolution that involve reactions like $^{12}$C+$^{12}$C \cite{Jia13} or  $^{16}$O+$^{16}$O and refer to
Iliadis \cite{Iliadis2007} for a detailed discussion. These advanced burning phases lead to a nuclear
statistical equilibrium and end up in the region of most tightly bound nuclei, the ``iron peak" elements (Fig.~\ref{f:zuni}).

Heavier elements are produced primarily by neutron captures (s-- and r--processes) or photo--disintegration (p--process) \cite{Thi11,AT99}. 
For the sake of simplicity, we will shortly present these different processes, but often, one  can find isotopes with
mixed origins. 

Nearly half of the elements heavier than iron are produced by slow successive neutron captures ( s-process) 
followed by $\beta^-$ decays. The s--process occurs mainly in Asymptotic--Giant--Branch (AGB) stars of low and intermediate mass (M $<$ 8~M$_\odot$)
and during the helium burning phase in massive stars (M $>$ 8~M$_\odot$). The two identified neutron sources for this process are 
the reactions $^{13}$C($\alpha$,n)$^{16}$O and $^{22}$Ne($\alpha$,n)$^{25}$Mg. 
The first reaction is the main neutron source in low mass AGB stars (0.8--3~M$_\odot$) where 
the s-elements of the main component having a mass 90 $\leq$A$\leq$ 209 are produced \cite{Gali98} in the He-rich intershell at temperatures 
around 10$^8$ K. The second reaction is the main neutron source  in AGB stars of intermediate mass  (3~M$_\odot$ $<$ M $\leq$ 8~M$_\odot$), 
and massive stars where the s-elements of the weak component having a mass 58 $\leq$A$\leq$ 88 are produced at 
temperatures around 2.2--3.5$\times10^8$~K. 
Thanks to the various direct and indirect studies of $^{13}$C($\alpha$,n)$^{16}$O, the cross section of this 
reaction is now sufficiently well established (see \S~\ref{s:c13an}) which is not the case of  $^{22}$Ne($\alpha$,n)$^{25}$Mg  
whose cross section at the energy of astrophysical interest is still very uncertain. 
Concerning the cross sections of the (n,$\gamma$) reactions involving stable isotopes, most of them are very well known experimentally 
\cite{Kapp06}. This is less true for reactions at the branching points (competition between beta--decay and neutron capture rates) which involve radioactive isotopes such as $^{59}$Fe, 
$^{79}$Se and $^{95}$Zr.

The other half of the heavy elements are produced by the r--process \cite{AGT07} i.e. rapid neutron captures that require a very high neutron flux. 
Constraints, besides solar system isotopic abundances, now come from elemental abundances observed at the surface of metal--poor stars 
\cite{Bee05}.  Nevertheless, the actual site of the r--process is not yet firmly established. Core collapse supernovae,
in particular within the neutrino driven wind expelled by the proto--neutron star (Farouqi et al. \cite{Far10}; and references therein), have long been 
the preferred option but is now challenged by neutron star merger models \cite{Gor11}. The latter has recently received support from the observation 
of late-time near-infrared  emission following a short-duration gamma-ray burst. This emission was interpreted to be linked to a significant production of 
r--process material in the merger of compact objects, that gave rise to the gamma-ray burst \cite{Tan13}.
Nuclear networks for r-process nucleosynthesis involve thousands of nuclei on the neutron rich side of the chart of nuclei (see e.g. Fig.~15 in Ref.~\cite{AT99}), 
probably extending
to the drip--line (neutron star mergers), for which masses and decay properties are needed together with tens of thousand rates 
(n-capture, lifetimes, fission, neutrino induced reactions,....). Except for a few selected measurements, this problem requires massive input from theory.
Hence, we will not discuss this process further but emphasize that the main issue is to identify its astrophysical site(s).

There remain a few, proton--rich (or equivalently neutron poor), under abundant isotopes (see e.g. Fig.~3 in Ref.~\cite{AG03})
that are bypassed by the s-- and r--processes and originate from the p-- (or $\gamma$--)process \cite{AG03,Rau13}. 
Even though its astrophysical site(s) is(are) not definitively identified, one can safely state that it
typically operates from s-- and r--process seed nuclei that undergo photo--disintegration, 
at temperatures of a few GK. It proceeds mainly by ($\gamma$,n), and to a lesser extent by ($\gamma$,p) and ($\gamma,\alpha$) reactions followed by
subsequent capture reactions. Here again, reaction rates are mostly dependent on theory (Hauser--Feshbach model) but
can benefit from dedicated experiments \cite{Rau13}.

\subsection{Non-thermal nucleosynthesis}
\label{s:ntherm}

Non-thermal ion populations extend in kinetic energy up to a few GeV per nucleon in strong solar flares and exceed 10$^{20}$ eV in total energy for the highest-energy cosmic rays (CR) detected, very probably of extragalactic origin. Nuclear reactions involving such high-energy particles during their propagation change  the abundance pattern of both the energetic particles and the matter of the interaction medium. Despite the usually low densities of non-thermal particles and the ambient medium, non-thermal nucleosynthesis may be important locally and even globally for isotopes not produced in stars (e.g., Li, B and Be). 

Historically, the most important example is certainly the production of lithium, beryllium and boron (LiBeB) in fusion and spallation reactions of CR protons and $\alpha$ particles with interstellar carbon, nitrogen, oxygen (CNO) and helium. The p,$\alpha$ + CNO and $\alpha$ + $\alpha$  cross section measurements elucidated the origin of LiBeB when it was shown that CR nucleosynthesis could produce sufficient quantities of LiBeB in approximately correct ratios to explain today's abundances. Although there are still important questions concerning LiBeB nucleosynthesis, a more detailed account is out of the scope of this review and can be found in  \cite{RFH70,Rev94,VCA00,Pra12}. An example of local non-thermal nucleosynthesis in metal-poor halo stars is  in-situ $^6$Li production by solar-like flares, that has been proposed as an alternative to Big-Big nucleosynthesis to the observed high $^6$Li abundances \cite{VThi}.

However, the interest in studies of energetic particles and their interactions lies not only in their contribution to nucleosynthesis, but may also reveal their origin, teach us about acceleration mechanisms and  provide information about the propagation medium. In the last years nuclear reaction data have been obtained relevant to two axes of solar flare and CR observations: (1) direct observations of energetic particle  spectra and composition with balloon- or space-borne instruments; (2) remote observations of energetic-particle induced electromagnetic emission. We will only discuss briefly the first subject and shall concentrate on the studies centered at gamma-ray line emission in nuclear reactions, where  a large part of recent results were obtained at the Orsay tandem Van-de-Graaff accelerator.

\section{Nuclear reaction data for astrophysics}
\label{s:react}

Among all the astrophysical environments, there are basically three thermodynamical conditions in
which nuclear physics plays a predominant role. These are $i$) dense matter, 
$ii$) medium in local thermodynamical equilibrium (LTE) and $iii$) diluted medium.    
The first one is found in the interior of neutron stars or white dwarfs, and also occurs
during the core collapse of supernovae.  The important nuclear--physics inputs are e.g.
the equation of state of neutron dense matter, neutrino interaction cross sections, or
pycnonuclear reaction cross sections. This is beyond the scope of this review and we
refer to  \cite{EOS,SN,Ich93} for reviews. The second regime occurs when the 
density  is low enough so that the velocity distributions of the ions can be described by
Maxwell-Boltzmann distributions. Reaction products, including photons,  are readily 
thermalyzed and do not escape. This describes correctly the conditions prevailing
in the interior of most stars, including most of their explosive phases. 
Thermonuclear reaction rates as a  function of temperature, are the required quantities.
The third process occurs in diluted environments such as the interstellar medium
where the mean free path of accelerated particles is so long that they do not reach
LTE. Interestingly, in these diluted media, produced gamma rays can escape and
eventually be detected. Cross sections, to be folded with process--dependent
velocity distributions are the required inputs.

\subsection{Thermonuclear reaction rates}
\label{s:rates}
We consider here a medium that is in local thermodynamical equilibrium so that
the distribution of ion velocities/energies follows a Maxwell--Boltzmann (M.-B.) distribution,
\begin{equation}
\phi_\mathrm{MB}(v)vdv=
\sqrt{8\over{\pi\mu}}{{1}\over{(kT)^{3/2}}}e^{-E/kT}EdE
\label{q:mb}
\end{equation}

while the photons follows a Planck distribution,
\begin{equation}
dn={8\pi\over{({\hbar}c)^3}}{1\over{e^{E/kT}- 1}}E^2dE
\label{q:planck}
\end{equation}
corresponding to the same temperature $T$.
In such conditions, one defines the thermonuclear reaction rate by: 
\begin{equation}
N_A<\sigma v> =
N_A\int_0^\infty\sigma\phi_\mathrm{MB}(v)vdv
\label{q:taux}
\end{equation}
in cm$^3$s$^{-1}$mole$^{-1}$ units where  $N_A$ is Avogadro's number 
(mole$^{-1}$). We summarize here a few results, to be used in this review, and
refer to \cite{Iliadis2007,NACRE,rateval10a} for a detailed treatment.

\begin{figure}
\resizebox{0.45\textwidth}{!}{%
  \includegraphics{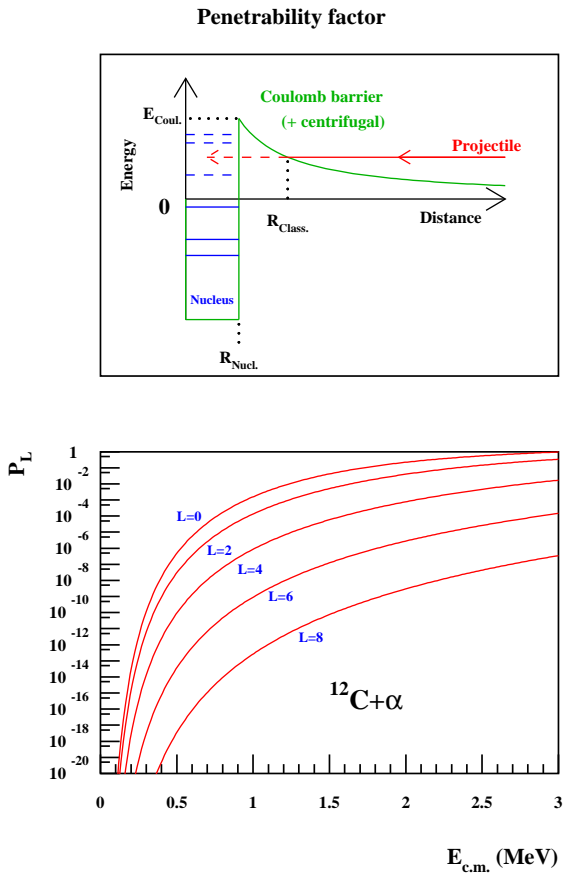}
}
\caption{Coulomb barrier penetrability scheme (upper panel) and penetrability in the $\alpha+^{12}$C channel
as a function of energy and orbital angular momentum L (lower panel).}
\label{f:pen}       
\end{figure}

Except for the important neutron capture  (s-- and r--processes) and photon induced
reactions ($\gamma$--process),  nuclear reactions involve charged particles in the initial state
and available kinetic energies are generally well below the Coulomb barrier (Fig.~\ref{f:pen}):
\begin{equation}
E_{Coul.}\approx{{Z_{1}Z_{2}e^2}\over{R}}=
1.44{{Z_{1}Z_{2}}\over{R ({\rm fm})}}\;{\rm (MeV)}
\label{q:coul}
\end{equation}
so that the energy dependence of the cross section is dominated by the tunneling effect
through the barrier. The  Coulomb plus centrifugal barrier penetration probability is given by (Fig.~\ref{f:pen}): 
\begin{equation}
P_\ell(E)={kR\over{F^2_\ell(\eta,kR})+G^2_\ell(\eta,kR)}
\label{q:pen}
\end{equation}
where $F$ and $G$ are the Coulomb functions, $k= \sqrt{2\mu E}/\hbar$ is the wave number,
$\ell$ the orbital angular momentum and 
\begin{equation}
\eta\equiv{{\textstyle Z_{1}Z_{2}e^2}\over{\textstyle \hbar v}}
\end{equation}
the Sommerfeld parameter.
To account for this strong energy dependency of the cross section, it is customary to introduce
the astrophysical \sfac:
\begin{equation}
\sigma(E)\equiv {\textstyle{S(E)}\over{E}}
\exp\left(-2\pi\eta\right)\equiv {\textstyle{S(E)}\over{E}}
\exp\left(-\sqrt{{E_G}\over{E}}\right)
\label{q:se}
\end{equation}
where $E_G$=
$(0.989Z_cZ_pA^{1\over2})^2$~(MeV) is the Gamow energy.
So that Eq.~\ref{q:taux} leads to
\begin{equation}
N_A<\sigma v> \propto
\int_0^\infty S(E)\exp\left(-{E\over{kT}}-\sqrt{{E_G}\over{E}}\right)dE,
\label{q:tause}
\end{equation}
where the argument in the exponential has a maximum, around which it can be approximated by a Gaussian: 
\begin{equation}
\exp\left(-{E\over{kT}}-\sqrt{{E_G}\over{E}}\right)\sim
\exp\left(-\left({{E-E_0}\over{{1\over2}\Delta}}\right)^2\right)
\label{q:integ}
\end{equation}
centered at\footnote{In nuclear astrophysics, it is usual to use 
 $T_9$, the temperature in units of GK.
} 
\begin{equation}
E_0 = 0.122(Z_1^2Z_2^2A)^{1\over3}T^{2\over3}_9\; {\rm MeV},
\label{q:e0}
\end{equation}
with a full width at 1/$e$ given by
\begin{equation}
\Delta=0.2368(Z_1^2Z_2^2A)^{1\over6}T^{5\over6}_9\; {\rm MeV}.
\end{equation}
that defines the {\em Gamow window}. 
When calculating a thermonuclear reaction rate, and in the case of a {\em slowly varying} \sfac, 
the dominant contribution to the integral comes from this energy range. This {\em window}
is generally used to guide experimentalist. Note, however, that when the cross section is
dominated by resonance contributions, the  Gamow window gives a good indication, but should   
be used with care \cite{gamow}.  
Resonances in the cross sections can lead to orders of magnitude increase in the thermonuclear 
reaction rate: their localization and the determination of their parameters are hence of the utmost
importance.

Since the pioneering works of Fowler and collaborators \cite{FCZ}, the importance of providing stellar modelers with databases 
of thermonuclear reaction rates has been recognized. 
They were obtained from compilations of experimental nuclear data with some theoretical input,
and were first presented
in the form of tables and analytical approximations. A first transition occurred between the last 
Caughlan \& Fowler paper \cite{CF88} and the NACRE \cite{NACRE} evaluation (recently partly updated \cite{NACRE2}), which incorporated
several improvements. The most important improvement was that for all reactions, not only  a ``recommended" reaction 
rate was given, but also ``low" and ``high" rates were provided, reflecting the rate uncertainties.
However, these uncertainties were not obtained by a rigorous statistical treatment and a second 
transition occurred with the Iliadis and collaborators \cite{rateval10a,rateval10b,rateval10c,rateval10d} 
evaluation.  In these works, thermonuclear rates are obtained by Monte Carlo calculations, sampling 
input data (resonance energies, strength, partial widths, spectroscopic factors, upper limits,...) according
to their associated uncertainties and {\em probability density functions} (PDF). For instance \
resonance energies, strengths  or reduced widths are expected to follow, respectively  
normal, lognormal or Porter-Thomas PDF (see Refs.~\cite{rateval10a,astrostat}).         
For each value of the temperature, a Monte Carlo calculation of the reaction rate is performed, 
sampling all input parameters according to their uncertainties and associated  PDF. The result
is a distribution of rate values at that temperature. 
The median and associated 68\% confidence intervals are calculated  by taking respectively the 
0.5,  0.16 and  0.84 quantiles of this rate distribution
(see \cite{rateval10a} for details). 
It has been found, that these Monte Carlo rate distributions can be approximated  
by lognormal functions:
\begin{equation}
f(x) = \frac{1}{\sigma \sqrt{2\pi}} \frac{1}{x} e^{-(\ln x - \mu)^2/(2\sigma^2)}  \label{lognormalpdf}
\end{equation}
(with $x\equiv{N_A}\langle\sigma{v}\rangle$ for short).
This is equivalent to the assumption that  $\ln(x)$ is Gaussian distributed with expectation value $\mu$ and variance $\sigma^2$ (both functions of temperature).
The lognormal distribution allows to cope with large uncertainty factors (${\equiv}e^\sigma$) 
together with ensuring that the rates remain positive.
If these parameters are  tabulated as a function of the temperature, they can be used to perform subsequent
Monte Carlo nucleosynthesis calculations within astrophysical simulations (see \S~\ref{s:sensiv} and Ref.~\cite{astrostat}).

At present, such pieces of information are only available in the ``STARLIB" database which can be found online \cite{STARLIB},
with the possibility of running  Monte Carlo calculations of reaction rate with one's own input data.  The NACRE databases 
\cite{NACRE,NACRE2} provide ``low" and ``high" rates reflecting the uncertainties and are included in the online databases
of BRUSLIB \cite{BRUSLIB1,BRUSLIB2,BRUSLIB3} and REACLIB \cite{JINA}.  
Up to now, we have implicitly assumed that reaction rates were derived from experimental data, but this applies only
to the first stages of H and He burning. For other stages, in particular for the r--process, most rates are obtained from theory
(e.g. \cite{TALYS2}) and uncertainties are not provided in databases.

To be complete, we mention that these thermonuclear rates have to be corrected for $i$) electron screening that lower the Coulomb barrier at 
low energy \cite{Ass87} and $ii$) the thermal population of excited states of the target nuclei at high temperature.

\subsection{Non-thermal reactions}
\label{s:nonth}

Here, we consider interactions of two distinct particle populations, where the kinetic energies of one component are largely superior to the other. This is for example the case for cosmic rays interacting with the gas and dust of the interstellar medium and for solar flares where particles accelerated in the corona interact in the solar atmosphere. Typical particle energies considered here extend from a few MeV into the GeV range in solar flares and the GeV-TeV range for cosmic rays. The energy range of cosmic rays extends of course largely beyond the GeV-TeV range, but those extreme energies belong more to the domain of astroparticle physics, and will not be discussed here. Thermal energies are typically far below the eV range for the interstellar medium, and even for strong solar flares, where temperatures may rise to several tens of million Kelvin, the ambient particle energies are largely below the MeV range. It is therefore safe to suppose the target at rest in the calculations and limit reaction rate integrations to the projectile energy.

\begin{figure} 
\resizebox{0.45\textwidth}{!}{%
  \includegraphics{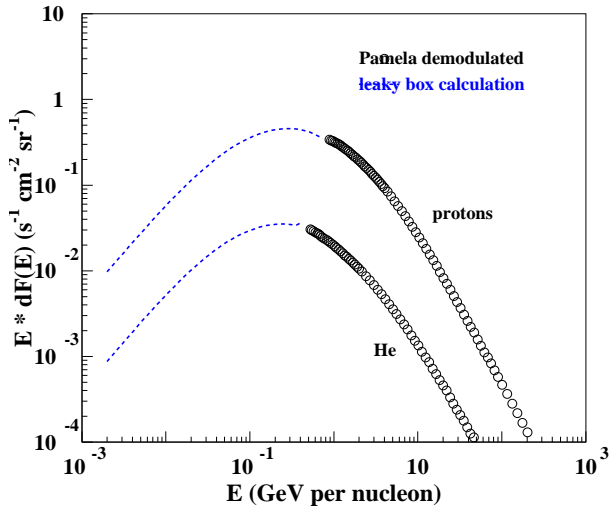}
}
\caption{Local interstellar proton and He spectra. Symbols are data from the satellite-borne PAMELA experiment \cite{Pamela}, corrected for solar modulation with a force-field model \cite{Hinda13}. The extrapolation to lower energies, shown by the dashed lines is done with a model for galactic CR propagation \cite{Hinda13}. The CR fluxes have been multplied with E to emphasize the CR intensity per logarithmic bin.  }
\label{f:crspec}       
\end{figure}

As in the case of thermonuclear reactions  the product of cross section $\mathit \sigma(E)$ and what we shall call ``interacting effective particle flux" $\mathit dL\Phi(E)$ determines the important energy range for the calculation of  a particular rate. $\mathit dL\Phi(E)$ is the product of the effective path length in the interaction medium and particle number density at energy $E$. In the case of CR interactions  with the interstellar medium (``thin target" mode), the CR density distribution can be supposed to be in a steady state and $\mathit dL\Phi(E)$ is then simply proportional to the CR flux spectrum $\mathit dF(E)$. Local interstellar CR proton and He spectra are displayed in  Fig. \ref{f:crspec}. Above about 10 GeV per nucleon, both spectra show the typical power-law behaviour  $\mathit dF(E) \propto E^{-s}$ with $\mathit s\sim2.7$ that is expected for propagated CR nuclei accelerated by diffuse shock-acceleration in e.g. supernova remnants \cite{Dru83,Lon11}. 

Particle acceleration takes place in impulsive solar flares mainly in the  low-density solar corona of active regions by magnetic reconnection events.  Part of the energetic particles are then precipitated along magnetic field lines to the denser chromospheric and photospheric regions of the solar atmosphere where they induce  emission of secondary particles and electromagnetic radiation, heat the ambient matter and eventually are absorbed \cite{Hua}.  Then the interacting effective particle flux  $\mathit dL\Phi(E)$ results in the stopping process of an injected particle spectrum $\mathit  dI(E_0)$, given by (``thick target" mode):

\begin{equation}
dL\Phi(E)~=~ \frac{\rho}{dE/dx}~\int_{E}^{\infty}~dI(E_0)~dE_0~ 
\label{eq:sflare}
\end{equation}
where  $\rho$ is the ambient matter density and $\mathit dE/dx$ the stopping power. $\mathit \rho$ would typically be given in atomic number density and $\mathit dI(E_0)$ as the number of injected particles per energy, which results in units of [atoms cm$^{-2}$ MeV$^{-1}$] for $\mathit dL\Phi(E)$. Multiplication with $\mathit \sigma(E)$ and integration over $\mathit E$ gives then directly the number of interactions.  Particle losses in nuclear collisions are not included here, being usually negligible for the particle energies prevailing in solar flares where electronic stopping dominates completely the energy-loss process. 

\begin{figure}
\resizebox{0.45\textwidth}{!}{%
  \includegraphics{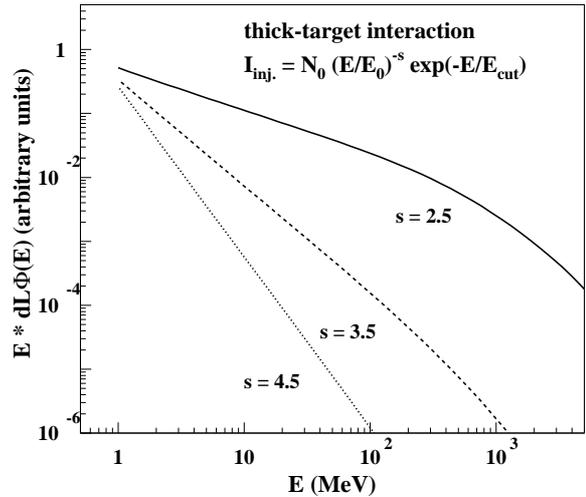}
}
\caption{Interacting effective proton flux $\mathit dL\Phi(E)$ in a thick-target composed of 90\% H and 10\% He for injected power-law spectra with an exponential cutoff $\mathit E_{cut}$ = 10 GeV. The fluxes have been multiplied with $E$ to emphasize the intensity per logarithmic bin. }
\label{f:flspec}       
\end{figure}

Examples of $\mathit dL\Phi(E)$  for injected power-law particle spectra with energy cutoff  are presented in Fig. \ref{f:flspec}. The most important energy ranges for nuclear reactions induced by CRs and in solar flares are in the GeV and MeV ranges, respectively. Depending of course also on the shape of the cross section functions, data are often needed in a very wide range,  from reaction threshold to hundreds of GeV per nucleon for CRs and to hundreds of MeV per nucleon for  solar flares. It is worthwhile to mention that there is no steady state in solar flares: the reaction rate is strongly dependent on the temporal behaviour of e.g. the acceleration process that usually shows short-time ($\sim$1 min) burst-like behaviour. An explicit time-dependent treatment for e.g. the 2.223-MeV neutron-capture line on H  \cite{Hua87,Mur07}  or for the emission of long-lived radioactive species \cite{Solrad} can provide additional valuable information on the flare geometry and properties of the solar atmosphere.

\section{Identifying important reactions}
\label{s:ident}

Experimental nuclear astrophysics is driven by the need to determine cross sections of
{\em important reactions}. Many were identified early during the development of the
discipline and have been measured to a good accuracy (see Figs. 1--60 in 
Ref. \cite{rateval10d}), or are still under investigation because of experimental 
limitations (e.g. \cago, having an extremely low cross section). In addition, new important reactions
have been recently identified thanks to new observations or improved model studies.
New observations can open up a new field (e.g. gamma--ray astronomy, \S~\ref{s:obs26}--\ref{s:crem}) that requires
improved knowledge of previously overlooked reactions or point out discrepancies
(e.g. between primordial lithium and CMB observations, \S~\ref{s:li}) that  require nuclear
physics attention. With the progress in computing power, it is now possible to perform
thousands of calculations with the same astrophysical model and  parameters, but with
different  reaction rates, including Monte Carlo sampling of rates, to identify potential key reactions. 
In this review, we will mainly concentrate on these newly identified reactions. 

\subsection{New observations}

\subsubsection{New Li, D and CMB observations}
\label{s:li}

Observations of the anisotropies of the cosmic microwave background (CMB) by the WMAP \cite{WMAP9} and more recently the 
Planck \cite{Planck14} space missions have enabled the extraction of cosmological parameters with an unprecedented precision.  
In particular, the baryonic density of the Universe, which was the last free parameter in big bang nucleosynthesis (BBN) calculations, 
is now measured with $\approx$ 1\% precision \cite{Planck14}). Standard BBN predictions can now be precisely compared with
primordial abundances deduced from observations. 

The primitive lithium abundance is deduced from observations of low metallicity stars in the halo of our Galaxy where the lithium 
abundance is almost independent of metallicity, displaying the so-called Spite plateau \cite{Spite82}. 
This interpretation assumes that lithium has not been depleted at the surface of these stars, so that the presently observed 
abundance (Li/H = $(1.58 \pm 0.31) \times 10^{-10}$ \cite{Sbo10} in number of atoms relative to hydrogen) can be assumed to 
be equal to the primitive one. BBN calculations using the CMB deduced baryonic density give $(4.94^{+0.38}_{-0.40})\times10^{-10}$ 
\cite{Coc14b,CV14}, a factor of $\approx$3 above observations. This is the so-called ``lithium problem" whose solution \cite{Fie11}  can come from
stellar physics and/or exotic physics but first, nuclear physics solutions have to be excluded (see \S~\ref{s:be}).

A few years ago, observations \cite{Asp06}  of \six\ in a few metal poor stars had suggested the presence of a plateau, at 
typically  \six/\sep$\approx$1\% or \six/H$\approx10^{-11}$, 
leading to a possible pre-galactic origin of this isotope. 
This is orders of magnitude higher than the BBN predictions of  \six/H$\approx1.3\times10^{-14}$ \cite{dag10}: this was the second lithium problem.
Later, the observational \six\  plateau has been questioned  
due to line asymmetries which were neglected in previous abundance analysis. 
Presently, only one star, HD84937, presents a  \six/\sep\  ratio of the order of 0.05
\cite{Lin13} and there is no remaining evidence for a plateau at very low metallicity.

Deuterium is a very fragile isotope, easily destroyed after BBN. Its most primitive abundance is determined from the observation of very few
cosmological clouds at high redshift, on the line of sight of distant quasars. 
The observation of about 10  quasar absorption systems  gave the weighted mean abundance of deuterium 
D/H = $(3.02 \pm 0.23) \times 10^{-5}$~\cite{Oli12}. However, recently, observations of Damped Lyman-$\alpha$ (DLA) 
systems at high redshift show a very small dispersion of values leading to a more precise average value : 
D/H = $(2.53 \pm 0.04) \times 10^{-5}$ \cite{Coo14},  marginally compatible with BBN predictions of 
$(2.64^{+0.08}_{-0.07})\times10^{-5}$\cite{Coc14b,CV14}. If a 1.6\% precision in observations is confirmed,
more attention should be paid to some nuclear cross sections (\S~\ref{s:bbn}).

\subsubsection{New $^{26}$Al observations}
\label{s:obs26}

\begin{figure}[b]
\resizebox{0.45\textwidth}{!}{%
  \includegraphics{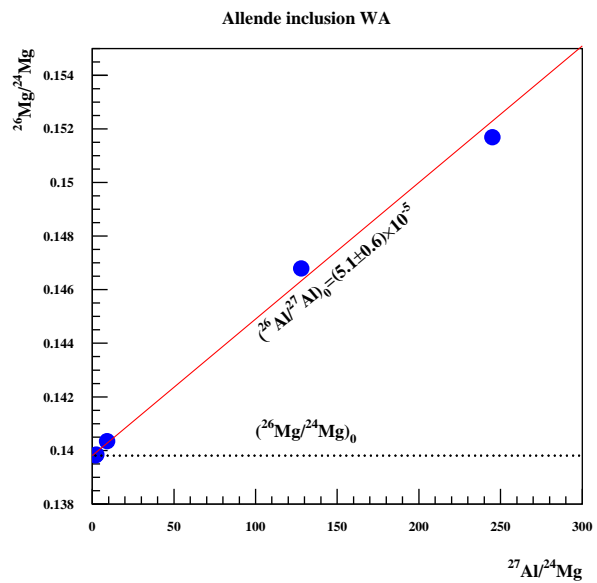}
}
\caption{Al--Mg isochron: different minerals (Melilite, Anorthite, Spinel, Pyroxene) having different chemical compositions, in particular Al/Mg ratios \cite{Lee76},
allow for the determination of the initial $^{26}$Al/$^{27}$Al isotopic ratio.
}
\label{f:meteorite}       
\end{figure}

Before its observation by its gamma-ray emission, evidence of $^{26}$Al decay products in meteorites was observed in calcium-aluminum rich inclusions (CAIs) from the Allende
meteorite as an excess of its daughter nuclei ($^{26}$Mg) with respect to the stable $^{24}$Mg isotope \cite{Lee76}. 
The linear correlation between the $^{26}$Mg/$^{24}$Mg and $^{27}$Al/$^{24}$Mg isotopic ratios  (Fig~\ref{f:meteorite}) 
yields an initial value of 5.3$\times10^{-5}$ for the $^{26}$Al/$^{27}$Al ratio \cite{Jac08}. 
The content of $^{26}$Al in these CAIs demonstrates that this short-lived nucleus was indeed present at the birth of the Solar system.

Contrary to most observations in other wavebands that are sensitive to element abundances, gamma-ray astronomy provides isotopic information through the characteristic gamma-ray signature of radioactive isotopes. There is also the penetrating nature of gamma rays that makes them less sensitive to interstellar absorption and the insensitivity of radioactive decay to the ambient conditions. A gamma-ray line flux is therefore often a direct measure of the radioisotope activity and thus the isotope abundance if the distance is known. The long-lived $^{26}$Al$_{g.s.}$ ($\tau$ = 1.0$\times$10$^6$ years) was the first detected cosmic radioisotope. Its 1.809-MeV decay line was observed by the HEAO-3 satellite more than 30 years ago from the inner Galaxy  \cite{HEAO} and confirmed later on by several other instruments. These observations provided an estimation of the galactic $^{26}$Al content, but could not establish  the sources of $^{26}$Al because of angular resolution and sensitivity limits  \cite{PraDi}.  

The origin of the observed $^{26}$Al are nucleosynthesis sites with efficient $^{26}$Al production and ejection into the interstellar medium before its decay. The main production mechanism is the $^{25}$Mg(p,$\gamma$)$^{26}$Al reaction in high-temperature environments (T $\gap$ 50~MK) with sufficient abundances of  H and Mg. Those conditions are met in AGB and Wolf-Rayet stars, where convection and massive stellar winds disperse the nucleosynthesis products of hydrostatic hydrogen burning. Other important sources of galactic  $^{26}$Al are the carbon and neon shells of massive stars  releasing  the synthesized radioactive isotopes during subsequent supernova explosions and - probably of less importance - explosive hydrogen burning in classical novae (see  \cite{PraDi} for a more detailed account of $^{26}$Al nucleosynthesis). 

A breakthrough in the observation of galactic $^{26}$Al came with the CGRO and INTEGRAL satellites.  The Compton imaging telescope Comptel aboard CGRO \cite{Sch93} made the first maps of $^{26}$Al emission with good angular resolution ($\sim$4$^{\circ}$).  They show irregular extended emission along the galactic plane  with brighter spots that favor massive star origin \cite{PraDi,Die95,Obe96,Plu01}. More recently, the high-resolution gamma-ray spectrometer SPI  of the INTEGRAL mission \cite{SPI} measured fluxes, redshifts and widths of the $^{26}$Al line from different locations. They included the inner Galaxy and some massive-star groups like the Cygnus, Orion and Sco-Cen regions \cite{Die06,Wan09,Mar09,Vos10,Die10,Kre13,Die13}. The redshifts measured by INTEGRAL/SPI demonstrate that the $^{26}$Al source regions corotate with the Galaxy, specifying their distribution in the Galaxy. The deduced radial velocities, however, exceed the velocities expected from galactic rotation and may hint to some very specific emission process of freshly synthesized $^{26}$Al. Kretschmer et al.  \cite{Kre13} proposed  massive stars  that are situated in the leading edges of spiral arms ejecting their nucleosynthesis products preferentially towards the inter-arm regions.  The total amount of live $^{26}$Al in the Galaxy could be established to be of $\sim$2 solar masses.

The total amount of $^{26}$Al and its distribution pose interesting constraints on the $^{26}$Al yields of massive stars and novae. This holds even more for the individual groups with well-known populations of massive stars, where detailed stellar models of massive-star nucleosynthesis can be confronted with the actual $^{26}$Al content.  These observations naturally  triggered considerable activity in experimental  nuclear astrophysics  to determine more accurate yields of reactions relevant to $^{26}$Al nucleosynthesis.

\subsubsection{New observations related to energetic-particle populations}
\label{s:crem}

\noindent
(a) Cosmic rays
\vspace{0.15 cm}

There is an impressive record of new CR spectra and composition data, from H to Sr and for electrons, positrons and antiprotons, obtained in the last decade from dedicated experiments on high-altitude balloons, satellites, the space shuttle and the international space station. A complete review will not be given here, a compilation of publications and data since 1963 can be found in \cite{CRdat}. Most relevant for nuclear astrophysics are probably the recent data of ATIC \cite{ATIC2}, TRACER \cite{TRACER}, CREAM \cite{CREAM} and ACE/CRIS \cite{ACE-CRIS} instruments that  provide high-precision CR compositions and spectra at $\sim$0.2 - 10$^5$  GeV per nucleon for elements  up to Fe, Ni,  while the TIGER instrument \cite{TIGER} yielded abundance data for elements up to Sr above $\sim$2.5 GeV per nucleon. Still more precise data are expected from the AMS-02 experiment on the international space station \cite{AMS02}.

\begin{figure}[h]
\resizebox{0.45\textwidth}{!}{%
  \includegraphics{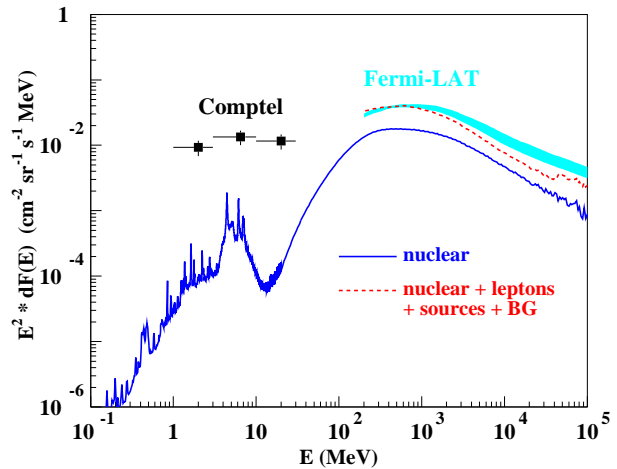}
}
\caption{Predicted gamma-ray emission due to nuclear interactions from the inner Galaxy 
(within -80$^{\circ}$$\leq$l$\leq$80$^{\circ}$, -8$^{\circ}$$\leq$b$\leq$8$^{\circ}$, in respectively Galactic longitude and latitude) 
with a LECR component added to the standard CRs (full blue line). The LECR properties have been adjusted such that  the mean CR ionization rate of the inner Galaxy deduced from H$_3^+$ observations (see text) and the Fermi-LAT observations (cyan band) \cite{Ack12} at E = 1 GeV are simultaneously reproduced.  This example is for shock-accelerated LECRs  with an exponential cutoff E$_{c}$ = 45 MeV per nucleon (see \cite{Hinda13} for more details). The dashed red line shows the total calculated emission when adding leptonic contributions, point sources and extragalactic gamma-ray background that were taken from \cite{Ack12}. Also shown are the Comptel data \cite{Comptel} from (-60$^{\circ}$$\leq$l$\leq$60$^{\circ}$, -10$^{\circ}$$\leq$b$\leq$10$^{\circ}$).}
\label{f:lecr}       
\end{figure}

The LAT instrument on the Fermi satellite \cite{Atw09}, launched in 2008 and featuring much-improved sensitivity and angular resolution with respect earlier missions, has enabled a big step forward in the observation of the high-energy gamma-ray sky.  The diffuse galactic emission in the Fermi-LAT energy band (30 MeV to several hundred GeV)  is dominated by $\pi^0$-decay gamma rays from the interaction of CR nuclei with interstellar matter, the largest contribution coming from proton-proton and proton-$\alpha$ particle reactions with energies in the GeV range. Fermi-LAT observations therefore trace the spectra and densities of light CR nuclei in the Galaxy. Examples of observations include local molecular clouds \cite{Ack12a,Ack12b,Yan14}, supernova remnants \cite{Tho12,Ack13,Tib13,Piv13,Yua13,Han14}, superbubbles \cite{Ack11} and general diffuse emissions throughout the Galaxy \cite{Abd09,Ack12} (an example is shown in Fig. \ref{f:lecr}). A complete account of CR-relevant Fermi publications can be found in \cite{Fermipub}.

These direct CR observations and  CR-induced gamma-ray emissions put stringent constraints on the CR origin and propagation that are fully utilized in modern CR transport models like Galprop \cite{Galprop}. Galprop calculates local CR spectra and composition after propagation of a given galactic source distribution and has also implemented CR-induced electromagnetic emissions from the radio to the high-energy gamma-ray band. Taken together, these new observations and CR modeling have furnishhed in a broadly consistent picture of CR rigidity-dependent diffusion in our Galaxy with a CR halo extending a few kpc above and below the galactic thin disk (see e.g. \cite{TIGER,Str07,Put10,Mau10,Tro11}).  The CR composition and gamma-ray observations indicate an origin  closely tied to massive stars, with shock waves in supernova remnants as the most likely sites of CR acceleration \cite{Pra12b,Bla14} at GeV-TeV energies. 

This progress must be accompanied by providing an accurate nuclear reaction network in those codes, calling for cross sections that have an accuracy in the ten percent range or better, comparable to observations. While the calculation of $\pi^0$ production and decay in nuclear collisions  has recently been updated \cite{kam06}, precise fragment production cross sections exist only for a part of abundant CR nuclei and often in a limited energy range.  The most important needs are probably cross sections for  heavier nuclei e.g. Fe-Sr interacting with H and He and a much better coverage of cross sections above a few GeV per nucleon for practically all nuclei. 

\vspace{0.25 cm}

\noindent
(b) Low-energy cosmic rays
\vspace{0.15 cm}

The observations described above have largely contributed towards a consistent picture of galactic cosmic rays above a few hundred MeV per nucleon. Below these energies, however, no direct observation is possible inside the heliosphere because of solar modulation, the suppression of low-energy cosmic-ray (LECR) flux due to the action of the wind streaming out from the sun \footnote{The Voyager 1 spacecraft may have recently crossed the heliospheric boundary and may now observe the local interstellar particle spectra \cite{Voyager}, but this conclusion has been questioned \cite{Voy14}.}.  Likewise, the high-energy gamma-ray observations with Fermi-LAT probe CR spectra above about 1 GeV per nucleon only. However, the CR energy density is probably dominated by lower-energy particles and  there is evidence that at least some regions of our Galaxy contain an important LECR component.

There are in particular three recent observations suggesting important LECR fluxes:

(i) Observations of the molecule H$_3^+$ in diffuse interstellar clouds indicate a mean CR ionization rate of molecular hydrogen in our galaxy of $\mathit \zeta_2$ = 3-4$\times$10$^{-16}$ s$^{-1}$ \cite{mcc03,ind09,ind12}). When taking typical cosmic-ray spectra obtained by extrapolating the locally observed CR spectrum to lower energies, simple or more sophisticated galactic propagation models yield mean ionization rates that fall short by about a factor of 10.  The authors of  \cite{mcc03,ind09,ind12} concluded that a distinct low-energy galactic CR component, probably localized production in e.g. weak shocks, must be responsible for the extra ionization. 

(ii) Very recent millimeter-line observations of molecular species in dense interstellar clouds close to the supernova remnant W28 indicate a cosmic-ray ionization rate much larger ($\ge \sim 100$) than the standard value in dense galactic clouds, with the most likely interpretation of these observations being a locally-confined hadronic LECR component in the range 0.1 - 1 GeV, accelerated in the supernova remnant \cite{Vau14}. 

(iii) Another indication of an enormous flux of low-energy ions has been deduced from X-ray observations of the 6.4-keV Fe K$\alpha$ line in the Arches cluster \cite{Arches}. There, in a nearby molecular cloud a CR energy density of about 1000 times the local CR energy density was estimated from the observations, dominantly due to LECRs. 
We note, however that the recent detection of a variation of the X--ray non--thermal emission in the Arches cloud \cite{Cla14} is difficult to explain with a model of LECRs.

\vspace{0.25 cm}

Supposing that LECRs were primarily hadrons, besides contributing significantly to the LiBeB synthesis, they would be responsible for considerable emission of nuclear gamma-ray lines from collisions with atomic nuclei of  the interstellar medium. Actually, it has been shown that the intensity of some strong lines and even more the total nuclear gamma-ray line emission in the 1-8 MeV band from the inner Galaxy would be largely in the sensitivity limits of next-generation gamma-ray telescopes for most of the ion-dominated LECR scenarios \cite{Hinda13}. Figure \ref{f:lecr} shows an example of predicted nuclear gamma-ray emission of CRs containing such a low-energy component. A future observation of this  emission would be the clearest proof  of an important LECR component  in the Galaxy and probably the only possible means to determine its composition, spectral and spatial distribution. From the nuclear side, gamma-ray line cross sections for the total emission in the 1-8 MeV band are required. This applies in particular to a component that is a superposition of thousands of weak lines that form a quasi-continuum and for which no individual cross section data exist.

\vspace{0.25 cm}
\noindent
(c) Solar flares
\vspace{0.15 cm}

\begin{figure}
\resizebox{0.45\textwidth}{!}{%
  \includegraphics{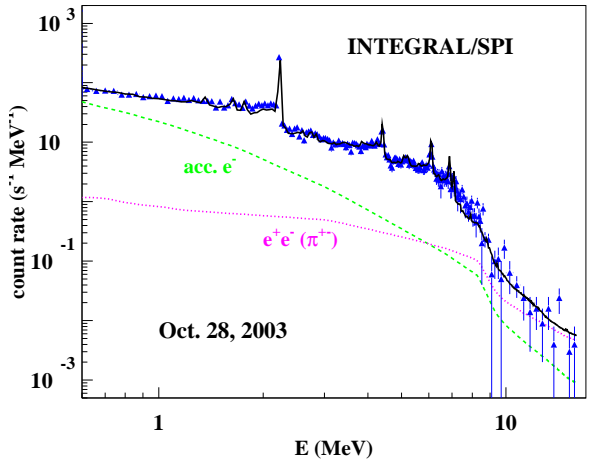}
  }
\caption{INTEGRAL/SPI  spectrum of the Oct. 28, 2003 X-class flare. Symbols present the observed dead-time corrected count rates in the Ge detectors during the most intense phase of the flare lasting about 10 min \cite{Octobs}. The full line shows the calculated spectrum with energetic proton and $\alpha$-particle properties extracted from the narrow line intensities, shapes and temporal evolution and otherwise impulsive solar-flare composition \cite{Octflare,Octmeas,Roorkee}. The bremsstrahlung contributions of accelerated electrons and pion-decay leptons are shown by the dashed and dotted lines, respectively.}

\label{f:flare}       
\end{figure}

Observations of solar-flare gamma-ray emission benefit since the launch of RHESSI \cite{RHESSI} in 2002 and INTEGRAL in 2003 from the high-resolution Ge detectors  that are onboard these spacecraft. RHESSI is dedicated to the observation of high-energy phenomena on the Sun and together with good energy resolution provides also imaging at the few arcsecond level. It has observed several tens of solar flares with gamma-ray emission (see e.g. \cite{Shi09}), obtaining  spectra from a few keV to typically 17 MeV. Another highlight was certainly the observation of slightly different interaction sites at the solar flare foot points for high-energy electrons and ions \cite{Hur06}. The gamma-ray spectrometer SPI onboard INTEGRAL \cite{SPI}, although it was not designed for solar-flare studies, observed the gamma-ray emission above $\sim$1 MeV of several strong X-class solar flares \cite{Octflare,Harris}. Analysis of the high-resolution spectra of both instruments required new  studies of gamma-ray line production in nuclear reactions, in particular detailed line-shape calculations. The weak-line quasi-continuum component already mentioned above is also important here. The gamma-ray spectrum of the Oct. 28, 2003 solar flare as observed by INTEGRAL/SPI is shown in Fig. \ref{f:flare}.
  
\vspace{0.5 cm}

\subsection{Sensitivity studies}
\label{s:sensiv}

Important reactions were mostly identified by direct inspection of a limited nuclear network.
For instance, if one is interested in solar $^7$Be core abundance and associated neutrino emission, 
the inspection of the pp III branch 
in hydrostatic hydrogen burning points to the importance of the $^3$He($\alpha,\gamma)^7$Be production
and $^7$Be(p,$\gamma)^8$B destruction reactions.  
It seems natural to extend this deduction to explosive hydrogen burning in novae and $^7$Be associated
gamma ray emission or BBN and the \sep\ problem. However, in these two cases, temperatures are high enough
to photodisintegrate efficiently $^8$B, blocking its  destruction by $^7$Be(p,$\gamma)^8$B for which the cross
section becomes inessential. 
This example points out the limitations of educated guessing in this domain.

It is hence essential to perform sensitivity studies, varying rates within network calculations, 
to find those reactions that influence the abundance of isotopes of interest.  
This leads, as we shall see, to find influential reactions that seem, at first sight, totally unrelated with
the observed effect.
One first step is to vary each reaction rate in the network, by a given factor,
or better within the rate uncertainties when available\footnote{Most often, rate uncertainties are not
available and would require much effort to evaluate. It is more convenient to use first a constant,  
overestimated, uncertainty factor and postpone the evaluations after the important reactions have been 
found.}
and calculate the effects on nucleosynthesis
or energy generation [see examples in \S~\ref{s:novae} (novae) and \ref{s:bbn} (BBN) or Refs. \cite{Bra12,Par13} (thermonuclear supernovae),
Iliadis et al. \cite{Ili11} ($^{26}$Al in massive stars), \cite{Nis14x} (r-process)].  
Nevertheless, in this way, one may overlook {\em chains} of reactions, 
whose uncertain cross section could, if changed in conjunction, cause an effect not observed   
when changing one of these reaction cross sections. 
To overcome these limitations, 
the second step in sensitivity analyses is to search for correlations between isotopic yields
and reaction rates and select those reactions which have the highest correlation coefficient
as was done by \cite{Par08} for X--ray bursts and by \cite{Coc14b} for BBN (see  \S~\ref{s:bbn}).

\subsubsection{Novae}
\label{s:novae}

It is interesting to start this discussion on sensitivity studies with nova 
nucleosynthesis, because a nova is the only explosive
astrophysical site for which all reaction rates could soon be derived from experimental data only \cite{JHI06}.
Novae are thermonuclear runaways occurring at the surface of a white dwarf by the accretion
of hydrogen rich matter from its companion in a close binary
system\cite{Sta72,Geh98,JH98,JH07,PJS14}. 
Material from the
white dwarf [$^{12}$C and $^{16}$O (CO nova) or $^{16}$O, $^{20}$Ne plus some 
Na, Mg and Al isotopes (ONe nova)] provides the seeds for the operation of the CNO cycle and
further nucleosynthesis. Novae are supposed to be the source of galactic $^{15}$N and 
$^{17}$O and to
contribute to the galactic chemical evolution of $^7$Li and $^{13}$C. In addition they
produce radioactive isotopes that could be detected by the gamma--ray emission that follow their $\beta^+$ decay to an excited state,
$^7$Be$(\beta^+)^7$Li*  (478~keV),  $^{22}$Na($\beta^+)^{22}$Ne* (1.275~MeV) and $^{26}$Al($\beta^+)^{26}$Mg* (1.809~MeV), while
positron annihilation ($\le$511~keV) only follow $^{18}$F$(\beta^+)^{18}$O decay.
Other constraints can come from a few silicon carbide (SiC) or graphite (C) presolar grains found in
some meteorites and that are of possible nova origin. 
Laboratory measured isotopic ratios, in particular those of C, N and Si can be compared to 
nova models \cite{JHA04}. 
The yields of these isotopes depend strongly on the hydrodynamics of the explosion but
also on nuclear reaction rates involving stable and radioactive nuclei.

The identification of important reactions for nova nucleosynthesis have followed the
progress in computing power. First explorations were done in one zone models with
constant temperature and density (e.g. Wiescher and Kettner \cite{WK82}), a crude 
approximation for  an explosive process, but which provided valuable insights for 
key reactions and rate uncertainties. More elaborate models \cite{Bof93} assumed an exponential
decrease of temperature and density and two zones to take into account the convection time scale,
essential during nova explosions or a semi-analytic model of the temperature and density 
profile time evolution \cite{Mac83,Coc95} to explore nova nucleosynthesis.
These nuclear rate sensitivity studies are now superseded by post processing studies \cite{Ili02} of 
temperature and density profiles, or even better using hydrodynamic simulations. 
Indeed, tests of the sensitivity to single reaction rate variation have been done using  the 1-D 
hydrocode (SHIVA) \cite{JH98} to evaluate the impact of nuclear uncertainties 
in the hot-pp chain\cite{Be96}, the hot-CNO cycle\cite{F00}, the Na--Mg--Al\cite{NaAl99} and
Si--Ar\cite{SiAr01} regions.
In this way, the temperature and density profiles, their time evolution, and the effect 
of convection time scale were taken into account. 
Because of the much longer computational time involved compared to BBN, no {\em systematic} 
sensitivity study  has been performed so far with 1-D hydrocode. 
Multidimensional hydrodynamic simulations are being successfully developed \cite{Cas11} but  
require drastically more computational power.  
To partially circumvent these limitations,  a systematic sensitivity study has been done using time 
dependent temperature and density profiles from 1-D hydrocode and post-processing nucleosynthesis
calculations \cite{Ili02}.
Following these sensitivity studies, the nuclear reaction rates whose uncertainties strongly affect the nova
nucleosynthesis have been identified. We summarize below some of these results, emphasizing those
which motivated  nuclear physics experiments that we describe in Sections \ref{s:f18p} and \ref{s:o17p}.

The hot--CNO cycle deserves special attention as it is the main source of energy for
both CO and ONe novae and is the source for the production of $^{13}$C, $^{15}$N, 
$^{17}$O (galactic chemical evolution) and $^{18}$F (gamma--ray astronomy).
The positrons produced in the $\beta^+$ decay of $^{18}$F annihilate and are the dominant 
source of gamma rays during the first hours of a nova explosion\cite{GG98}.  
Through a series of hydrodynamical calculations, using available \cite{NACRE} or evaluated upper 
and lower limits of reaction rates, major nuclear uncertainties on the
production of $^{17}$O and $^{18}$F were pointed out \cite{F00}.  
They correspond to the $^{18}$F(p,$\alpha)^{15}$O, $^{17}$O(p,$\alpha)^{14}$N, 
$^{17}$O(p,$\gamma)^{18}$F and $^{18}$F(p,$\gamma)^{19}$Ne reaction rates,
in decreasing order of importance. 
The $^{17}$O(p,$\gamma)^{18}$F reaction leads to the 
formation of $^{18}$F from the $^{16}$O seed nuclei through the
$^{16}$O(p,$\gamma)^{17}$F($\beta^+)^{17}$O(p,$\gamma)^{18}$F chain while  $^{18}$F(p,$\alpha)^{15}$O and 
$^{17}$O(p,$\alpha)^{14}$N divert the flow reducing both the $^{18}$F and $^{17}$O yields.
The proton capture reaction cross sections on $^{17}$O
can nowadays be considered known to sufficient precision at nova energies, in particular
thanks to the decisive breakthrough made in Orsay and TUNL (\S~\ref{s:o17p}). On the contrary, those involving the radioactive
$^{18}$F still suffer from significant uncertainties associated with the $^{19}$Ne spectroscopy (\S~\ref{s:f18p}).
Even though some nuclear reaction rates are still uncertain, leaks from the CNO cycle
are negligible at novae temperatures. In particular, experimental data on
the $^{15}$O($\alpha,\gamma)^{19}$Ne and 
$^{19}$Ne(p,$\gamma)^{20}$Ne reactions rates \cite{rateval10b}, 
are now sufficiently known 
to rule out any significant nuclear flow out of the CNO cycle. Hence, production of heavier
elements depends on the presence of $^{20-22}$Ne, $^{23}$Na,  $^{24-26}$Mg and $^{27}$Al
in ONe white dwarfs.

The decay of $^{22}$Na ($\tau_{1/2}$ = 2.6 y) is followed by the emission of a 1.275 MeV $\gamma$--ray. 
Observations of this gamma--ray emission have  only provided upper limits that are compatible 
with model predictions. In ONe novae, $^{22}$Na comes from the transmutation of $^{20}$Ne, starting by the 
$^{20}$Ne(p,$\gamma)^{21}$Na reaction. Important reactions were identified \cite{NaAl99,Ili02} to be 
$^{21}$Na(p,$\gamma)^{22}$Mg and $^{22}$Na(p,$\gamma)^{23}$Mg, while 
photodisintegration of $^{22}$Mg that is important at nova temperatures, prevents
further processing.

The ground state of $^{26}$Al decays to $^{26}$Mg, which emits a 1.809~MeV
gamma ray. Due to its long lifetime of $\tau_{1/2}$=0.717 My, it can accumulate in the Galaxy.
Production of $^{26}$Al by novae (and AGB stars) is now considered to be subdominant compared
to sites such as massive stars (Wolf-Rayet phase and core-collapse supernovae).
However, it is important as its gamma ray emission has been observed by different satellites (\S~\ref{s:obs26}). 
For novae, the major nuclear uncertainties affecting its production were
identified to be the $^{25}$Al(p,$\gamma)^{26}$Si and  $^{26g.s.}$Al(p,$\gamma)^{27}$Si reactions\cite{NaAl99,Ili02}.
The $^{26}$Si isotope can either decay to the short lived 
228~keV isomeric level of $^{26}$Al or be destroyed by subsequent proton capture.
In either case, it bypasses the long lived ground 
state of $^{26}$Al. (At nova temperatures, the isomer and ground state in $^{26}$Al have to be
considered as separate species\cite{Therm}.)

No significant amount of elements beyond aluminum are normally found in the composition of 
white dwarfs. 
The production of ``heavy elements", i.e. from silicon to
argon, rely on the
nuclear flow out of the Mg-Al region through $^{28}$Si and subsequently 
through $^{30}$P whose 
relatively long lifetime ($\tau_{1/2}$= 2.5~m) can halt the flow unless the 
$^{30}$P(p,$\gamma)^{31}$S reaction is
fast enough \cite{SiAr01}. This reaction is also important to calculate the silicon isotopic ratios.
The results can 
be compared to the values measured for some presolar grains that may have a nova 
origin\cite{JHA04}.

This series of 1-D hydrodynamical calculations,  followed by post--processing works,
(see  Parikh, Jos\'e \& Sala, \cite{PJS14} for a review) have led to the identification of reactions
[in particular $^{17}$O(p,$\gamma)^{18}$F, $^{17}$O(p,$\alpha)^{14}$N, $^{18}$F(p,$\alpha)^{15}$O,
$^{21}$Na(p,$\gamma)^{22}$Mg, $^{22}$Na(p,$\gamma)^{23}$Mg,
$^{25}$Al(p,$\gamma)^{26}$Si  and $^{30}$P(p,$\gamma)^{31}$S], that deserved further 
experimental efforts. Much progress has been made (e.g. \S~\ref{s:o17p}) but work is still needed
concerning the $^{18}$F(p,$\alpha)^{15}$O (\S~\ref{s:f18p}), $^{25}$Al(p,$\gamma)^{26}$Si  and 
$^{30}$P(p,$\gamma)^{31}$S reactions\footnote{Following the ``Classical Novae in the Cosmos",
Nuclei in the Cosmos XIII satellite workshop held in ATOMKI, Debrecen,
a special issue of The European Physical Journal Plus will be devoted to the evaluation of the $^{30}$P(p,$\gamma)^{31}$S  
reaction rate.
}.

\subsubsection{Big bang nucleosynthesis}
\label{s:bbn}

It is interesting to discuss sensitivity studies in the context of big bang nucleosynthesis (BBN). Not because
it is the first nucleosynthetic process to take place but because, in its standard version,
all parameters of the model are fixed and the thermodynamic conditions (density and temperatures)
can be calculated exactly from known ``textbook" physics. In particular, compared to
stellar nucleosynthesis, there are no complications like stratification, mixing by convection or diffusion, 
and most reaction cross sections can be measured directly at the required energy, and
are not affected by electron screening. Furthemore, even with the largest network, computing
time is of the order of a fraction of a second, allowing extensive Monte Carlo calculations. 

\begin{figure}
\resizebox{0.45\textwidth}{!}{%
 \includegraphics{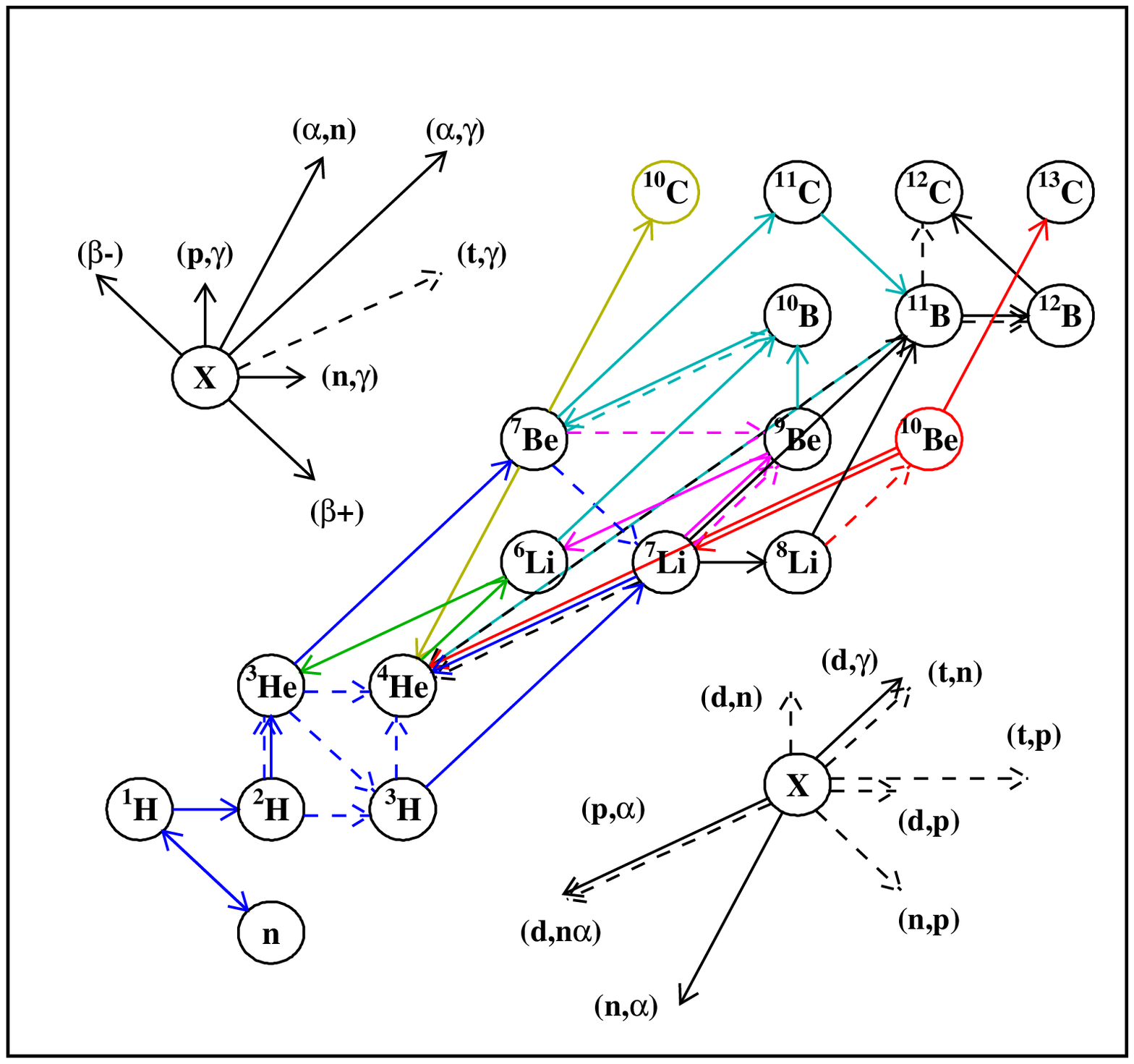}
}
\caption{(Color online) Nuclear network  of the most important reactions in BBN, up to\sep\ (blue), 
including \six\ (green), $^{10,11}$B (light blue), \neu\ (pink) and up to CNO (black and red).
The red arrows represents the newly found reactions that could affect CNO production.
The yellow arrows indicate the \bedp\ and \bery+\tro\ reactions that were considered as possible
extra \bery\ destruction mechanisms.}
\label{f:bbnet}       
\end{figure}

Table~\ref{t:sensib} lists the 12 main reaction for BBN up to \sep. They are also shown in Fig.~\ref{f:bbnet}.
The table also displays the sensitivity of the calculated abundances, ($Y_i$ with $i$ = \hli)  with respect to a change in the 12 
reaction rates by a constant arbitrary factor (1.15), defined as ${\partial\log}Y/{\partial\log}<{\sigma}v>$   \cite{CV10} (see also Refs.~\cite{Cyb04,Den07}).
It shows that some of these reactions (e.g. $^3$H($\alpha,\gamma)^7$Li) are not important anymore
because at CMB deduced baryonic density, \sep\ is produced through $^7$Be that decays to \sep\ after the end of BBN. 
Naturally, $^7$Be yield is sensitive to the  $^3$He($\alpha,\gamma)^7$Be (production) and $^7$Be(n,p)$^7$Li (destruction)
reaction rates, but unexpectedly, the highest sensitivity is to the $^1$H(n,$\gamma)^2$H rate! This is a first example of the
usefulness of sensitivity studies to identify influential reactions, otherwise unexpected. We will not discuss any further 
the uncertainties associated with these main reactions, as they are small and cannot, for sure, solve the lithium problem.
We just point out that if the new observations of deuterium are confirmed, high precision ($\sim$1\%) will be required on 
the main cross sections involved in Deuterium destruction \cite{DiV14}. The most recent measurements concerning
these cross sections have been done directly at LUNA ($^2$H(p,$\gamma)^3$He) \cite{Cas02} and TUNL 
($^2$H(d,n)$^3$He and $^2$H(d,p)$^3$H) \cite{Leo06}. These latter were very recently,  determined using the Trojan Horse 
Method \cite{Tum14} and theoretically through ab-initio calculations \cite{Ara11}. 

\begin{table}
\caption{\label{t:sensib}Abundance sensitivity \cite{CV10}: ${\partial\log}Y/{\partial\log}<{\sigma}v>$ at CMB deduced baryonic density.} 
\label{tab:1}       
\begin{center}
\begin{tabular}{ccccc}
\hline\noalign{\smallskip}
\noalign{\smallskip}\hline\noalign{\smallskip}
Reaction &  $^4$He & D & $^3$He & $^7$Li  \\ 
\noalign{\smallskip}\hline\noalign{\smallskip}
n$\leftrightarrow$p & -0.73 & 0.42 & 0.15 & 0.40\\ 	
$^1$H(n,$\gamma)^2$H & 0.005 & -0.20 & 0.08 & 1.33\\ 
$^2$H(p,$\gamma)^3$He & $<$0.001	& -0.32 & 0.37 & 0.57  \\
$^2$H(d,n)$^3$He & 0.006 & -0.54 & 0.21 & 0.69   \\
$^2$H(d,p)$^3$H    & 0.005 & -0.46 & -0.26 & 0.05   \\
$^3$H(d,n)$^4$He & $<$0.001 & 0 & -0.01 & -0.02   \\
$^3$H($\alpha,\gamma)^7$Li & $<$0.001 & 0 & 0 & 0.03 \\
$^3$He(n,p)$^3$H & $<$0.001 & 0.02 & -0.17 & -0.27  \\	
$^3$He(d,p)$^4$He & $<$0.001 & 0.01 & -0.75 & -0.75  \\
$^3$He($\alpha,\gamma)^7$Be &  $<$0.001 & 0 & 0 & 0.97  \\
$^7$Li(p,$\alpha)^4$He & $<$0.001 & 0 & 0 & -0.05  \\
$^7$Be(n,p)$^7$Li & $<$0.001 & 0 & 0& -0.71  \\	
\noalign{\smallskip}\hline
\end{tabular}
\end{center}
\end{table}

The reactions in Table~\ref{t:sensib} represent the near minimum network needed for BBN calculations up to \sep.
It is interesting to extend it to incorporate a priori negligible reactions, but whose rates may not be firmly established,
and reactions involved in the production of  sub-dominant isotopes. 
Systematic sensitivity studies have been performed by varying  one rate at a time by given factors to search for
a solution to the lithium problem, and the path for  \six, \neu, \onz\ and CNO nucleosynthesis. 
For instance, it was found  \cite{Coc04,be7dp} that the most promising reaction for \sep\ (\bery) destruction was
\bedp,  and to a lesser extent \bery+\tro\ channels, whose rates were unknown at BBN energy. 
It triggered several experimental and theoretical studies (see \S~\ref{s:be}).  The most influential reaction
for the production \cite{Ioc07,CNO12} of  sub-dominant isotopes (\six\ to CNO), displayed in Fig.~\ref{f:bbnet},
were obtained in the same way. Surprisingly, it was found in that study that the $^7$Li(d,n)2\qua\ reaction
rate has no impact on  \sep\ nor \deu\ {\em final} abundance but does influence the CNO (\carb) final one!  
The explanation is that even though the  \sep\  {\em final abundance is left unchanged}, 
the \sep\ abundance reaches a peak value at $t\approx$200~s (Fig.~15 in ref.~\cite{CNO12}),
before being destroyed efficiently by $the ^7$Li(p,$\alpha$)\qua\ reaction.
The effect of an increased $^7$Li(d,n)2\qua\ reaction rate is to lower this peak value, with as a consequence,
a reduced  feeding of the chains of reactions $^7$Li(n,$\gamma)^8$Li($\alpha$,n)$^{11}$B  followed by d or n 
captures on \onz\ that lead to CNO isotopes.
(Note however that the uncertainty on the $^7$Li(d,n)2\qua\ reaction rate \cite{Boy93} is small
enough not to influence CNO production.)

\begin{figure}
\resizebox{0.45\textwidth}{!}{%
  \includegraphics{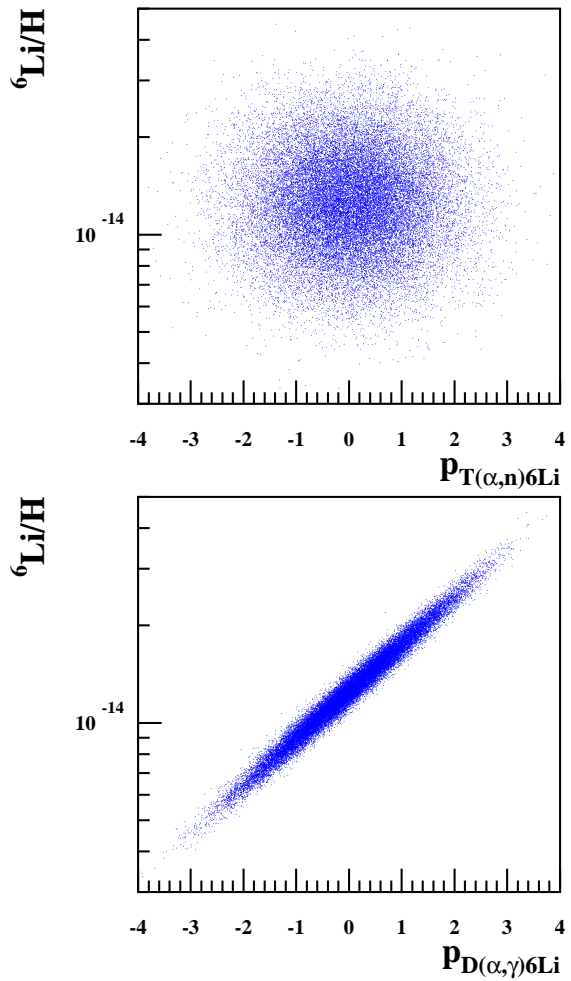}
}
\caption{Scatter plots of \six\ yields versus random enhancement factors applied to
reaction rates in the context of BBN showing no ($C_{^6\mathrm{Li,T}(\alpha,\mathrm{n})^6\mathrm{Li}}\approx0$, top panel) or strong 
(($C_{^6\mathrm{Li,D}(\alpha,\gamma)^6\mathrm{Li}}\approx1$, bottom panel) correlation to respectively T($\alpha$,n)$^6$Li and \zdag\ reactions 
(data from Ref.~\cite{Coc14b}).}
\label{f:licor}       
\end{figure}

\begin{figure}
\resizebox{0.45\textwidth}{!}{%
  \includegraphics{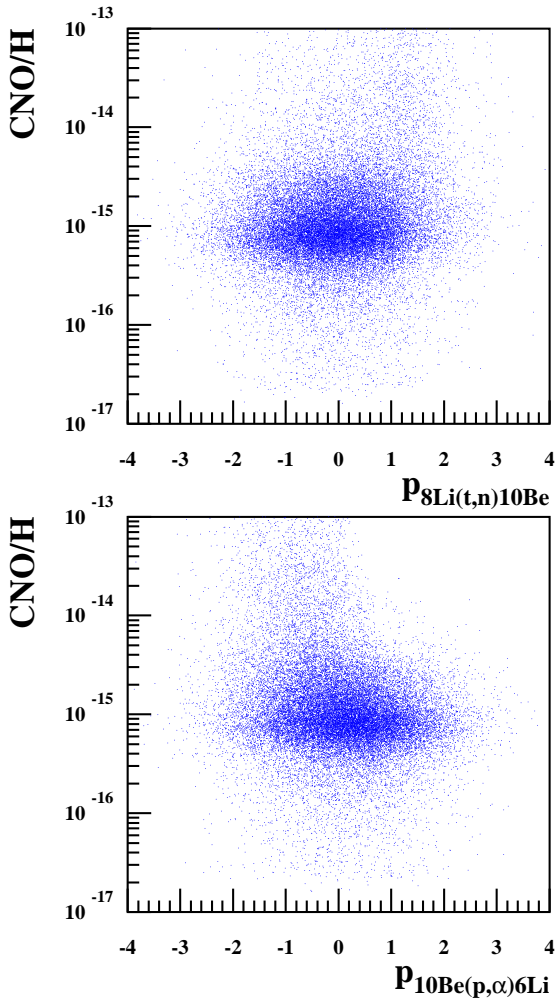}
}
\caption{Scatter plots of CNO/H yields versus random enhancement factors ($p_k$) applied to
reaction rates showing weak correlation  with respectively $^8$Li(t,n)$^{10}$Be (top panel)  and 
$^{10}$Be(p,$\alpha)^6$Li (bottom panel) reactions 
(data from Ref.~\cite{Coc14b}).}
\label{f:cnocor}       
\end{figure}

It has recently been recognized that traditional sensitivity studies, in which only one reaction is varied while the others are held constant
as discussed above cannot properly address all the important correlations between rate uncertainties and nucleosynthetic predictions. 
Sensitivity studies can be improved by performing Monte Carlo calculation, and searching for such correlations  \cite{Par08}. 
To start with, we follow the prescription of \cite{STARLIB}. 
Namely the reaction rates $x_k$ $\equiv{N_A}\langle\sigma{v}\rangle_k$, (with $k$ being the index of the reaction), 
are assumed to follow a lognormal distribution:
\begin{equation}
x_k(T)=\exp\left(\mu_k(T)+p_k\sigma_k(T)\right)
\label{q:ln}
\end{equation}
where  $p_k$ is sampled according to a {\em normal} distribution of mean 0 and variance 1 (Eq.~(22)  of \cite{STARLIB}).
$\mu_k$ and $\sigma_k$ determine the location of the distribution and its width which are tabulated as a function of $T$.
First, by taking the  quantiles of the Monte Carlo calculated distributions of final isotopic abundances one obtains, not
only their median values but also the associated confidence interval.
Second, the (Pearson's) correlation coefficient (e.g. in Ref.~\cite{Feller}) between isotopic abundance $y_{j}$  and reaction rate random enhancement factors ($p_k$ in Eq.~\ref{q:ln}) can be calculated as:
\begin{equation}
C_{j,k}=\frac{Cov(y_j,p_k)}{\sqrt{Var(y_j)Var(p_k)}}.
\label{q:cor}
\end{equation}
Illustrative examples are given in Fig.~\ref{f:licor} and \ref{f:cnocor}. They refer to \six, whose nucleosynthesis is simple:
it is produced by \zdag\ (\S~\ref{s:breakup}) and destroyed by $^6$Li(p,$\alpha)^3$He (Fig.~\ref{f:bbnet}), other reaction
playing a negligible role (e.g. Ref.~\cite{Van00}). Hence, as anticipated, the correlation coefficient is $C_{^6\mathrm{Li,D}(\alpha,\gamma)^6\mathrm{Li}}\approx1$
for the production reaction and $C_{^6\mathrm{Li,T}(\alpha,\mathrm{n})^6\mathrm{Li}}\approx0$ for a reaction of negligible contribution
to \six\ production. 
More interesting examples are shown in Fig.~\ref{f:cnocor}: a weak [anti--]correlation ($C\approx$[-0.25]+0.20) 
between the CNO/H and [$^{10}$Be(p,$\alpha)^6$Li ] $^8$Li(t,n)$^{10}$Be  
reaction rates. These two (among a total of 6, all related to $^{10}$Be) {\em were not  previously identified in simple sensitivity studies} \cite{CNO12}.
For this chain of reactions pivoting around $^{10}$Be to be efficient, higher rates for  $^{10}$Be producing reactions,
in conjunction with lower rates for  $^{10}$Be destruction reactions, are needed. This finding could not be obtained when varying one reaction at a time.

Finally, after considering all, 2$^6$ combinations of high and low rates, four previously 
overlooked reactions are found to be important (in red in Fig.~\ref{f:bbnet}) and could lead to
a significant increase of CNO production in BBN \cite{Coc14b}.

\begin{figure}
\resizebox{0.45\textwidth}{!}{%
  \includegraphics{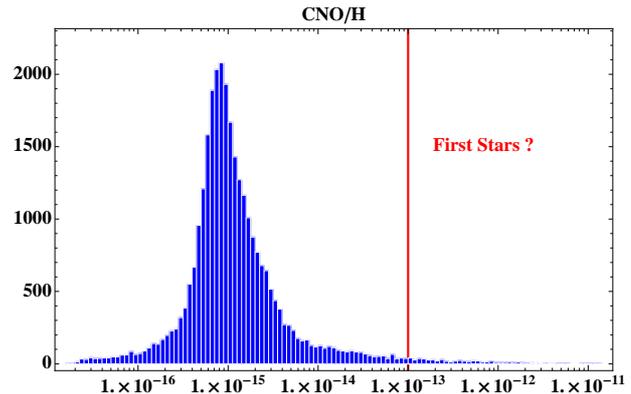}
}
\caption{CNO/H distribution as a result of a Monte Carlo calculation sampling
$\approx$ 400 reaction rates. About 2\% of the events lie to the right of the vertical
line that corresponds to the minimum value that can affect the evolution of some first stars   
(data from Ref.~\cite{Coc14b}).}
\label{f:cnohist}       
\end{figure}

Last, but not least, Monte Carlo calculations allow  extracting confidence limits from
the distribution of calculated abundances values. For instance Fig.~\ref{f:cnohist}
displays the distribution of CNO/H values that allows to extract a 68\% confidence interval
of $\rm{CNO/H}=(0.96^{+1.89}_{-0.47})\times10^{-15}$ \cite{Coc14b,CV14}, essentially related to
the reactions that were identified and whose rates were re-evaluated \cite{CNO12}. 
However, the right tail of the distribution, extends to values way off the above interval:
this is the effect of the newly identified reactions, from the analysis of correlations, 
when their rates happen to be simultaneously and favorably changed in the Monte Carlo 
sampling.

In conclusion, we have seen that, in BBN,  {\em systematic} sensitivity studies, changing one reaction rate at
a time, can detect important reactions that have unexpected effects. For instance, the $^1$H(n,$\gamma)^2$H 
reaction affects \sep\ production. On the contrary,  while the $^7$Li(d,n)2\qua\ reaction does not influence the
\sep\ or \qua\ abundances, it affects \carb\ production.  
The analyses of correlations between abundances and reaction rates,  obtained by Monte Carlo calculations,
can allow discovering other essential reactions. 
Obviously, BBN is a favorable candidate for such studies as the {\em standard} model has no more
free parameters and calculations are fast, but it would be desirable to extend these kind of analyses to other
nucleosynthesis sites. 
Besides BBN, Monte Carlo sensitivity studies have been essentially achieved for  X-ray bursts \cite{Par08}, 
but with only minor differences with simpler analyses, and for novae  in a pioneering study \cite{Hix03}.
It is likely, that such studies will require the development of new tools, to identify
correlations between abundances and rates beyond the simple calculation of correlation coefficients.
Figure~6 in Ref.~\cite{astrostat} (a review on statistical methods in nuclear astrophysics) display a few
illustrative examples.


\section{Direct measurements}
\label{s:direct}

One of the main characteristics of the nuclear reactions involved in primordial and stellar nucleosynthesis is the low 
energy where they occur (between a few keV to a few MeV).  When they involve charged particles,
because of the Coulomb barrier effect (\S~\ref{s:rates}), the cross sections, 
can be very small, ranging from hundreds of pico--barn to femto--barn. 
These features make 
the direct measurements at stellar energies very challenging since the expected count rates decrease 
dramatically with decreasing energy. For such measurements, it is necessary to use high beam currents together with targets that can withstand them.

For some reactions, the expected experimental yields are so small that measurements become hopeless unless the background
produced by the environment and the beam can be reduced to acceptable levels. Consequently, progress in direct measurements 
comes essentially from underground laboratories e.g.  LUNA \cite{LUNA}. Exceptions to this are in the domain of explosive burning 
(e.g. novae), where the cross sections are much larger, but often require radioactive ion beams \cite{RIB} of adapted nature, energy and intensity. 
As examples for such challenging experiments, we describe in the following recent results that were obtained in 
$^{17}$O(p,$\alpha$), $^{17}$O(p,$\gamma$) and $^{18}$F(p,$\alpha$)$^{15 }$O studies. 
The $^{18}$F(p,$\alpha$)$^{15 }$O measurements require a $^{18}$F radioactive beam, only available since the mid-90's.
The importance of the $^{17}$O(p,$\alpha$) and $^{17}$O(p,$\gamma$) cross sections for novae was overlooked, 
before sensitivity studies were made.

\subsection{The $^{18}$F+p reactions}
\label{s:f18p}

Due to the unknown contributions of low energy resonances,
the $^{18}$F(p,$\alpha)^{15}$O reaction was recognized as the
main source of uncertainty for the production of  $^{18}$F in novae (\S~\ref{s:novae}). 
The source of uncertainties comes from the poorly known spectroscopy of the $^{19}$Ne
compound nucleus, as compared to its mirror $^{19}$F displaying a high level density, 
that makes the identification of analog states challenging \cite{Utk98,Nes07}.
Only two resonances have their properties (strength, partial widths, spin and parity) unambiguously 
determined by direct measurements \cite{Coz95,Gra97,Bar02}, at  LLN and ORNL. 
They are located at  $E_r$=330 keV ($J^\pi$=3/2$^-$) and
665 keV ($J^\pi$=3/2$^+$ ). In spite of its high energy,
the second resonance plays an important role due to its large width (44~keV), so that its tail gives a significant
contribution in the range of interest for novae (50--350~keV). 
Possible interferences with other 3/2$^+$ lower lying resonances remain
a major source of uncertainty.
It is hence, extremely important to determine whether they are constructive or destructive by direct 
off--resonance measurements \cite{Bar02,Cha06,f18pa,Bee11} performed at energies down to 250~keV,
and
compared to R-matrix calculations. 
The comparison with the spectrum in $^{19}$F suggests that several levels are missing in the 
$^{19}$Ne spectrum \cite{Nes07}, while spins and parities of the observed levels are still
a matter of debate \cite{Utk98,dp05,Ade12,Lai13}.
In particular, it is essential to identify the low energy 3/2$^+$ resonances that can interfere with the  
665 keV one \cite{f18pa,Cha06}; this issue is not yet settled.

Rather than discussing the important but complicated spectroscopy of levels close to threshold, 
we will concentrate here on two levels 
of interest that were first predicted by theory, then possibly experimentally confirmed.   
These are two 1/2$^+$ ($\ell$=0) broad levels, predicted by microscopic\cite{Duf07} calculations,
one at $\approx$1 MeV above threshold and the other below. If they exist they would lead to a significant
contribution in the relevant energy range, especially the lowest energy one, above the threshold.
Indeed, experiments performed at LLN \cite{Dal09} and GANIL \cite{Mou12}  support the presence of such a 
broad state at  $E_r\approx$1.3~MeV, motivating the search for its subthreshold predicted counterpart.    

Direct measurements of $^{18}$F(p,$\alpha$)$^{15}$O and $^{18}$F(p,p)$^{18}$F cross sections were 
carried out at the GANIL-SPIRAL facility \cite{Mou12}. The $^{18}$F radioactive ions were produced by bombarding 
a thick carbon target by a 95 MeV.A primary beam of  $^{20}$Ne. The $^{18}$F ions were then ionized in an ECR ion source and postaccelerated with the
CIME cyclotron to an energy of 3.924 MeV.A . The obtained beam intensity was about 2$\times$10$^4$~pps. 
The $^{18}$F beam was degraded to an energy of 1.7 MeV.A using a 5.5$\pm$0.3 $\mu$m gold foil and then sent to a 
CH$_2$ polymer target of 55$\pm$4 $\mu$m thickness. This thickness was enough to stop the beam and allow the light ions to 
escape. The emitted protons and $\alpha$ particles from the $^{18}$F(p,p)$^{18}$F and $^{18}$F(p,$\alpha$)$^{15}$O 
reactions respectively, as well as the emitted $^{12}$C ions from  $^{18}$F($^{12}$C,$^{12}$C)$^{18}$F scattering reaction 
were detected in a 50 mm $\times$ 50 mm double--sided silicon detector located downstream of the target. 
The identification of the different emitted particles was achieved thanks to the energy versus time of flight measurement  \cite{Mou12}.

\begin{figure}
\resizebox{0.45\textwidth}{0.27\textheight}{%
  \includegraphics{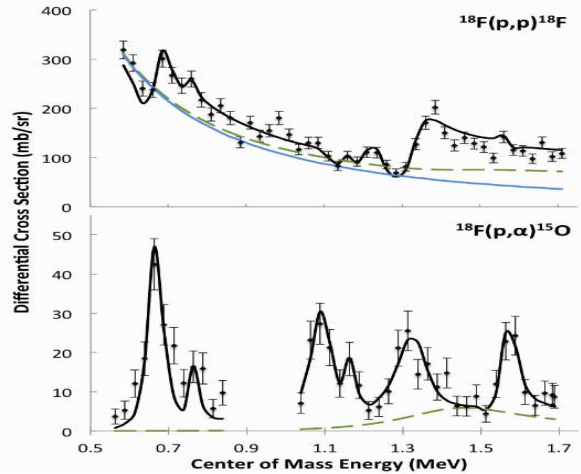}}
\caption{Differential cross sections of $^{18}$F(p,p)$^{18}$F and $^{18}$F(p,$\alpha$)$^{15}$O reactions as a 
function of center-of-mass energy. The curves represent R--matrix calculations, with (in dashed green) the contribution of the 1/2$^+$ broad level. 
[Reprinted figure with permission from D. J. Mountford, A. St J. Murphy,  et al., \zprc\  {\bf 85} 022801(R) (2012) \cite{Mou12}. 
Copyright 2012 by the American Physical Society.]}
\label{f:18Fspectra}       
\end{figure}

The measured excitation functions in center of mass energy for the $^{18}$F(p,$\alpha$)$^{15}$O  and  $^{18}$F(p,p)$^{18}$F reactions 
are displayed in Fig.~\ref{f:18Fspectra}. 
Several resonant structures are observed and the most important one, at 655 keV, belongs  to the 
well known 7076 keV J$^\pi$=3/2$^+$ state in $^{19}$Ne. 
Seven resonances in total were identified and their parameters were deduced from a $\chi^2$ minimization R-matrix fit of 
the measured excitation functions of both the $^{18}$F(p,p)$^{18}$F and $^{18}$F(p,$\alpha$)$^{15}$O reactions 
\cite{Mou12}. An overall agreement was found  between the derived parameters 
and associated error bars of the populated states in $^{19}$Ne and those deduced in previous measurements. 
Details of the analysis, error bars estimation and comparison of the obtained resonance parameters 
with previous results are given in  \cite{Mou12}. 

The strong structure observed in Fig.~\ref{f:18Fspectra} at E$_{c.m}$=1.2-1.4 MeV is well described by the two previously observed states at 
7624 keV J$^\pi$=3/2$^-$ and 7748 keV J$^\pi$= 3/2$^+$ of $^{19}$Ne when including an additional 
broad resonance at E$_{c.m}$=1455 keV displayed as the dashed green line in Fig.~\ref{f:18Fspectra}. Without this state, the R-matrix fit of
the whole data is substantially worse and the deduced parameters for other populated resonances deviate considerably from literature values. 
With the inclusion of the broad resonance, the best fit to the data corresponds to J$^\pi$=1/2$^+$ state at an excitation energy 
of 7870$\pm$40 keV with a partial proton width of 55$\pm$12 keV and an $\alpha$-partial width of 347$\pm$92 keV. These results are in agreement with observed state in \cite{Dal09} and the predicted state by Dufour and Descouvemont \cite{Duf07}. 
Hence, the presence of this broad state above the threshold in the measured data supports the prediction by Dufour and Descouvemont \cite{Duf07} of an additional subthreshold broad state. The latter can contribute substantially at novae temperatures, enhancing thus the rate of $^{18}$F destruction. This will lead to less $^{18}$F in nova ejecta and consequently to a reduced detectability distance.

This reaction, with the unsettled questions of the $^{19}$Ne level scheme around the $^{18}$F+p threshold and interferences
is still the subject of intense experimental investigation.
Note also that it was one of the first to be investigated by transfer reaction (\S~\ref{s:trans}) with {\em a radioactive ion beam}
\cite{dp03,Ade12}.

\subsection{The $^{17}$O+p reactions}
\label{s:o17p}

The $^{17}$O(p,$\alpha)^{14}$N and $^{17}$O(p,$\gamma)^{18}$F reactions were also identified
as sources of uncertainties for the production of $^{18}$F and $^{17}$O in explosive hydrogen burning (novae \S~\ref{s:novae}) and 
oxygen isotopic ratios following hydrostatic hydrogen burning (e.g. red giant stars).

At low energy,  in hydrostatic hydrogen burning,
it is the 65~keV resonance which is important. Its strength, in the (p,$\alpha$) channel, $\omega\gamma=4.7\pm0.8$ neV \cite{Han99}, 
was first directly measured at TUNL by Blackmon et al. \cite{Bla95}. Due to the limited signal statistics and high background,
a more elaborate statistical analysis was performed \cite{Han99} that led to the value quoted above. 
Preliminary results from a direct LUNA measurement \cite{Bru14} confirms this result.   
Note that this resonance has been also investigated by indirect techniques, previously by DWBA in Orsay \cite{Lan89} and 
later with the Trojan Horse Method in Catania \cite{Ser10}, with results in agreement with the direct measurements.

Before decisive progress was made in Orsay and LENA, the uncertainty on these rates at explosive hydrogen burning temperatures
came from the resonance, at that time unobserved, around 190~keV. This introduced an additional factor 
of $\sim$10 uncertainty on the production of $^{18}$F \cite{F00}. 
The NACRE rates were based on experimental data for this resonance which were found to be inaccurate (energy and total width).
These inaccuracies were discovered during experiments that were performed at the PAPAP (Orsay) \cite{PAPAP}, CENBG (Bordeaux) and LENA (North Carolina) small 
accelerators. A decisive result was obtained from a DSAM experiment performed at the 4 MV Van de Graaff accelerator of the 
CENBG laboratory using  $^{14}$N($\alpha,\gamma)^{18}$O to feed the corresponding level in $^{18}$O.
They found $i$) that
the upper limit for its lifetime $<$~2.6~fs \cite{Cha05} was much smaller than the value (15$\pm$10~fs \cite{Rol73}) previously
assumed, and, $ii$) that the resonance energy needed to be re-evaluated to E$^{lab}_{R}$=194.1$\pm$0.6 keV  \cite{Cha05}. 
Measurement of the $^{17}$O(p,$\gamma$)$^{18}$F and \cite{Fox04,Fox05,Cha07} and  $^{17}$O(p,$\alpha$)$^{14}$N 
\cite{Cha05,Cha07,New07}  resonance strengths soon followed, both at Orsay and LENA.

 \begin{figure}
\resizebox{0.45\textwidth}{!}{%
  \includegraphics{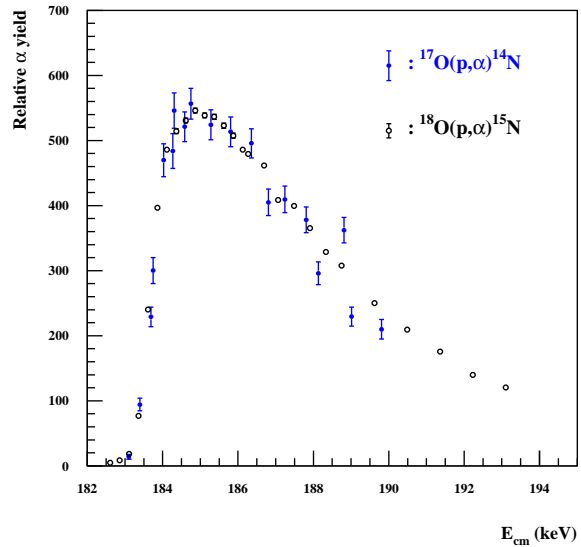}
}
\caption{Excitation functions for the new resonance at E$_R^{lab}$=194.1 keV in the $^{17}$O(p,$\alpha$)$^{14}$N reaction and the 
well-known $^{18}$O(p,$\alpha$)$^{15}$N  resonance at E$_R^{lab}$=150.9 keV. Data for the latter resonance were normalized and 
shifted in energy to be compared with those obtained for $^{17}$O(p,$\alpha$)$^{14}$N reaction.}
 \label{f:spectra}       
\end{figure}

In particular, experiments were conducted at the electrostatic accelerator PAPAP \cite{PAPAP} of the CSNSM laboratory  
using the thick target yield direct technique and the activation method \cite{Cha05,Cha07}. 
The $^{17}$O(p,$\alpha$)$^{14}$N  measurement consisted in sending a proton beam of 60-90 $\mu$A intensity on a water--cooled $^{17}$O target implanted in a thick 0.3 mm Ta backing. The emitted $\alpha$-particles were detected with 4 silicon detectors of 3 cm$^2$ active area placed at 14 cm from the target at four different laboratory angles. 
The strength of the $^{17}$O(p,$\alpha$)$^{14}$N  resonance was determined relative to the well-known resonance at 
 E$^{lab}_{R}$=150.9 keV in $^{18}$O(p,$\alpha$)$^{15}$N which was also measured at the PAPAP accelerator.
 The obtained value $\omega\gamma_{p\alpha}$=1.6$\pm$0.2 meV  \cite{Cha05} was found to be well above the upper limit ($\leq$0.42 meV\cite{Rol73}) used in the NACRE
 evaluation \cite{NACRE}.
 This result was swiftly confirmed by Moazen et al. \cite{Moa07} and Newton et al. \cite{New07}. 
 The measured excitation functions in laboratory energy for the new resonance at E$_R^{lab}$=194.1 keV in the $^{17}$O(p,$\alpha$)$^{14}$N reaction and the well-known $^{18}$O(p,$\alpha$)$^{15}$N  resonance at E$_R^{lab}$=150.9 keV are displayed  in Fig.~\ref{f:spectra}.

The strength of the $^{17}$O(p,$\gamma$)$^{18}$F resonance at E$_R^{Lab}$=194.1 keV  
was obtained from activation measurements  performed  on two $^{17}$O targets 
irradiated at two proton incident energies  E$_p$=196.5 keV (on resonance) and E$_p$=192.7 keV (off resonance). The $\beta^+$ activity of the produced $^{18}$F was measured with two Ge detectors placed on opposite sides to detect in coincidence the two 511~keV $\gamma$-rays 
coming from positron-electron annihilation. 
The total number of $^{18}$F nuclei produced at high beam energy was found of about 
one order of magnitude larger than at 192.7 keV. This is most probably due to the excitation of the $^{17}$O(p,$\gamma$)$^{18}$F 
resonance at E$_R^{lab}$=194.1 keV while the $^{18}$F production at E$_p$=192.7 keV is due to interference between the direct capture (DC) process and the low-energy tail of the studied resonance. To extract the resonance strength at  E$_R^{lab}$= 194.1 keV, a small 
contribution coming from the DC process (4.3$\pm$2.2)\% \cite{Cha07} was subtracted from the $^{18}$F total production at E$_p$=196.5 keV 
and a small correction (4$\pm$2)$\%$ taking into account of a possible backscattering of $^{18}$F from the target was also applied. 
From the weighted mean of the (p,$\alpha$) and (p,$\gamma$) measurements at E$_p$=196.5 keV, a value of 717$\pm$60 was obtained 
for $\omega\gamma_{p\alpha}$/ $\omega\gamma_{p\gamma}$.  
The resulting resonance strength $\omega\gamma_{p\gamma}$ = 2.2$\pm$0.4~$\mu$eV \cite{Cha07} was found to be about a factor two larger than the value 
measured by Fox et al. \cite{Fox04}: $\omega\gamma_{p\gamma}$ = 1.2$\pm$0.2 ~$\mu$eV .  

After these measurements, performed almost simultaneously at LENA\cite{Fox04,Fox05} and in Orsay\cite{Cha05,Cha07}, subsequent experiments have confirmed 
and improved these results (see \cite{rateval10b} and references therein) and  \cite{DiL14,Kon12,Hag12,Sco12}.
In particular, the inconsistency on the strength between LENA and Orsay experiments  has now been solved thanks to the high precision 
measurement made at LUNA \cite{DiL14} ($\omega\gamma_{p\gamma}$ = 1.67$\pm$0.12 ~$\mu$eV). 
Indeed, thanks to the strong reduction  of the background induced by cosmic rays, the sensitivity in LUNA was greatly improved allowing the observation of several additional 
$\gamma$-ray transitions, contrary to previous works where only two primary transitions could be observed, which according to LUNA's work yield about 65\% of the total strength. This explains the fact that  the Fox et al. \cite{Fox04} result for the strength of the resonance at E$_{CM}$=183~keV is 28\% lower than LUNA's  and in disagreement with the Chafa et al. 
\cite{Cha07} result even when considering the large error bars of the latter. 
Concerning the direct capture component, the results obtained  with DRAGON in TRIUMF \cite{Hag12}, S$_{DC}$=5.3$\pm$0.8 keV-b, and in Notre-Dame \cite{Kon12}, S$_{DC}$=4.9$\pm$1.1 keV-b were found to be higher than Fox et al.  \cite{Fox04} (S$_{DC}$=3.74$\pm$1.68 keV.b) probably due to non observed transitions in \cite{Fox04} and hence to low evaluation of the resonances strengths at E$_{CM}$=557 keV and 677 keV. They are also in good agreement with the precise value obtained by LUNA  \cite{Sco12}, S$_{DC}$=4.4$\pm$0.4 keV-b. The comparison to Chafa et al. results is not conclusive due to the very large uncertainty of the S$_{DC}$ evaluation, S$_{DC}$=6.2$\pm$3.1 keV-b.  

\begin{figure}
\resizebox{0.45\textwidth}{!}{%
  \includegraphics{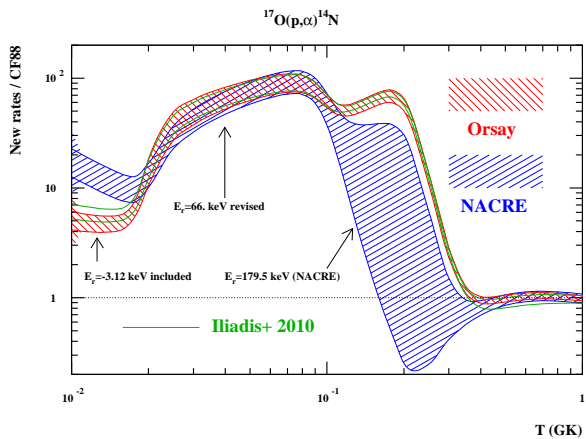}
}
\caption{Relative uncertainty on the $^{17}$O(p,$\alpha)^{14}$N reaction rate obtained in Orsay \cite{Cha05,Cha07},
compared to previous \cite{NACRE} and present \cite{rateval10b} evaluations. (Rates are normalised to CF88.)}
\label{f:o17pa}
\end{figure}

\begin{figure}
\resizebox{0.45\textwidth}{!}{%
  \includegraphics{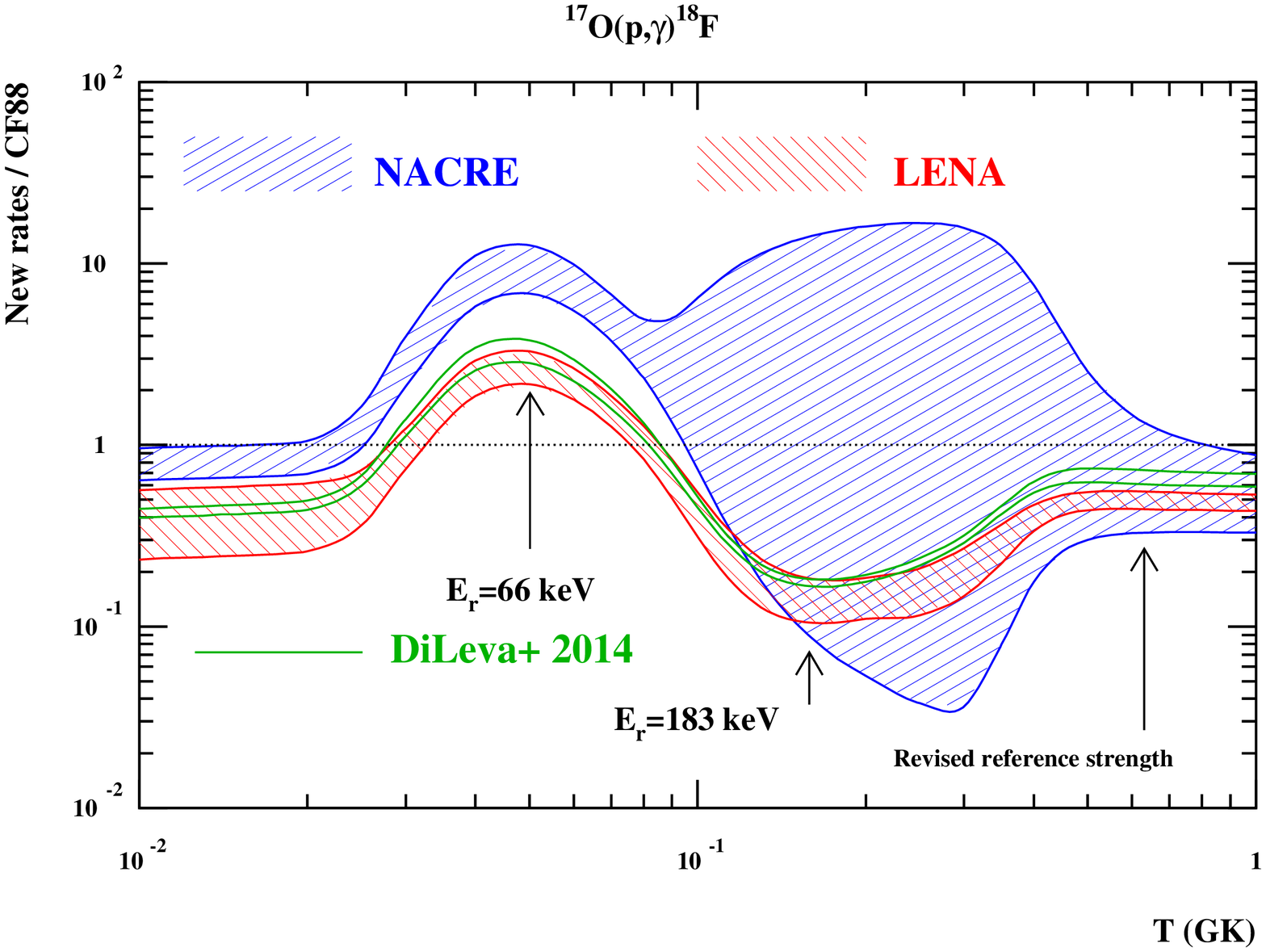}
}
\caption{Relative uncertainty on the $^{17}$O(p,$\gamma)^{18}$F reaction rate obtained at LENA \cite{Fox04,Fox05}, compared to previous \cite{NACRE} evaluation
and newest rate from Di Leva et al. \cite{DiL14}. (Rates are normalised to CF88.)}
\label{f:o17pg}
\end{figure}

Thermonuclear rates of the $^{17}$O+p reactions were calculated \cite{rateval10b,DiL14} using the present results by the Monte Carlo technique \cite{rateval10a}.   
The new calculated rates 
reduce the previous uncertainties by order of magnitudes at temperatures between 0.1-0.4 GK. The new evaluated uncertainties are reasonably small,
 in particular for the $^{17}$O(p,$\alpha$) reaction rate which is now well established. 
Figures~\ref{f:o17pa} and ~\ref{f:o17pg} display the evolution of 
the $^{17}$O+p rates since  the CF88 \cite{CF88} and  NACRE \cite{NACRE} compilations, until the last
evaluation \cite{rateval10b,DiL14}, with  the Orsay or LENA rates \cite{Fox04,Cha05,Fox05,Cha07}.     
They show that both rates are now known with sufficiently good accuracy for
nova applications.

\section{Indirect methods}
\label{s:indirect}

As mentioned above, direct measurements at stellar energies are very difficult and often impossible. Hence, direct measurements 
are usually performed at higher energies and then extrapolated down to stellar 
energies using R--matrix calculations. However, these extrapolations are not always free of problems. In some cases, they can even lead 
to wrong results because they do not take into account the contributions of a possible unseen low--energy resonance, as in 
$^{22}$Ne($\alpha$,n)$^{25}$Mg \cite{Kapp94}
or a possible sub--threshold resonance, like in $^{13}$C($\alpha$,n)$^{16}$O \cite{Pell08} 
and $^{12}$C($\alpha$,$\gamma$)$^{16}$O \cite{OULE12}. The effect of these resonances may 
change the extrapolated S-factor at astrophysical energies by a huge factor (sometimes orders of magnitude).  
The other problem concerning direct measurements is related to the radioactive nature of the nuclei 
involved in reactions occurring in explosive sites (novae, supernovae, X-ray bursts,...) and those involved 
in (n,$\gamma$) radiative captures (in r-process and sometimes in s-process). The intensities of the radioactive beams are often low, 
rarely exceeding 10$^5$ to 10$^6$ pps while for nuclei with relatively long half life, making targets with enough 
atoms per cm$^2$ is very difficult. Hence the direct measurements of such reactions are very difficult and 
challenging and in case of r-process reactions it is currently impossible. To bypass these difficulties (sub--threshold resonances, radioactive nuclei,...) indirect methods such as transfer reactions \cite{Sat83}, Coulomb dissociation \cite{BaurB86}, ANC method \cite{Mukh99} and Trojan Horse Method \cite{Baur86,Spit99,Spi13} 
are good alternatives (see \cite{Tri14} for a general review on indirect methods).
In these methods, the experiments are usually performed at high energies implying higher cross sections and the conditions are relatively 
less stringent than in direct measurements (target thickness and composition, high background, ...). However, these methods are model dependent. They depend on the uncertainties relative to the different parameters used in the model. Hence, there are two sources of errors, experimental and theoretical. 
But the global uncertainty on the measured cross section can be reduced by combining different  methods.   

In this review, we will focus mainly on the Coulomb dissociation and transfer reaction methods. 

\subsection{Theoretical methods}

\subsubsection{Coulomb dissociation method}
\label{s:breakup}

Coulomb dissociation can be considered as the equivalent of the inverse process of radiative capture, photo--disintegration. 
These two last reactions are related by the detailed balance theorem through the following equation: 

\begin{equation}
\sigma_{photo}/\sigma_{capture}=\frac{(2J_A+1)(2J_x+1)}{2(2J_B+1)}\frac{k^2_{cm}}{k^2_\gamma} 
\end{equation} 

where k$_{cm}$ is the wave number associated to A+x and k$_\gamma$ is the wave number associated to the photon.

In most cases, the wave length of the photon is larger than the one of the A+x system. Hence, the ratio $k_{cm}/k_\gamma$ 
is always larger than 1. This implies a photo--disintegration cross section much larger than the one of the radiative capture reaction of interest. 

The Coulomb dissociation (CD) method involves impinging a high energy nuclei
beam on a high-Z target, e.g. lead. The interaction of the high velocity incident nuclei 
with the intense Coulomb field of the lead target allows for virtual excitations in the 
form of virtual photons. These photons are absorbed by the incident nuclei, which then disintegrate 
into two fragments. If we assume that the excitation of the projectile is purely electromagnetic, then the Coulomb 
dissociation cross section is related to the photodissociation cross section via:  

\begin{equation}
\frac{d^2\sigma}{d\Omega dE_{rel}}=\frac{1}{E_{rel}+Q} \sum_{\pi\lambda}\frac{dn_{\pi,\lambda}}{d\Omega}\sigma^{photo}_{\pi,\lambda}
\end{equation} 

where dn$_{\pi,\lambda}$/d$\Omega$ is the virtual photon flux for the different multipolarities. 

By knowing precisely the number of created virtual photons and measuring the
Coulomb dissociation cross section (by detecting in coincidence the two emitted fragments), 
one can experimentally deduce the photodissociation cross section. Since the
cross section for the capture process and the (time-reversed) process
are related by the theorem of detailed balance, one can deduce the radiative capture cross section of interest.  

The first interest of the method is the amplification coming from the important number
of created virtual photons which leads to a high Coulomb dissociation cross
section. The second interest of the method comes from 
the use of high energy beam which implies on one hand the use of thick targets and on the other hand 
a forward focus of the fragments leading to a better detection efficiency. 

However, this method has also drawbacks. The most important ones come from the simultaneous contributions of E1, M1 
and E2 multi--polarities to the virtual photon spectrum which may contribute differently 
to the Coulomb dissociation cross section and the radiative capture one, the possible interference with the nuclear breakup and the 
possible post-acceleration effects. All these effects should be taken into account in the analysis of breakup experiments. 
This method was used in the study of various reactions of astrophysical interest, 
$^{12}$C($\alpha$,$\gamma$)$^{16}$O \cite{FLE05,Tat98}, $^{11}$C(p,$\gamma$)$^{12}$N \cite{LEF95,Moto}, $^{13}$N(p,$\gamma$)$^{14}$O  \cite{MOT91},
$^7$Be(p,$\gamma$)$^8$B (see \cite{SCHU03} and references therein) and D($\alpha,\gamma$)$^6$Li \cite{dag10,Kie91}. 

\subsubsection{Transfer reaction method}
\label{s:trans}

The transfer method where one or many nucleons are exchanged between the target and projectile is often used in 
nuclear structure to determine the energy position, spin and the orbital occupancy of various nuclei. It is also used in nuclear astrophysics 
to study the partial decay widths of nuclear states involved in resonant reactions. 

\begin{figure}
\resizebox{0.45\textwidth}{!}{%
  \includegraphics{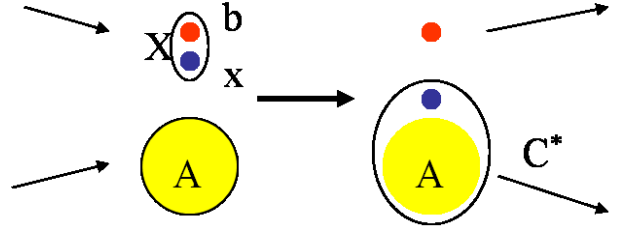}
}
\caption{Sketch of a transfer reaction}
\label{f:transfer}       
\end{figure}

To study a resonant reaction x+A$\rightarrow$C*$\rightarrow$B+y and measure 
the partial decay width $\Gamma_x$ of the state of interest in C* into the entrance 
channel, one can populate the excited states of C by transferring the light particle x (Fig.~\ref{f:transfer}) which can be a nucleon or a cluster of nucleons from the nucleus X to the nucleus A. This will feed the valence states of the final nucleus C, hopefully with no perturbation of the core, which is why it is called 
one step direct transfer reaction. The other part of the projectile b will continue its movement and will be detected. 
By measuring 
the emitted angle and energy of the particle b, one can deduce the energy of the excited state that was populated in C from kinematics and by 
comparing the shape of the measured angular distributions 
to those predicted by the distorted Born approximation theory (DWBA), one can deduce the angular orbital momentum $l$ of the populated state. 

The theoretical direct transfer cross section is calculated using the DWBA formalism and it is given by the following matrix element:

\begin{equation}
(\frac{d\sigma}{d\Omega})_{DWBA}\propto|<\chi_fI^C_{xA}(r_{xA})|V|I^X_{bx}(r_{bx})\chi_i>|^2
\end{equation}

Where $\chi_{i,f}$ are the distorted wave functions of the initial and final states, V is the transition transfer operator, 
$I^{C}_{xA}(r_{xA})$ is the overlap function of the final bound state C formed by A+x and 
$I^{X}_{bx}(r_{bx})$ is the overlap function of the bound state X formed by b+x. 

The radial part of these last functions is given by the following product:

\begin{equation}
 I^{\alpha}_{\beta\gamma}(r_{\beta\gamma})=S^{1/2}\varphi_{\beta\gamma}(r_{\beta\gamma})
\end{equation}

Where $S$ is the spectroscopic factor, $\varphi_{\beta\gamma}$(r) is the radial wave function of the bound state C or X 
with $\alpha$ being the final bound state C or the bound state X, $\beta$ being the transferred particle x and $\gamma$ being
A or b respectively. 

The spectroscopic factor of the different populated states in C expresses the overlap probability between 
the wave functions of the entrance channel A+x and the final state C. It can be extracted from the ratio of the measured differential 
cross section to the one calculated by DWBA:

\begin{equation}
(\frac{d\sigma}{d\Omega})_{exp}=S_xS'_x(\frac{d\sigma}{d\Omega})_{DWBA}
\end{equation}

One can see in the latter formula that there are two spectroscopic factors, $S_x$ for the final bound state of interest in the exit channel 
and $S'_x$ for the bound state in the entrance channel. Hence, by knowing one of the spectroscopic factors 
it is possible to extract the other one. Once the spectroscopic factor of the state of interest is extracted, one can then determine the reduced 
decay width using the following formula \cite{BECH78}:

\begin{equation}
 \gamma_x^2=\frac{\hbar^2 R}{2\mu}S_{x}{|{\varphi}(R)|}^2
\label{q:reduced}
\end{equation}

where $\varphi$(R) is the radial wave function of the bound state C formed by A+x, calculated at a channel radius R 
where $\varphi$(R) has its asymptotic behavior. 
The partial decay width $\Gamma_x$ is then given by \cite{Rolf88}: 
\begin{equation}
\Gamma_x=2P_l\gamma_x^2
\end{equation}
where $P_l$ is the Coulomb and centrifugal barrier penetrability (Eq.~\ref{q:pen}). 

The transfer DWBA differential--cross--section calculations depend on the optical potential parameters describing the wave functions 
of relative motion in the entrance and exit channels and on  the potential well parameters describing the interaction of the transferred 
particle with the core in the final nucleus. If the transfer reaction is performed at sub-Coulomb energies, then the dependence on the DWBA calculations on the potential parameters is greatly reduced.  This particular case of transfer reactions is called Asymptotic Normalization Coefficient  (ANC) method. 
This method relies on the peripheral nature of the reaction process that makes the calculations free from the geometrical parameters (radius,diffusivity) of the binding potential of the nucleus of interest and less sensitive to the entrance and exit channel potentials. The ANC method was extensively used for direct proton-capture reactions of astrophysical interest where the binding energy of the captured charged particle is low \cite{Trib05} and for reactions where the capture occurs through loose subthreshold resonance states \cite{Brun99,John06,John09}. 

Note that one can also extract from the usual transfer reaction, the ANC 
describing the amplitude of the tail of the radial overlap function at radii beyond the nuclear interaction radius (r $>$R$_N$), 
via the expression  \cite{Mukh99}: 
\begin{equation}
 {C}^2=S_x\frac{R^2 \varphi^2(R)}{{W}^2(kR)} 
\end{equation}
where $W$ is the Whittaker function, describing the asymptotic behavior of the loosly bound-state wave function. 
But in this case, the ANC is dependent on the well--potential parameters.

\subsection{Experimental examples}
\label{s:expe}

\subsubsection{Study of D($\alpha$,$\gamma$)$^6$Li through $^6$Li high energy breakup}
\label{s:dag}

One of the most puzzling questions discussed these last ten years in the astrophysics community
is related to the origin of the observed $^6$Li in very old halo stars
\cite{Asp06}. Indeed, as it was mentioned in \S~\ref{s:li}, the abundance plateau of the observed $^6$Li was
found to be unexpectedly high compared to the $^6$Li BBN
predictions. Hence, many scenarios were proposed to solve this
puzzle; e.g. the pre-galactic production of $^6$Li \cite{Rol05}, or production of $^6$Li
by late decays of relic particles \cite{Jed06}. Before seeking exotic solutions
to the lithium problem, however, it was important to improve the standard BBN calculations by considering 
the key nuclear reactions involved in the $^6$Li formation. 
According to calculations of Vangioni--Flam et al.  \cite{Van00}, the most dramatic effect is observed for D$(\alpha,\gamma)^6$Li 
whose huge uncertainty of about a factor 10 \cite{NACRE} on the cross section 
at the energies of astrophysical interest (50 keV $\leq$ $E_{cm}$ $\leq$ 400 keV) induces an uncertainty  
of a factor of $\approx$20 on the primordial $^6$Li abundance. This uncertainty originates from the discrepancy between
the theoretical low-energy dependence of the S-factor \cite{Kha98} and the only existing experimental data 
at BBN energies \cite{Kie91} obtained with the Coulomb break-up technique of $^6$Li  at 26 A MeV. 
Hence, a new precise measurement of the cross section of the  D$(\alpha,\gamma)^6$Li reaction was performed at GSI
using Coulomb dissociation (CD) of $^6$Li at high energy 150 A MeV \cite{dag10}.
A $^{208}$Pb target with a 200 mg/cm$^2$ thickness was bombarded by a primary
$^6$Li beam of 150 A MeV energy. 

The angles and positions as well as the energy losses of the outgoing particles,
D and $^4$He, were measured by two pairs of silicon strip detectors placed at distances of 15 and 30 cm, 
respectively,  downstream from the target. Deuteron and alpha momenta were analyzed
with the Kaos spectrometer which has a large angular and momentum acceptance and 
were detected in two consecutive multi-wire chambers followed by a plastic-scintillator 
wall made with 30 elements. 

The opening angle $\theta_{24}$  between the two fragments was deduced from their position 
measurement in the DSSDs. The deuteron and $^4$He momenta,   P$_d$ et P$_{^4He}$,  were determined
from their trajectories reconstructed by using their measured positions in the SSDs and in the 
multi-wire chambers behind Kaos spectrometer. From the measured opening angle between the fragments and their momenta, the 
relative energy E$_{rel}$ between the deuteron and the $\alpha$ particles in the c.m. system could be reconstructed. 
The details of the experiment and analysis are given in \cite{dag10}.                                                                                                                                                                                                                                                                                                                                                                                   

\begin{figure}
\resizebox{0.45\textwidth}{!}{%
  \includegraphics{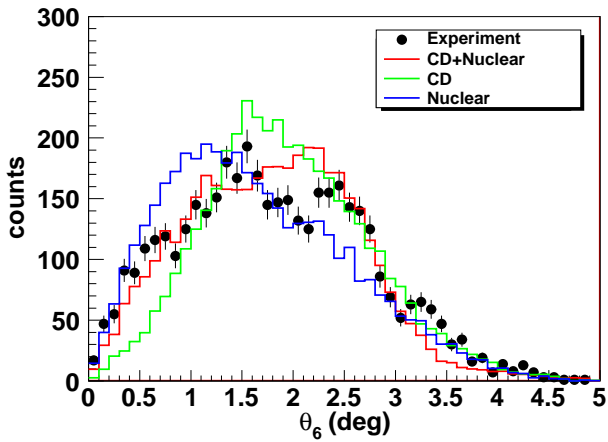}
}
\caption{Angular distribution of the excited $^6$Li$^*$ after reaction summed over E$_{rel}$ values up to 1.5 MeV. 
The experimental data are compared to the simulation results where is considered a pure nuclear contribution (blue curve), 
a pure Coulomb contribution (green curve) and an interference between the two (red curve). 
The simulated count number is normlaized to the experimental one.}
\label{f:theta6}       
\end{figure}

A comparison of the results with the theoretical predictions convoluted by the experimental acceptance and resolution 
was performed. The breakup calculations were performed with CDXSP code \cite{dag10} where Coulomb and  
nuclear contributions are considered. These new CDXSP calculations of S. Typel show a dominant nuclear contribution to the 
$^6$Li breakup \cite{dag10} contrary to Shyam et al. predictions \cite{Shy94}.  
In Fig.~\ref{f:theta6}, is displayed the angular distribution of the excited $^6$Li$^*$ after reaction which is, according 
to CDXSP calculations, the observable most sensitive to the reaction mechanism. The black points depict the measured  
$\theta_6$ angular distributions and the histograms, the predicted ones convoluted  by the experimental acceptance using GEANT simulations for pure Coulomb (CD) and pure nuclear interactions as well as combined (CD+nuclear) interaction. 
In this figure, the calculation which reproduces the best the observed structures in the experimental data is the one where 
the interferences between the Coulomb contribution and the nuclear one is taken into account (red curve). This shows, clearly, that
the Coulomb-nuclear interference is at play and the interference sign considered is correct.

Usually, in Coulomb dissociation experiments \cite{SCHU03}, the astrophysical S-factors of the reaction of interest are deduced 
by scaling the theoretical astrophysical S-factors by the ratio of the measured to simulated differential cross sections. 
In this experiment, the extraction of the S-factors is not possible because of the interference between the Coulomb and the nuclear components. Given that the calculations of CDXSP model take well into account such mechanisms and describe 
well the various measured observables in this experiment \cite{dag10}, one can then conclude 
that the model is reliable as well as the calculated astrophysical S-factors S$_{24}$ of D($\alpha$,$\gamma$)$^6$Li used in it \cite{dag10}. 

\begin{figure}
\resizebox{0.45\textwidth}{!}{%
  \includegraphics{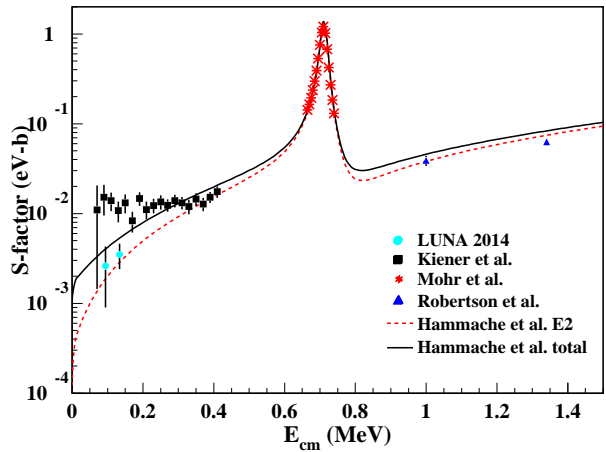}
}
\caption{Astrophysical S-factors $S_{24}$ of the E2 and total (E2+E1) deduced from this work. 
Black points are data from CD measurements of Kiener et al. \cite{Kie91}, red points are those coming from direct measurements. 
\cite{Moh94,Rob81} and blue points are from LUNA \cite{And14}.}
\label{f:Sfactor-6Li}       
\end{figure}

The astrophysical S-factors S$_{24}$ of the E2 component and total (E1+E2) deduced in this work are displayed in Fig.~\ref{f:Sfactor-6Li}  together with 
the previous CD data of Kiener et al. \cite{Kie91}, the direct data of Mohr et al. \cite{Moh94}, 
Robertson et al. \cite{Rob81}, and the very recent LUNA \cite{And14} data.
Note that the E1 component considered in the calculation of total S$_{24}$ is not constrained by the GSI experimental data which are sensitive only to the E2 component \cite{dag10}. 

The good agreement between the GSI results  for the E2 component (red curve) and the direct measurements 
is an indication of the relevance of the performed calculations and the quality of the experiment. Moreover, the very good agreement observed 
between GSI total S$_{24}$ factors and the latest direct data coming from LUNA  \cite{And14} experiment gives also strong confidence in the GSI calculated 
E1 component and so on the whole GSI data. GSI results were found to be in agreement with various theoretical works \cite{dag10}
In  Fig.~\ref{f:Sfactor-6Li}, one can see that Kiener et al. \cite{Kie91} results are in disagreement with E2 GSI results. This is due to the large contribution of the nuclear component  \cite{dag10} which was not taken into account in the analysis of Coulomb dissociation data at 26 A MeV. 

A calculation of the new $^6$Li reaction rate was performed using GSI total S$_{24}$ factors and then 
introduced in the BBN model of Coc et al. \cite{Coc04} to evaluate the primordial $^6$Li
abundance as a function of the baryonic density of the Universe. 
The obtained value \cite{dag10} at the baryonic density deduced from WMAP observations is 1000 times less than 
the observations of Asplund et al. \cite{Asp06}. The results of this experiment \cite{dag10}, which reduce 
significantly the uncertainties surrounding the cross section of 
D($\alpha$,$\gamma$)$^6$Li reaction, exclude definitely the primordial origin of the observed $^6$Li. 
This conclusion is supported by the very recent observations of Lind et al. \cite{Lin13} which indicate an absence of 
$^6$Li in the old halo galactic stars except in HD8493 star, as it was mentioned in 3.1.1

\subsubsection{Search of resonant states in $^{10}$C and $^{11}$C and the $^7$Li problem}
\label{s:be}

The main process for the production of the BBN $^7$Li is the decay of $^7$Be which is 
produced by the reaction $^3$He($^4$He, $\gamma$)$^7$Be, as it was mentioned in \S~\ref{s:bbn}. 
Direct measurements of this reaction were performed by several groups in order to improve its knowledge and no satisfactory answer to 
the cosmological $^7$Li problem was achieved \cite{DIL09,Ade11,deB14}. 
The same conclusion stands for the main reaction which destroys $^7$Be, $^7$Be(p,n)$^7$Li followed by $^7$Li(p,$\alpha$)$\alpha$. 
These two reactions are well studied and their cross section are known better than few percents according to Descouvement et al. 
BBN compilation  \cite{Des04}.

Moreover, many experimental attempts to explain the $^7$Li  anomaly by studying 
other key nuclear reaction rates such as $^7$Be(d,p)$^9$B \cite{Coc04,Cyb09} did not lead to successful 
conclusions  \cite{{Ang05},{Malle11},{Kirs11}}. What about missing resonances in other secondary reactions involving  
$^7$Be or $^7$Li not yet studied? That is what was investigated recently by some authors \cite{Cha10,Bro12,Civ13}. 
From their exploration of the potential resonant destruction channels of both $^7$Li and $^7$Be via  n, p, d, $^3$He, t and $^4$He, 
two candidates, besides the $^7$Be+d case already mentioned above, looked promising to solve partially or totally the lithium problem. The 
two candidates are a potentially existing resonant states close to 15 MeV excitation energy  in $^{10}$C  \cite{Cha10}  and between 
7.793 MeV and 7.893 MeV in $^{11}$C \cite{Bro12,Civ13} compound nuclei formed by $^7$Be+$^3$He and  $^7$Be+$^4$He  respectively. 

A search of missing resonant states in $^{10}$C and $^{11}$C was investigated through $^{10}$B($^3$He,t)$^{10}$C and 
$^{11}$B($^3$He,t)$^{11}$C cha\-rge-exchange reactions respectively \cite{Ham13}. The two ($^3$He,t) charge--exchange reactions 
were induced on 90 $\mu$g/cm$^2$ enriched $^{10}$B target and a  250 $\mu$g/cm$^2$ self-supporting natural boron target, respectively, irradiated by a $^3$He beam of 35 MeV energy delivered by the Tandem accelerator of the Orsay Alto facility. The emitted reaction products were detected at the focal plane of Split-Pole spectrometer by a position-sensitive gas chamber, a $\Delta$E proportional gas-counter and a plastic scintillator. 
The tritons coming from $^{10}$B($^3$He,t)$^{10}$C  were detected at four different angles in the laboratory system while those coming from $^{11}$B($^3$He,t)$^{11}$C were detected at two angles. 

\begin{figure}
 \resizebox{0.45\textwidth}{0.24\textheight}{%
  \includegraphics{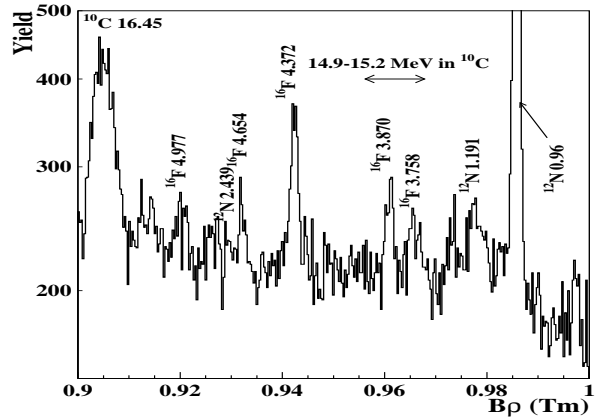}
}
\caption{Triton B$\rho$ spectrum measured at $\theta$=10$^\circ$ (lab) in the excitation energy 
region from 14 to 16.5 MeV. The excitation energy (MeV) of $^{10}$C levels are indicated as well as those of  $^{12}$N and   $^{16}$F
coming from a substantial $^{12}$C and  $^{16}$O contamination of the target. The unlabeled peaks correspond to unidentified heavy contamination.
}
\label{f:Spectre-10C}       
\end{figure}

 A B$\rho$ position spectrum of the tritons produced by the reaction $^{10}$B($^3$He,t)$^{10}$C at $\theta_{lab}$= 10$^\circ$ is 
 displayed in Fig.~\ref{f:Spectre-10C} for the excitation energy region between 14 and 16.5 MeV of astrophysical interest.  The only isolated and well populated peaks observed in the energy region of astrophysical interest between 14.9 and 15.2 MeV belong to the unbound states at 3.758 and 3.870 MeV excitation energy of $^{16}$F coming from the contaminant $^{16}$O($^3$He,t) reaction. 
 
A detailed study of the background in the region of interest taking into account the width of an hypothetical state, as well as its populating cross section lead to
the conclusion that any 1$^-$ or 2$^-$ state of $^{10}$C in the excitation energy region around 15 MeV should have very likely, if present, a total width larger 
than 590 keV to escape detection. 

\begin{figure}[h]
\resizebox{0.45\textwidth}{0.24\textheight}{%
\includegraphics{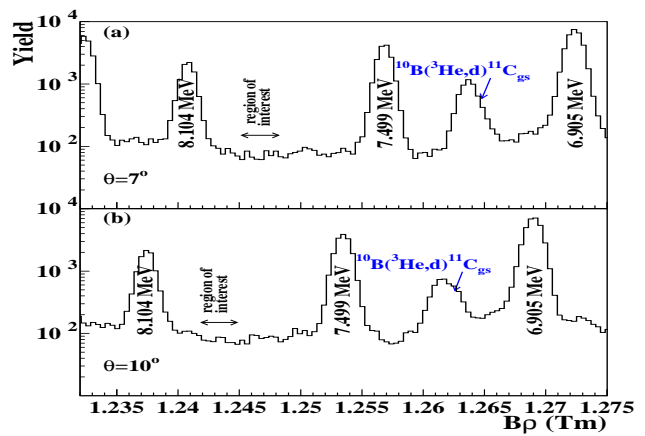}
}
\caption{$^{11}$B($^3$He,t)$^{11}$C B$\rho$ spectra measured at $\theta$=7$^\circ$ (a) and 10$^\circ$ (b) 
in the excitation energy region of interest close to 8 MeV.  Excitation energies of $^{11}$C levels are indicated.
The double arrow indicates the astrophysical region of interest.
}
\label{f:Spectre-11C}       
\end{figure}

Concerning the $^{11}$B($^3$He,t)$^{11}$C measurements,  spectra obtained at 7$^\circ$ (lab) and 10$^\circ$ (lab) 
are shown in Fig.~\ref{f:Spectre-11C} for the energy region of interest. One can see that all the known states of $^{11}$C are well populated. 
On the other hand, no new state of $^{11}$C is observed  in the excitation energy region between 
7.499 and 8.104 MeV: the very small observed peaks are due to statistical fluctuations \cite{Ham13}.

Reaction rate calculations for the two only open channels, $^7$Be($^3$He,$^4$He)$^6$Be and
$^7$Be($^3$He,$^1$H)$^9$B, were performed \cite{Ham13} assuming a 1$^-$ state in the compound nucleus $^{10}$C having 
a total width equal to the lower limit deduced from the Orsay work, 590 keV, and 200~keV in case the differential charge-exchange cross 
section is three times smaller than the expected minimum one. 
The calculated rates were included in a BBN nucleosynthesis calculation and were found to have no effect on the primordial $^7$Li/H abundance. 

In conclusion, the results of the Orsay work exclude the two reactions $^7$Be+$^3$He and $^7$Be+$^4$He as solution to the 
cosmological $^7$Li problem. If one takes into account the conclusions of the experimental works 
\cite{{Ade11},{Des04},{Ang05},{Malle11},{Kirs11}} concerning the other important reaction channels for the synthesis and destruction of $^7$Be and thus of $^7$Li,  the results of the Orsay work exclude a nuclear solution to the  $^7$Li problem. This does not exclude that sub-dominant reactions may marginally affect
the \sep\ production. For instance, the $^7$Be(n,$\alpha)^4$He channel is suppressed, with respect to the $^7$Be(n,p)$^7$Li one, due to parity conservation. 
However, since the origin of its rate \cite{Wag69} is unclear, due to the scarcity of experimental data \cite{Ser04}, this reaction could reduce the \sep/H ratio, but definitely,
far from enough.   
 
\subsubsection{Study of $^{13}$C($\alpha$,n)$^{16}$O reaction via ($^7$Li,t) transfer reaction}
\label{s:c13an}

Direct measurements at the astrophysical energy of interest, E$_{cm}\approx$ 190~keV, 
of $^{13}$C($\alpha$,n)$^{16}$O reaction, the main
neutron source for the s-process in AGB stars of 1-3 solar masses (see section \ref{s:advanced}), 
is extremely difficult because the cross section
decreases drastically when the incident $\alpha$ energy decreases.
Thus, direct measurements \cite{Drot93} have only been
performed down to 260~keV too far away from the energy
range of interest. R-matrix extrapolations \cite{{Desc87},{Hale97}} of the cross sections measured at
higher energies have then to be performed, including the contribution of
the 6.356~MeV, 1/2$^+$,  state of the compound nucleus $^{17}$O, which lies 3 keV below the
$\alpha$+$^{13}$C threshold. The contribution of this sub--threshold state strongly depends on
its ${\alpha}$-spectroscopic factor, S$_\alpha$. 
However, the results of previous studies
of this contribution using ($^6$Li,d) transfer reaction \cite{Kub03,Keel03} and ANC \cite{John06} 
measurements lead to different conclusions.

\begin{figure}[h]
\resizebox{0.5\textwidth}{!}{%
\includegraphics{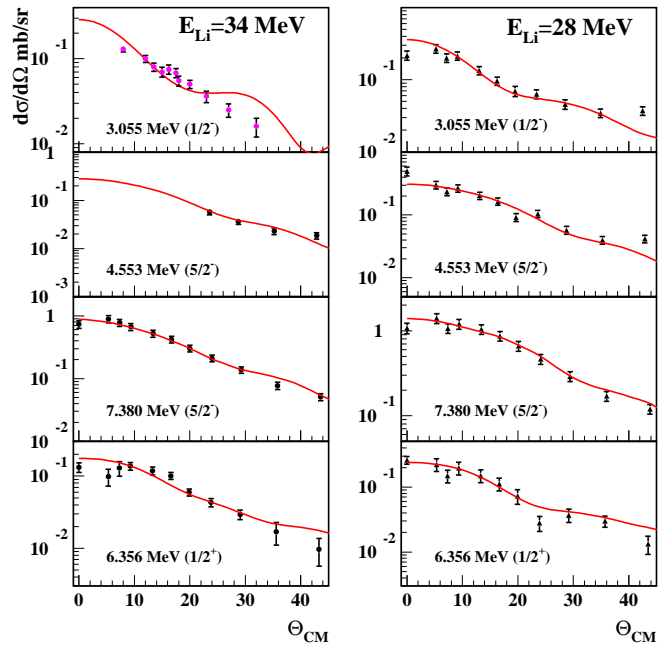}
}
\caption{Experimental differential cross sections of the
$^{13}$C($^7$Li,t)$^{17}$O reaction obtained at 28 and 34 MeV,
compared with finite-range DWBA calculations normalized to the data.
}
\label{f:dsigmadomeg}       
\end{figure}

A new investigation of the effect of the sub--threshold resonance on the 
astrophysical S-factor was performed through a determination 
of the alpha spectroscopic factor of the 6.356 MeV state using the transfer
reaction $^{13}$C($^7$Li,t)$^{17}$O at two different incident energies and an improved DWBA analysis. 
The experiment \cite{Pell08} was performed using a $^7$Li$^{3+}$ beam provided by
the Orsay TANDEM impinging on a self-supporting enriched  $^{13}$C target. 
The reaction products were analyzed
with an Enge Split-pole magnetic spectrometer and the tritons were detected at angles ranging from 0 to 31 degrees in laboratory system. 
The experimental $^{13}$C($^7$Li,t)$^{17}$O differential cross
sections measured for the 6.356, 3.055, 4.55 and 7.38 MeV, at the
two incident energies of 34 and 28 MeV, are displayed in  Fig.~\ref{f:dsigmadomeg} together with the Finite-range DWBA calculations, using 
the FRESCO code \cite{Thom88}.  The data points displayed for the 3.055 MeV state in the 34 MeV left-column are from the 
measurements of Ref.~\cite{CLAR78} at 35.5 MeV. 

The $\alpha$-spectroscopic factors were extracted from the normalization of the finite-range DWBA curves to the experimental data. 
The good agreement between the DWBA calculations and the measured
differential cross sections of the different excited states of
$^{17}$O at the two bombarding energies of 28 MeV and 34 MeV
respectively, gives strong evidence of the direct nature of the
($^7$Li,t) reaction populating these levels and confidence in the
DWBA calculations. An S$_\alpha$ mean value of 0.29$\pm$0.11 is deduced
for the sub--threshold state  at 6.356 MeV in $^{17}$O, which is in good
agreement with that obtained by Keeley et al.  \cite{Keel03} and
those used earlier (S$_\alpha\approx$0.3--0.7) in the s-process
models.  The uncertainty on the extracted $\alpha$ spectroscopic
factor for the state of interest (6.356 MeV) was evaluated from the
dispersion of the deduced S$_\alpha$ values at the two incident
energies and using different sets of optical potentials in the
entrance and exit channels and different
$\alpha$-$^{13}$C well geometry parameters \cite{Pell08}.

The $\alpha$-reduced width $\gamma_\alpha^2$ of about 13.5$\pm$6.6 keV for the 6.356 MeV state was obtained using 
Eq.~\ref{q:reduced}. The calculation was performed at the radius R=7.5 fm where the Coulomb asymptotic behavior of the radial part of the
$\alpha$-$^{13}$C wave function is reached. 
\begin{figure}[h]
\resizebox{0.5\textwidth}{0.28\textheight}{%
\includegraphics{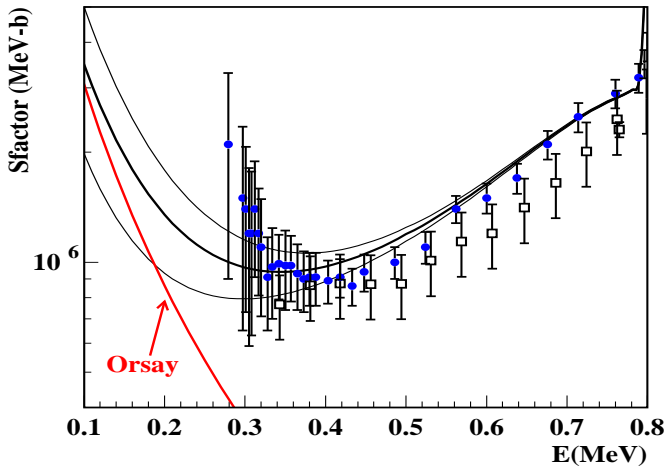}
}
\caption{Astrophysical $S$-factor for the
$^{13}$C($\alpha$,n)$^{16}$O reaction with $R$-matrix calculations.
The data points are taken from Refs \cite{{Drot93},{Brun93}}. The
contribution of the 6.356 MeV state is shown as red curve.
The thick black curve corresponds to the recommended $\gamma^2_{\alpha}$
values, and the thin black ones to the lower and upper limits.
}
\label{f:S-factor}       
\end{figure}

The contribution of the 1/2$^+$ state to the astrophysical S-factor when using this deduced
$\gamma_\alpha^2$ is shown in red  curve in  Fig.~\ref{f:S-factor}.
At the energy of astrophysical interest, $E_{cm}$=0.19 MeV, the contribution of this
sub--threshold state to the total S-factor is dominant ($\approx$ 70\%) \cite{Pell08}. This is much larger than what was obtained in 
Kubono et al. \cite{Kub03} (1.6$\%$) and Johnson et al. \cite{John06} (30$\%$) works and it confirms the 
dominant character of the sub--threshold state on the cross section of   $^{13}$C($\alpha$,n)$^{16}$O at the energies of astrophysical interest. 

The calculated $^{13}$C($\alpha$,n)$^{16}$O reaction rate, at
temperature T=0.09 GK important for the s-process in low mass AGB
stars was found to be in a good agreement with  NACRE compilation adopted value but the range of allowed values is significantly
reduced in this work \cite{Pell08}. 
The Orsay result is confirmed by the work of Heil et al. \cite{Heil08} and very recently by the results of Guo et al. \cite{Guo12} and La Cognata et al. \cite{LACO12} where  the transfer reaction $^{13}$C($^{11}$B,$^7$Li)$^{17}$O and the Trojan Horse method were used respectively. 

\subsubsection{Study of $^{26}$Al(n,p)$^{26}$Mg and $^{26}$Al(n,$\alpha$)$^{23}$Na through $^{27}$Al(p,p')$^{27}$Al}

$^{26}$Al was the first cosmic radioactivity to be detected in the galaxy as well as one of the first extinct 
radioactivity observed in meteorite \cite{PraDi} (\S~\ref{s:obs26}). Its nucleosynthesis in massive stars is still uncertain due to 
the uncertainties surrounding the $^{26}$Al(n,p)$^{26}$Mg and  $^{26}$Al(n,$\alpha$)$^{23}$Na 
reactions \cite{Ili11}. The uncertainties on the rate of these two reactions are mainly due to the lack of 
spectroscopic information on the $^{27}$Al compound nucleus above the neutron and alpha 
thresholds for which no experimental data was available. 

The first experimental study of the \np\ and \na\ reaction rates was done using
the time reverse reactions $^{26}$Mg(p,n)\alr\ and \nas($\alpha$,n)\alr\ \cite{Ske87}. 
However, this method only provides the branching to the ground states and
has been superseded by direct measurements with \alr, radioactive targets~\cite{Tra86,Koe97,Sme07}.
However, the different experiments give results that are inconsistent within each other ~\cite{Ogi11} 
by a factor of up to $\approx2$.
The reaction rates used in stellar evolution calculations, based on the Hauser-Feshbach 
statistical approach rely on the level density, but since \alr\ ground state has $J^\pi$~=~5$^+$, 
the most important \als\ states have high spin such as 9/2$^+$ and 11/2$^+$ or 7/2$^-$ to 13/2$^-$ for s- or 
p-wave neutron capture, respectively. The level density of such high spin states may not be well reproduced 
in Hauser-Feshbach calculations.
This is why the spectroscopy of neutron-unbound levels in $^{27}$Al was investigated through $^{27}$Al(p,p')$^{27}$Al 
reaction which was studied at the Tandem/ALTO facility using a proton beam at 18 MeV \cite{Bena14a}.

The (p,p') measurement was induced on a self-supporting $^{27}$Al target of 80 $\mu$g/cm$^2$ thickness and the emitted particles 
were detected in the focal plane of the split-pole spectrometer at 10$^\circ$, 40$^\circ$ and 45$^\circ$.  
A careful focal-plane detector calibration was performed using a low--excitation energy measurement populating well known 
isolated states. 
A series of overlapping spectra covering \als\ excitation energies from the ground 
state up to about 14~MeV were obtained by changing the magnetic field. 

\begin{figure}[h]
\resizebox{0.5\textwidth}{0.27\textheight}{%
\includegraphics{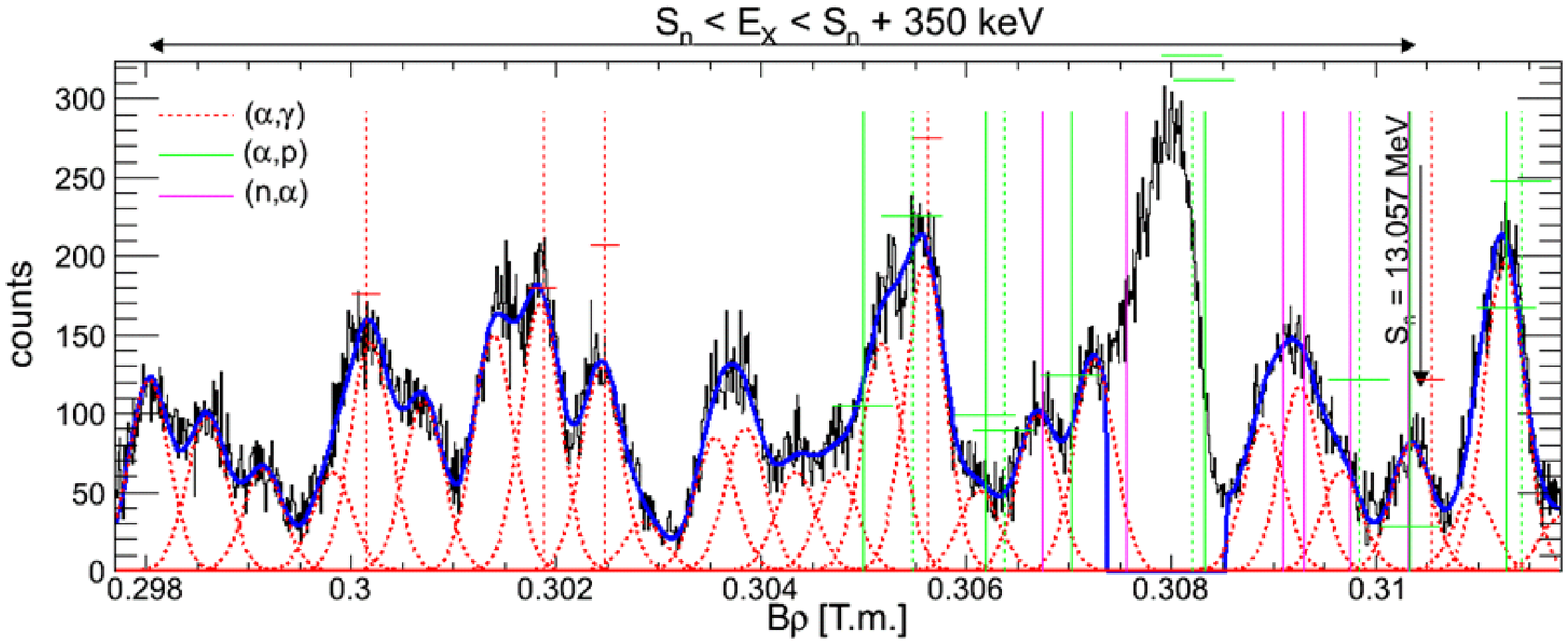}
}
\caption{Proton B$\rho$ rigidity spectrum measured at $\theta$=40$^\circ$. Excitation energies within about 350 keV above the 
$^{26}$Al+n threshold are displayed. Seet text for curve and vertical lines description.} 
\label{f:27Al}       
\end{figure}

States, up to excitation energies of around 14 MeV in $^{27}$Al, have been populated. A small part of the measured spectrum 
corresponding to the excitation energies within about 350 keV above the $^{26}$Al+n neutron threshold is displayed in Fig.~\ref{f:27Al}.
The spectrum was deconvoluted after background subtraction. Few of the measured states were observed in previous experiments. 
A very good agreement is obtained for the states measured with the $^{23}$Na($\alpha$,$\gamma$)$^{27}$Al reaction \cite{Voig71} (red vertical line). Similar agreement 
is obtained with the data corresponding to the direct $^{26}$Al(n,$\alpha$)$^{23}$Na  reaction \cite{Smet07} (magenta vertical line) and a 
marginal agreement within the error bars is obtained for the states observed in $^{23}$Na($\alpha$,p)$^{26}$Mg measurement (green vertical line) \cite{Whit74}.

In total 30 states above the $^{23}$Na+$\alpha$ threshold and more than 30 states above the $^{26}$Al+n threshold have been 
observed for the first time \cite{Bena14a} and  their excitation energies have been determined with an uncertainty of 4~keV. 
The precise determination of the excitation energy of the $^{27}$Al states of astrophysical interest is important for the $^{26}$Al(n,$\alpha$)$^{23}$Na and $^{26}$Al(n,p)$^{26}$Mg reaction rate calculations. However, measurements of the branching ratios, 
partial widths and spins and paritites are also necessary to reduce the uncertainties in the $^{26}$Al(n,p)$^{26}$Mg and $^{26}$Al(n,$\alpha$)$^{23}$Na reaction rates calculations. Hence, coincidence measurements, coupling the Split-Pole spectrometer with 
three DSSSDs placed in a close geometry in the reaction chamber, were performed at the Tandem/ALTO facility \cite{Bena14b}; the analysis is in progress.

\subsubsection{Study of $^{25}$Al(p,$\gamma$)$^{26}$Si}
\label{s:al25pg}

In the explosive hydrogen burning of novae (\S~\ref{s:novae}), $^{26}$Al production by  $^{25}$Al($\beta^+ \nu$)$^{25}$Mg(p,$\gamma$)$^{26}$Al reaction 
is in competition with  $^{25}$Al(p,$\gamma$)$^{26}$Si($\beta^+ \nu$)$^{26}$Al$^m$. 
The latter synthesis path produces the short-lived isomer ($^{26}$Al$^m$, 228~keV above the ground state$, \tau_{1/2}$ = 6.3 s) that decays to the 
ground state of $^{26}$Mg, thus bypassing the 1.809-MeV emitting long-lived ground state of $^{26}$Al \cite{Coc95}. 
The reaction rate  of $^{25}$Al(p,$\gamma$)$^{26}$Si is dominated at nova temperatures by direct capture and resonance levels in $^{26}$Si, up to $\approx$500 keV above the proton emission threshold at E$_x$ = 5513.7~keV.  
Iliadis et al. \cite{Ili96} deduced from available spectroscopic data in $^{26}$Al and $^{26}$Mg and theoretical studies that the rate above 2$\times10^8$ K may be dominated by a resonance corresponding to a then unobserved 3$^+$ level at E$_x$ $\simeq$ 5.97 MeV. In that case proton capture on $^{25}$Al may well bypass to a large extent its beta decay and thus $^{26}$Al$_{g.s.}$ synthesis during nova outbursts. The lack of spectroscopic information on the properties of this and other states in $^{26}$Si above the proton emission threshold implies in any case a large reaction rate uncertainty.  Motivated by  the observations of $^{26}$Al described in section \ref{s:obs26}, a lot of experimental efforts were dedicated in the last ten years to reduce this reaction rate uncertainty  employing various indirect methods.  

Properties of astrophysicaly important excited levels of $^{26}$Si were obtained in several transfer reactions  using light ions \cite{Cag02,Par04,Bar06,Chi10,Mat10} or heavy ions \cite{Sew07} including radioactive $^{25}$Al beam \cite{Pep09,Che10} and by $\beta$-decay  \cite{Tho04}. In particular, in the excitation energy range E$_x$ = 5.5 - 6 MeV  four different levels were observed in these experiments. Furthermore, a more accurate reaction Q-value for the $^{25}$Al(p,$\gamma$)$^{26}$Si reaction was deduced in recent mass measurements \cite{Par05,Ero09,Kwi10}. A critical review of excitation energies and spin assignments was done in \cite{Wre09}, concluding at a consistent identification of the 3$^+$ level in the different experiments at a resonance energy of 412 keV. These studies permitted a significant reduction of the thermonuclear reaction rate uncertainty at nova temperatures.

\begin{figure}[h]
\resizebox{0.5\textwidth}{0.27\textheight}{%
\includegraphics{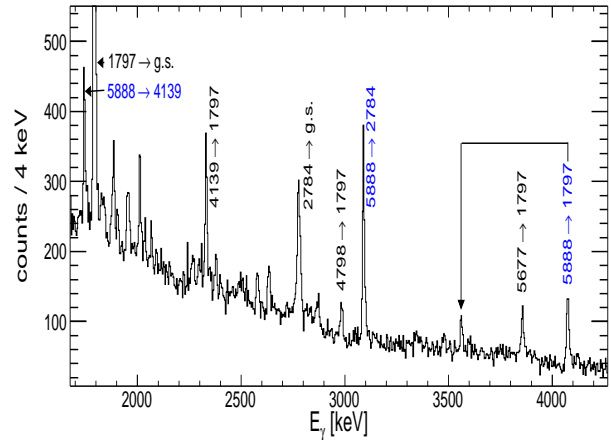}
}
\caption{Energy spectrum for one coaxial Ge detector in coincidence with the peak in the neutron-TOF spectrum corresponding to a level in $^{26}$Si at E$_x$ $\simeq$ 5.9 MeV. Transitions related to this level are labelled as well as subsequent gamma-ray decays. Figure adapted from Ref.~\cite{SerNIC10}.} 
\label{f:26Si}       
\end{figure}

An experimental study of the $^{24}$Mg($^3$He,n$\gamma$)$^{26}$Si reaction at the Orsay tandem facility found indications of a yet unobserved level at E$_x$ = 5.888 MeV  \cite{SerNIC10}.  In the experiment the neutron-TOF method with an efficient neutron detection setup was combined with high-resolution gamma-ray detectors for neutron-gamma coincidences. The new state  could be identified and attributed to $^{26}$Si in the gamma-ray spectrum coincident with a peak close to 5.9 MeV in the neutron-TOF spectrum. Fig. \ref{f:26Si} shows this gamma-ray spectrum where several gamma-ray transitions to known $^{26}$Si levels are clearly present. This state is plainly inside the Gamow window for explosive hydrogen burning in novae and may play an important role. Based on theoretical calculations, Richter et al. \cite{Ric11}  proposed that it could be the 0$^+_4$-state, which in that case would not be significant for the reaction rate. A very recent experiment suggests indeed a 0$^+$ assignment for this state \cite{Kom14}. 
Further studies, however, are required to assess the importance of this and other observed levels for the $^{25}$Al(p,$\gamma$)$^{26}$Si thermonuclear reaction rate.

\section{Experimental data for non-thermal reactions}
\label{s:non}

Contrary to reactions in thermonuclear burning, center-of-mass kinetic energies are usually well above the Coulomb barrier for charged particles and 
nuclear reaction cross sections, are generally above the micro-barn range, and therefore are more easily accessible for direct measurement. 
Thus, the experimental challenges here are not small counting rates but  the sometimes enormous number of open reaction channels that need to be studied. 
Secondly, cross section excitation functions must often be measured in a wide energy range, usually from the reaction threshold to e.g. a few tens of MeV for 
solar flare studies and up to TeV energies for cosmic-ray induced reactions. 

There are two principal axes where recent non-thermal reaction cross sections have been measured:

\begin{itemize}
\item The production of residual nuclei in cosmic-ray interactions.
\item Gamma-ray emission in energetic-particle interactions.
\end{itemize}

\subsection{Residual nuclei production in cosmic-ray interactions}

A very important class of observables are the fluxes of secondary CR particles, i.e. particles that are essentially created in collisions of CR nuclei with interstellar matter. Those include antiprotons, radioactive isotopes and nuclei with very low source abundances like LiBeB and the sub-Fe elements Sc,V,Ti,Cr,Mn (Fig.~\ref{f:zuni}) that are copiously produced in fragmentation reactions of  the abundant heavier nuclei CNO  and Fe, respectively. The fluxes of those secondary species or flux ratios  like $\bar{p}/p$, $^{10}$Be/$^9$Be, B/C and sub-Fe/Fe are key observables constraining the CR propagation parameters, like the diffusion coefficient and the galactic CR halo size.  

A recent dedicated measurement of fragmentation yields  has been done at the heavy-ion accelerator facility of GSI Darmstadt.  The spallation of $^{56}$Fe by protons, especially important for the sub-Fe CR fluxes, has been measured in the energy range 0.3 - 1.5 GeV per nucleon. It was done in reverse kinematics employing a liquid-hydrogen target where the forward-focusing permitted the detection and identification of all reaction products in several runs at the fragment separator FRS \cite{Nap04,Vil07}. In this experiment, data for particle-bound isotopes of elements Li to Co with cross sections exceeding 10$^{-2}$ mb have been obtained, that amount to  more than 150 different nuclei! Such data are valuable not only directly for CR propagation calculations, but also as a crucial test to cross section parameterizations or reaction codes that are required for the extrapolation of cross section data, and more importantly to estimate reaction cross sections for nuclei where no experimental data exist. 

\begin{figure}
\resizebox{0.45\textwidth}{!}{%
  \includegraphics{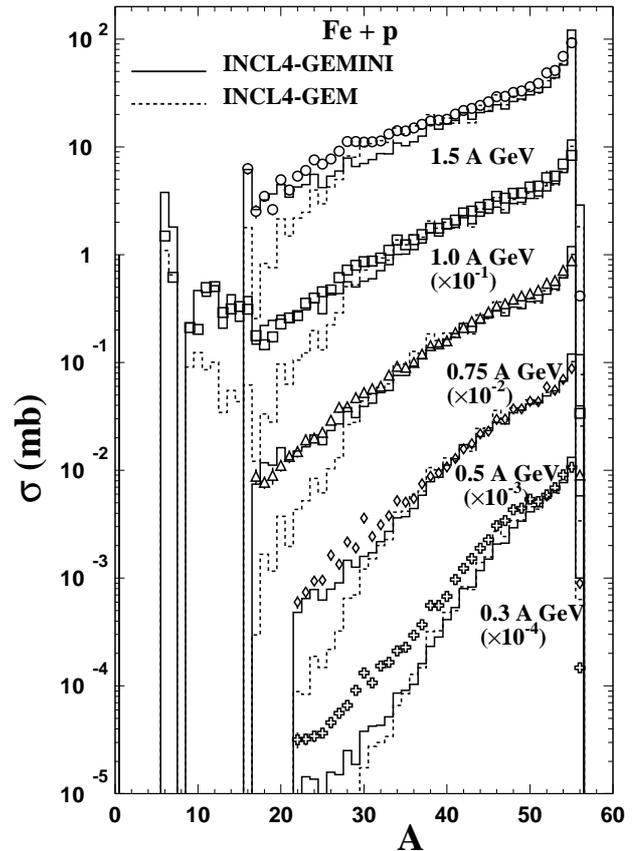}
}
\caption{Production cross sections of residual nuclei as a function of mass number A in the reaction Fe + p in the laboratory energy range 0.3 - 1.5 GeV per nucleon. Symbols: measured data at the fragment separator of GSI Darmstadt \cite{Nap04,Vil07}; full and dashed lines: calculated cross sections with INCL4 \cite{Bou02,Bou13} coupled to evaporation models GEMINI \cite{Cha88} and GEM \cite{Fur00}, respectively. Figure adapted from C. Villagrasa-Canton, A. Boudard et al. Physical Review C 75, 044603 (2007). Copyright (2007) by the American Physical Society. }
\label{f:FRS_GSI}       
\end{figure}

Partial cross section data of the GSI/FRS experiment and calculations with reaction codes for the Fe + p reaction are displayed in  Fig. \ref{f:FRS_GSI}. It demonstrates the ability of modern codes  for the intranuclear cascade stage (INC) coupled to codes treating the evaporation of excited nuclei after INC  to predict accurately the spallation fragment production in a wide range of masses and energy but also some shortcomings at the lowest reaction energy. The latter, however, concern mostly nuclei far away from the parent nucleus that have low production cross sections and therefore do not play an important role in the CR propagation.

\subsection{Total gamma-ray line emission in nuclear reactions}

The studies concerning the gamma-ray emission of LECRs and solar flares presented in \ref{s:crem} were made possible thanks to a longstanding effort of gamma-ray line production measurements. Ramaty, Kozlovsky and Lingenfelter \cite{RKL} presented a first comprehensive review of nuclear gamma rays produced in astrophysical sites by energetic particle interactions. Since then,  cross sections for the most intense lines in astrophysical sites have been measured in several dedicated experiments at tandem Van-de-Graaf and cyclotron accelerator laboratories from threshold up to about 100 MeV per nucleon, which is the most important energy range for solar-flare and LECR-induced gamma-ray emission. The strongest lines are from the ($\alpha,\alpha$) reaction populating excited states of $^7$Li and $^7$Be and from transitions of the first excited levels of $^{12}$C, $^{14}$N, $^{16}$O, $^{20}$Ne, $^{24}$Mg, $^{28}$Si, $^{32}$S and $^{56}$Fe populated in reactions with  protons and $\alpha$ particles. The latest compilation dedicated to solar flare studies contains thus about 180 cross section excitation functions for $\gamma$-ray line production in p, $\alpha$-particle induced reactions and also some for $^3$He-induced reactions \cite{MKKS}.

In Fig. \ref{f:lecr} in section  \ref{s:crem} the structure in the 0.1-10 MeV range is effectively studded with prominent narrow lines from energetic proton and $\alpha$-particle induced reactions, but the bulk of the emission is a broad continuum-like component. It is formed from the superposition of the same prominent lines strongly Doppler-broadened in energetic heavy-ion interactions and numerous weaker lines that form a quasi-continuum. A lot of progress concerning this weak-line quasi-continuum has been achieved recently in several experiments at the tandem Van-de-Graaff accelerator of IPN Orsay. These measurements  have been done in the last two decades with beams of protons, $^3$He and $\alpha$ particles. Cross section excitation functions for proton- and alpha-particle induced reactions on C, N, O, Ne, Mg, Si and Fe have been obtained in a typical energy range of a few MeV to 25 MeV for protons and 40 MeV for alpha particles \cite{CO98,Amel07,Hinda10}. The gamma-ray production in $^3$He-induced reactions on $^{16}$O, a potentially important signature of $^3$He acceleration in impulsive-type solar flares \cite{Man97}, has been studied in the range E$_{^3He}$ = 3 - 33 MeV \cite{flares03}. 

The experimental setup consisted typically of 4 or more large-volume high-purity Ge detectors equipped with BGO  shields for Compton suppression, placed around the target chamber in a wide angular range with respect the beam. Particular attention was given to the control and/or suppression of gamma-ray background in these mesurements. It consisted in the usual active background suppression by the BGO shielding, of effective shielding of the Faraday cup, typically sitting several meters downstream of the target behind a thick concrete wall, and regular monitoring of the room background and the determination of beam-induced background. For the latter, for each beam energy, an irradiation run with an empty target frame has been done. These measures resulted in high-statistics spectra with excellent signal-to-background ratio for the prominent lines and  providing the possibility to extract cross sections also for weaker lines down to the few mb range.

\begin{figure}
\resizebox{0.45\textwidth}{!}{%
  \includegraphics{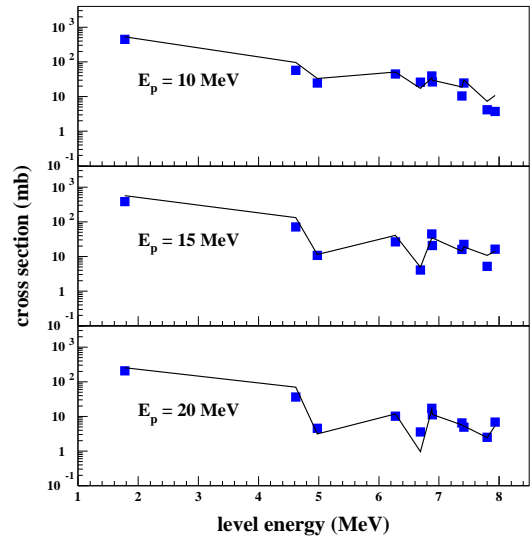}
}
\caption{Gamma-ray emission cross section of the first 11 excited states of $^{28}$Si in proton irradiation of a Si target at three different projectile energies deduced from gamma-ray line measurements (symbols).  Full curves connect the calculated emission cross sections of these states with the nuclear reaction code TALYS, including direct population of the state by inelastic scattering and by cascade transitions from higher-lying levels. It shows the results after the adjustment of collective band couplings to reproduce the experimental data simultaneously at the different proton energies. More details are found in  \cite{Hinda10}. Figure adapted from H. Benhabiles-Mezhoud et al., Physical Review C 83,  024603 (2011). Copyright 2011 by the American Physical Society.}
\label{f:Sipop}       
\end{figure}

The consequently large samples of gamma-ray line data for each nuclear reaction are very valuable for the test and parameter optimization of  nuclear reaction codes. This  is illustrated in the proton irradiation of a Si target where the  extracted gamma-ray line data permitted, among others, the determination of cross sections for the gamma-ray emission of the first 11 excited states of $^{28}$Si, up to E$_x$ = 7.933 MeV. Most states belong to collective bands and their population by inelastic proton scattering at the studied projectile energies depends essentially on the details of the band couplings. The experimental data were then used to complete the coupling schemes and adjust the deformation parameters of the collective bands of $^{28}$Si in  the nuclear reaction code TALYS \cite{Talys1}.  Figure \ref{f:Sipop} shows the measured data at three different proton energies and the result of calculations with TALYS  after adjusting the collective band couplings. As a total, cross section data for about 100 different excitation functions have been obtained in these experiment campaigns, which is to compare with the about 180 excitation functions included in the last compilation for accelerated-particle reactions in solar flares \cite{MKKS}.

\begin{figure}
\resizebox{0.45\textwidth}{!}{%
  \includegraphics{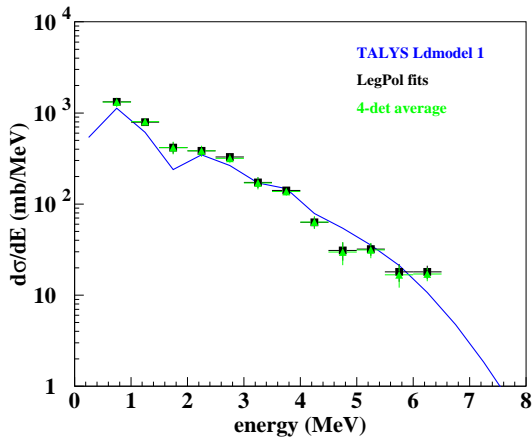}
}
\caption{Differential gamma-ray emission cross section as a function of energy in the proton irradiation of a Fe target at E$_p$ = 10 MeV. Symbols are experimental data from the Orsay tandem obtained by removing the Compton component in the detector spectra and cleared for ambient and beam-induced background.  Black squares show cross sections from Legendre polynomial fits of 4 detector data in range 67.5$^{\circ}$ - 157.5$^{\circ}$  with respect the beam direction, green triangles represent cross section averages of the 4 detectors. The full curve shows the prediction of the TALYS nuclear reaction code with default values for the optical model and nuclear structure parameters. More details can be found in \cite{Roorkee}.} 
\label{f:sigFe}       
\end{figure}

Calculations of total gamma-ray emission in light-par\-ticle induced nuclear reactions including the weak-line qua\-si-continuum have for a long time relied on estimations based on only one dedicated experiment \cite{RKL}. Recent data for the total gamma-ray emission in proton- and alpha-particle induced reactions in the Orsay experiments have been obtained  by subtracting completely all gamma-ray background components. Subtraction of ambient radiation and beam-induced gamma-ray background in these experiments was straightforward due to the availability of high-statistics spectra for all components and accurate beam charge determinations in the  Faraday cup. These subtractions remove completely all background not originating in the target. The remaining background in the spectra  is Compton scattering and pair production of target gamma rays in the detector and surrounding materials for beam energies below the neutron emission threshold. This background could be removed with the help of extensive simulations of the experiment set up with the GEANT code \cite{Geant} to enable spectrum deconvolution. In case of significant neutron production in the target irradiation, further modelisations of neutron interactions coupled to GEANT for the gamma-ray interactions can be used to subtract this neutron-induced background, easily recognizable by the characteristic triangular features in Ge detector spectra. 

An example of total gamma-ray emission in the p + Fe reaction at E$_p$ = 10 MeV is shown in Fig. \ref{f:sigFe}, where data could be extracted up to E$_{\gamma}$ = 6.5 MeV  \cite{Roorkee}. In this irradiation many hundred gamma rays can contribute from discrete transitions: the two main iron isotopes $^{54}$Fe and $^{56}$Fe have together more than 300 known levels below 8 MeV that may be excited, decaying by usually more than 2 different transitions, and (p,n), (p,$\alpha$) reactions may likewise contribute significantly to the gamma-ray emission. A small contribution is also expected from continuum transitions, induced by e.g. radiative proton capture. The gamma-ray emission calculation of TALYS \cite{Talys1} for this reaction is shown for comparison.  Although there seems to be a small underestimation of experimental data at $\gamma$-ray energies below a few MeV, the TALYS calculations reproduce reasonably the magnitude and shape of the cross section curve. Those studies and other relatively correct predictions of cross section excitation functions for gamma-ray lines in light-particle induced reactions  finally led to the inclusion of TALYS calculations  in the latest compilation, and in particular for the weak-line quasi-continuum \cite{MKKS}.

\subsection{Gamma-ray line shapes}

Another recent progress for non-thermal reactions in astrophysics due to recent experimental data concerns gam\-ma-ray line shapes. The exact shapes of prominent narrow lines in solar flares carry information on the accelerated proton-to-alpha-particle ratio and their energy spectra as well as to their directional distribution. The latter may be far from isotropic in the chromosphere  where most nuclear interactions take place in the presence of strong magnetic fields that extend up to the acceleration site in the corona. Recent line-shape studies concentrated on the 4.438-MeV line of $^{12}$C and the 6.129-MeV line of $^{16}$O. For both lines, a database of line shapes for proton and alpha-particle reactions with C and O is now available from the Orsay experiments \cite{CO98,Amel07,Hinda10} in a wide angular range and with a good coverage in projectile energies from threshold to about 25 MeV for proton and 40 MeV for $\alpha$-particle reactions. 

Together with the $^7$Li-$^7$Be lines from alpha-alpha reactions \cite{Sha03}, these lines are probably the best candidates for line shape studies in solar flares: (1) the emitting nuclei $^{12}$C and $^{16}$O are relatively light, meaning high recoil velocities and the relatively high gamma-ray energies lead to large Doppler shifts that are easily resolvable with high-resolution Ge detectors onboard gamma-ray satellites like RHESSI \cite{RHESSI} and INTEGRAL/SPI \cite{SPI}; (2) they are among the strongest prompt emission lines in solar flares; (3) line-shape calculations are facilitated by the negligible population of the emitting 4.439-MeV, $^{12}$C and 6.130-MeV, $^{16}$O levels  by gamma-ray cascades of higher-lying levels.

The first comprehensive study of the 4.438-MeV line shape in solar flares that was largely based on measured data, has been done at Orsay \cite{cshape}. In solar flares, this line is essentially produced by proton and $\alpha$-particle inelastic scattering off $^{12}$C and reactions with $^{16}$O. The measured line shapes and relative line intensities of 6 HP-Ge detectors placed at $\Theta$ = 45$^{\circ}$ - 145$^{\circ}$ in proton reactions with $^{12}$C could be fairly well reproduced with a simple parameterization of the  magnetic-substate population of the 4.439-MeV state after inelastic scattering similar to the method proposed in \cite{RKL} and with use of extensive optical-model calculations. Measured 4.438-MeV line shapes  in the $^{16}$O(p,p$\alpha$$\gamma$)$^{12}$C reaction could be nicely reproduced by adjusting the mean excitation energy in $^{16}$O before $\alpha$-particle emission and otherwise isotropic emission of the proton, $\alpha$ particle and $\gamma$ ray. These studies were later on, completed for $\alpha$-particle induced reactions and the 6.129-MeV line \cite{Octflare}.

\begin{figure}
\resizebox{0.45\textwidth}{!}{%
  \includegraphics{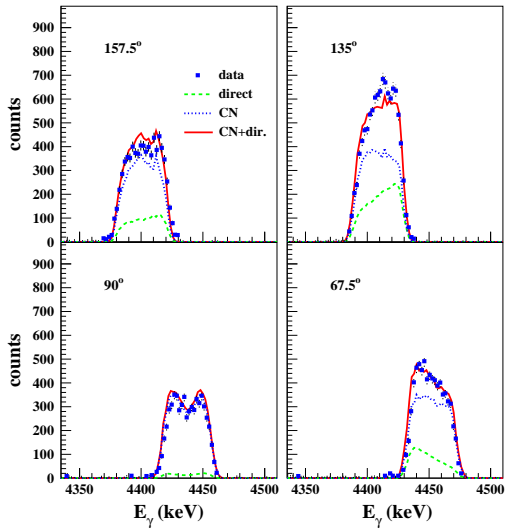}
}
\caption{Measured data of the 4.438-MeV line in the proton irradiation of a thin carbon target at E$_p$ = 6.5 MeV (symbols) and calculated line profiles. The solid red line shows the sum of the coumpound-nucleus component (CN) and the direct reaction contribution in the inelastic scattering reaction $^{12}$C(p,p'$\gamma$)$^{12}$C. } 
\label{f:cshape}       
\end{figure}

Since then, a new method has been developed, that aimed at a specific improvement of line-shape descriptions in the region dominated by compound-nucleus (CN) resonances. This is below about E$_{p,\alpha}$ = 15 MeV for proton and $\alpha$-particle  inelastic scattering  to the 4.439-MeV and 6.129-MeV states. It relies on optical-model calculations in the coupled-channels approach for the direct interaction component and explicit resonance calculations for the CN component \cite{Tours,Roorkee}. An example for the 4.438-MeV line  is shown in Fig. \ref{f:cshape}. The best reproduction of the measured line shapes was obtained by assuming a pure 3/2$^+$ resonance contribution for the CN component. For this proton energy E$_p$ = 6.5 MeV, the CN excitation energy is E$^x_{CN}$ = 7.94 MeV, where the only known suitable resonance in $^{13}$N is the 1.5-MeV broad 3/2$^+$ state at E$_x$ = 7.9 MeV. The  optical-model calculation in the coupled-channels approach was done with the ECIS code \cite{ECIS} and optical model parameters of \cite{Pea72} were taken from the compilation of Perey\&Perey \cite{Per76}. This work is in progress \cite{Tours,Roorkee} and will eventually replace the method described in \cite{cshape} for small projectile energies.

\section{Nuclear astrophysics and cosmology}

There are important aspects of cosmology, the scientific study of the large--scale properties of the Universe as a whole, for which nuclear physics 
can provide insight \cite{Raffelt1996,Uzan2009}. Here, we focus on the properties of the early Universe 
(big bang nucleosynthesis during the first 20 mn) and on the {\em variation of 
constants} over the age of the Universe.

\subsection{BBN as a probe of the early Universe}
\label{s:probe}

Now that the baryonic density of the Universe has been deduced from the 
observations of the anisotropies of the CMB radiation, there is no free parameter
in standard BBN.
The CMB radiation that is observed was emitted when the Universe became 
transparent $\approx3\times10^5$ years after the big bang. On the contrary,
the freeze-out of weak interactions between neutrons and  protons, and BBN,
occurred, respectively,  at a fraction of a second, and a few minutes after the big bang.
Hence, comparison between 
observed and calculated light--element abundances can be used to constrain
the physics prevailing in the first seconds or minutes of the  Universe \cite{Ioc09,Pos10}.

In fact a 10\% change in the expansion rate, within the first seconds after the big bang, would be sufficient to drive 
the \qua\ abundance out of the observational limits while providing little
help to the \sep\ discrepancy (\S~\ref{s:bbn}). The \qua\ yield is sensitive to the value of
the expansion rate ($H(t)$) at the time of n/p freeze-out, i.e., around 10$^{10}$~K
and 0.1 to 1~s after the big bang while the other isotopes are sensitive 
to its value 3 to 20 mn after.
The freeze-out occurs when the weak reaction rates $\Gamma_\mathrm{n{\leftrightarrow}p}$ become slower than the expansion rate i.e.:
\begin{equation}
H(t)\equiv
\sqrt{{8\pi\;G\;{\mathrm a}_R\;g_*(T)}\over6}{\times}T^2 \sim \Gamma_\mathrm{n{\leftrightarrow}p}\propto\frac{T^5}{\tau_\mathrm{n}},
\label{eq:expan}
\end{equation} 
where $\mathrm{a_R}$ is the radiation constant. Here, $g_*(T)$ is the the effective spin factor:
$g_*(T)\equiv2\sum_{i}{\rho_i}(T)/{\rho_\gamma}(T)$ where the $\rho_i$ ($i=\gamma$, e$^\pm$, and [anti--]neutrinos in standard BBN)
are the energy densities of the relativistic species \cite{Ioc09}.  
This factor varies slowly with temperature during BBN (3.36$\leq{g_*}\leq$10.25 in the standard model). 
There are several potential sources of deviation from the nominal expansion 
rate $H(t)$ as can be seen from  equation~\ref{eq:expan}.
For instance,  a deviation from General Relativity would affect the gravitational ``constant" $G$, and
new relativistic particles would modify the effective spin factor, $g_*(T)$, while the neutron lifetime, $\tau_\mathrm{n}$,
is sensitive to the Fermi constant, $G_F$.

There are various ways in which exotic particles can influence BBN \cite{Ioc09}.
The decay of a massive particle during or after BBN could affect the light element abundances 
and potentially lower the \sep\ abundance (see e.g.~\cite{Cyb13}). Neutrons, protons or photons
produced by these  decays may be thermalized but more likely have a non-thermal, high--energy
($\sim$1~GeV) distribution. Interestingly, some nuclear cross sections involved in  these
non-thermal processes are not known with a sufficient precision \cite{Cyb10}. If they can
be thermalized, it provides an extra source of neutrons that could alleviate the lithium problem \cite{Jed04,neutrons}. 
Another exotic source of thermalized neutrons could   
come from a ``mirror world" \cite{miroir} initially proposed to restore global parity symmetry.  
Long--lived (relative to BBN time scale) negatively charged relic particles, like the supersymmetric 
partner of the tau lepton, could form bound states with nuclei, lowering the Coulomb barrier and hence
catalysing nuclear reactions (see e.g.~\cite{Pos07,Kus08,Cyb12b}). Even though exotic, 
the interaction of these {\em electromagnetically} bound states with other nuclei can be treated by 
conventional nuclear physics theory.

\subsection{Variation of fundamental constants}

\begin{figure}
\resizebox{0.45\textwidth}{!}{%
  \includegraphics{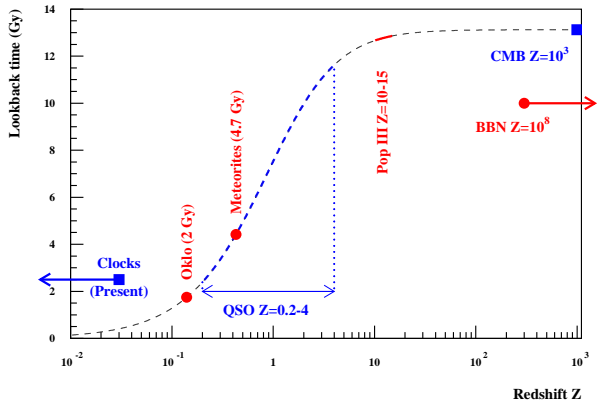}
}
\caption{Test of the variations of constants performed at different redshifts or lookback time 
(i.e. elapsed time until present). Those in red involve nuclear physics.}
\label{f:timez}       
\end{figure}

Experimental (or observational) tests of variations of a constant consists
in comparing quantities that have a different sensitivity to this constant (for reviews, see \cite{Uzan,Gar07}).
Tests involve atomic clocks, atomic absorption in quasar (QSO) spectra, the CMB, and 
nuclear physics (big bang nucleosynthesis, 
the triple--alpha reaction and stellar evolution, radioactivities in meteorites 
and the Oklo fossil reactor).
They are all interesting because they have different dependency to the
variation of constants and they probe variations on different cosmic time
scales (Fig.~\ref{f:timez}).  
In the following, we consider only those related
to nuclear physics, and in particular the triple--alpha reaction in stars, and BBN.
We will illustrate the effect of varying ``constants" like the fine structure constant,
the Fermi constant, the electron mass, on some nuclear reactions or decay but
leave aside the discussion of the {\em coupled} variations of these constants (e.g. \cite{Coc07})
which are beyond the scope of this review. Indeed, within theories like superstring theory,
the constants cannot be treated independently: their variations are related to each other
in a way that depend on the model. Following Fig.~\ref{f:timez}, from present to big bang,
we have the following constraints from nuclear physics.

\subsubsection{The Oklo nuclear fossil reactor}

At present, terrestrial uranium is mainly composed of $^{238}$U,  0.72\% of $^{235}$U and 0.0055\% of $^{234}$U. 
The $^{235}$U isotope has a half--life of 7.038$\times10^8$ years and decays by alpha emission. It is fissile through
the absorption of thermal neutron that can lead, within special conditions, to the controlled chain reaction at work in nuclear
reactors. One of the conditions is that uranium is enriched in $^{235}$U to a level of 3--4\%, another is that 
fission--produced neutrons are slowed down (``moderated") to take advantage of higher induced--fission cross section. 
In 1972, the French Commissariat \`a l'\'Energie Atomique discovered, in an uranium mine located at Oklo, in Gabon,
that a {\em natural} nuclear reactor had been operating two billion years ago, during approximately a million years 
(see \cite{Oklo} and references therein). 
This operation was made possible because, at a few $^{235}$U half--lives ago, its fractional abundance was
sufficiently high, and hydrothermal water acted as moderator. As a result, the ore displayed a depletion 
in $^{235}$U that was consumed by the chain reaction, and very peculiar isotopic rare-earth abundances. 
In particular $^{149}$Sm (samarium) was strongly depleted, an effect ascribed to thermal neutron absorption through 
the $E_R$ = 0.0973~eV resonance in $^{149}$Sm(n,$\gamma)^{150}$Sm.
As the neutron exposure time and energy distribution can be inferred from other rare-earth isotopic compositions,
the samarium isotopic ratios are sensitive to the $^{149}$Sm(n,$\gamma)^{150}$Sm cross section
and hence, to the position of the resonance. If its resonance energy can be related to 
the fine structure ($\alpha_\mathrm{em}$),and other constants, one can put limits on their variations \cite{Cstes02,Dav14},
typically $|\Delta\alpha_\mathrm{em}/\alpha_\mathrm{em}|<10^{-10}$ \cite{Cstes02} during the last 2~Gyr.

\subsubsection{Meteorites and $^{187}$Re}

The  $^{187}$Re isotope is of special interest for the study of possible variation of constants \cite{Uzan,re187} because of its 
very long lifetime, larger than the age of the Universe, and because of the high sensitivity of its lifetime
to the variation of constants.  
It is the most abundant (62.6\%) terrestrial rhenium isotope which 
$\beta^+$ decays to $^{187}$Os with a measured mean life of $\lambda_\mathrm{Lab}^{-1}$ =  
61.0$\times10^9$ years.
Its $\beta+$ decay rate can be approximated by $\lambda{\propto}G_F^2Q_\beta^3m_e^2$. 
Thanks to the very low value of $Q_\beta$ =  2.66 keV, the sensitivity of $\lambda$ to variations of  $Q_\beta$ is high. 
The imprint of $^{187}$Re decay since the birth
of the Solar System can be found in the isotopic composition of some meteorites. 
Indeed, one has:
\begin{equation}
\left.^{187}\mathrm{Os}\right|_\mathrm{Now}=\left.^{187}\mathrm{Os}\right|_\mathrm{Init}+
\left.^{187}\mathrm{Re}\right|_\mathrm{Now}\left[\exp{\left(\bar{\lambda}t_\mathrm{M}\right)}-1\right]
\end{equation}
where $t_\mathrm{M}$ is the age of the meteorite ($\approx$ the age of the Solar System) and
$\bar{\lambda}$ is the {\em averaged} $^{187}$Re decay constant assuming that $\lambda$ may 
have evolved over $\approx$ 4.6$\times10^9$. It shows that the present day $^{187}$Os versus $^{187}$Re meteoritic isotopic abundances (relative to 
stable $^{188}$Os) follow a linear dependence (an isochrone, as in Fig.~\ref{f:meteorite}  for $^{26}$Al) 
from which  $\left[\exp{\left(\bar{\lambda}t_\mathrm{M}\right)}-1\right]$ can be extracted.   
If the age of the meteorites $t_\mathrm{M}$ can be obtained by another dating method (U/Pb isotopes)
which is much less sensitive to the variation of $\alpha_\mathrm{em}$, then $\bar{\lambda}$ can be deduced and 
limits on $\Delta\alpha_\mathrm{em}$ (and of other constants) can be obtained from those on $\bar{\lambda}-\lambda_\mathrm{Lab}$  \cite{re187},
typically $|\Delta\alpha_\mathrm{em}/\alpha_\mathrm{em}|<10^{-6}$ \cite{re187} during the last 4.6~Gyr.

\subsubsection{Variation of constants and stellar evolution}
\label{s:popiii}

\begin{figure}[t]
\resizebox{0.45\textwidth}{!}{%
 \includegraphics{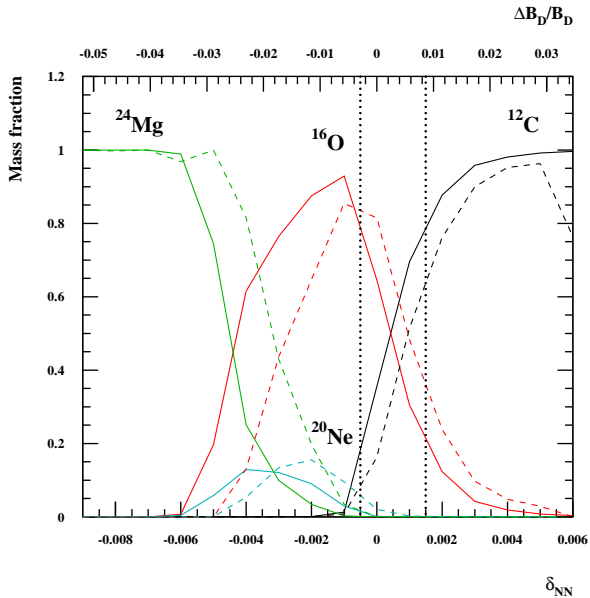}
}
\caption{Central abundances at the end of CHe burning as a function of the relative variation of the 
NN-inetraction,  $\delta_{NN}$, for 15 (solid) and 60 (dash) $M_\odot$ stars.}
\label{f:xoc}       
\end{figure}

The \aaag\ reaction is very sensitive to the position of the ``Hoyle state" (\S~\ref{s:heburn}, Fig.~\ref{f:aaag}). The corresponding
resonance width is very small (a few eV) as compared with the competing reaction
$^{12}$C($\alpha,\gamma)^{16}$O, dominated by broad ($\sim$100 keV) resonances
and subthreshold levels.
Small variations of the Nucleon--Nucleon--interaction ($\lap$1\%) induce small variations of the  position of the ``Hoyle state", but
huge variations (many orders of magnitude) on the 
triple--$\alpha$ reaction rate.
This effect was investigated for 1.3, 5, 15, 20 and 25 $M_\odot$ stars with
solar metallicity (Population I) by Oberhummer and collaborators \cite{ober,schl}. They
estimated that variations of more than 0.5\% and 4\% for the values of
the strong and electromagnetic forces respectively, would drastically reduce the stellar production of
either carbon or oxygen (depending on the sign of the variation). However, the stars that were
considered in this study, were born a few Gy ago, i.e. only at small redshift $z<1$.
Considering instead (see \cite{aaag} for more details) the very first generation of stars (Population III) 
extends the test to a much larger lookback time. 
These stars are thought to have been formed a few 
$10^8$ years after the big bang, at a redshift of $z\approx10-15$, and with {\em zero initial metallicity}.
For the time being, there are no direct observations of those  Population III stars but one may
expect that their chemical imprints  could be observed indirectly in the most metal-poor halo stars (Population II). 
Depending on the NN--interaction, helium burning results in a stellar core with a very different composition
from pure $^{12}$C to pure $^{16}$O (and even pure $^{24}$Mg), due to the competition between
the \aaag\ and $^{12}$C($\alpha,\gamma)^{16}$O reactions.   
As both C and O are observed in metal-poor halo stars, a $^{12}$C/$^{16}$O  abundance ratio close to
unity is required. To be achieved, the  relative variation of the NN--interaction ($\delta_{NN}$) should be less than $\approx10^{-3}$, 
which can be translated in limits on $\Delta\alpha_\mathrm{em}$ \cite{aaag} (Fig.~\ref{f:xoc}).

\subsubsection{Variation of constants and BBN}

\begin{figure}
\resizebox{0.45\textwidth}{!}{%
 \includegraphics{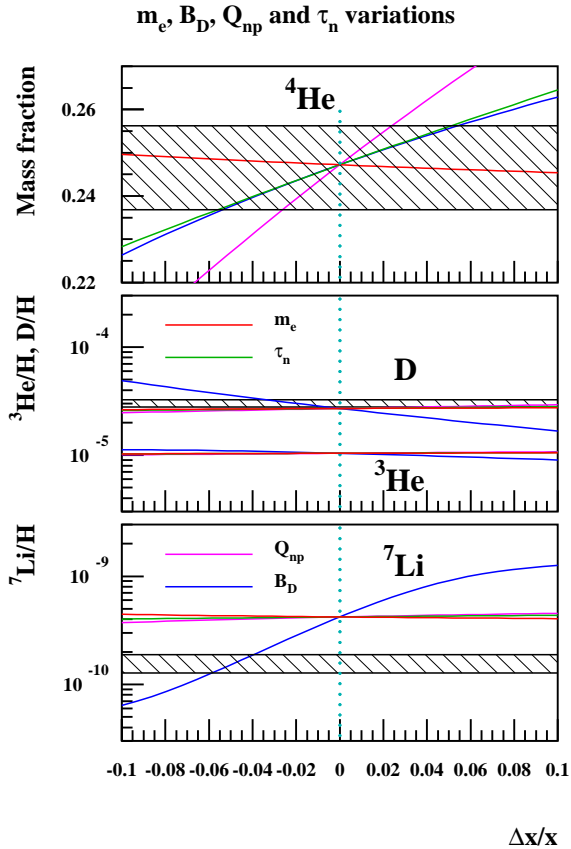}
}
\caption{Abundances of \qua\ (mass fraction), \deu, \tro\ and \sep\ as a 
function of a fractional change in the electron mass ($m_e$), the binding energy of deuterium ($B_D$),  the neutron--proton mass difference ($Q_{np}$), 
and the neutron lifetime ($\tau_n$). (Updated from Fig.~3 in Ref.~\cite{Coc07} with \cite{Ave13} re-evaluated \qua\ mass fraction.)}
\label{f:variall}       
\end{figure}

We have mentioned in \S~\ref{s:probe} that, at the BBN epoch,
gravity could  be different from General Relativity as in superstring theories.
That would affect the rate of expansion, through $G$ in equation~\ref{eq:expan}.
Here, we will illustrate the influence of the variations of constants on two key
nuclear reaction rates : n$\leftrightarrow$p and n+p$\rightarrow$d+$\gamma$.  
The first, together with the expansion rate, governs the production of \qua, while 
the second triggers further nucleosynthesis. The weak rates that exchange protons
with neutrons can be calculated theoretically and their dependence on $G_F$ (the Fermi constant),
$Q_{np}$ (the neutron--proton mass difference) and $m_e$ (the electron mass) is
explicit. For instance,  the $n{\rightarrow}p+e^-+\bar{\nu_e}$, neutron free decay displays a  
$G_F$, $q{\equiv}Q_{np}/m_e$ and $m_e$ explicit dependence.
The dependence of the n+p$\rightarrow$d+$\gamma$ rate \cite{And06} cannot be explicitly related
to a few fundamental quantities as for the weak rates but a modeling of its dependence on the
binding energy of the deuteron ($B_D$) has been proposed \cite{Dmi04}, although other prescriptions are possible \cite{Car13,Ber13}. 
Figure~\ref{f:variall} shows the dependence of the \hli, primordial abundances on
these four quantities, $m_e$, $Q_{np}$, $\tau_n$ and $B_D$. 
It shows that a variation of $B_D$ (i.e. the  n+p$\rightarrow$d+$\gamma$ rate)
has a strong influence on \sep, even reconciling calculations with observations.
However, even though the \qua\ abundance remains close to the lower observational limit,
Deuterium is overproduced with respect to observations.

We have seen (\S~\ref{s:popiii}) that Pop. III stars can put a limit on the NN-ineraction at $z\approx10-15$,
but at BBN time, $z\sim10^8$, it may be different. 
In particular, for $\delta_{_{\rm NN}}\gap7.52\times10^{-3}$, $E_{g.s.}(^8{\rm Be})$ (relative to the 2--$\alpha$ threshold, 
Fig.~\ref{f:aaag}) becomes negative i.e. \hui\ becomes stable.
In that case,  one has to consider two reaction rates, \aabe\ and \beac\ for a stable \hui, and one may expect
an increased \carb\ production, bypassing the $A$=8 gap.
However, calculations show that the carbon abundance has a maximum of C/H$\approx10^{-21}$ 
\cite{CNO12}, which is {\em six orders of magnitude} below the carbon abundance in SBBN \cite{CNO12,Coc14b}. 
Note that the maximum is achieved for $\delta_{_{\rm NN}}\approx0.006$ when \hui\ is still unbound so that contrary to
a common belief, a stable \hui\ would not have allowed the build--up of heavy elements during BBN. 

\section{Conclusions}

In this review paper, we have presented a few examples of experiments that we think are representative
of the field of nuclear astrophysics.
We want to emphasize that all have had significant impact on astrophysics, as summarized below.

Nova--related experimental results allow setting the expected gamma flux from close nova
explosions and are used to set triggering conditions for $\gamma$--ray astronomical observations in space. These nuclear physics results
have led to a drastic reduction of the maximum detectability distance of prompt gamma ray emission, dominated by
$^{18}$F decay, in nova explosions \cite{CapeTown}. Sensitivity studies have identified the most important reactions in nova 
nucleosynthesis and associated gamma ray emission from $^{22}$Na and $^{26}$Al and triggered
experimental studies that are still going on.

We also presented recent experimental studies relevant to reactions of accelerated particles in astrophysical sites. 
As for thermonuclear reactions, new observations were an important motivation for 
improving our knowledge of cross sections for energetic-particle induced nuclear reactions. 
This includes new measurements of the cosmic-ray composition and spectra and novel remote observations of 
emissions  induced by cosmic rays  in the interstellar matter as well as solar--flare gamma-ray emission with unprecedent
sensitivity and precision. The nuclear reaction studies for energetic particles have addressed particular points like  
fragmentation cross sections of important nuclei for cosmic-ray studies and  line-shape  calculations  for solar flares that are
directly applicable to astrophysics. This is accompanied by another important development: 
comprehensive nuclear  reaction codes like INCL-4 and TALYS that are an essential tool  for the study of energetic particles
in astrophysics.
 
With improved observations, cosmology has entered a precision era with big bang nucleosynthesis as a probe of new physics in early Universe. 
The ``lithium problem" remains a challenge for (astro-)physicists, but at least a nuclear solution is now excluded thanks to laboratory measurements. 
It is now clear from nuclear physics analyses that the high \six\ abundance observed in some stars 
could not be obtained by standard BBN. With improved precision on primordial deuterium observations, the cross sections for \deu\
destruction in BBN needs to be known with a similar precision ($\sim$1\%).

Finally, it is worth mentioning that almost all these experimental achievements  have been obtained at small scale facilities,
some of which are now dismantled \footnote{The  Louvain la Neuve RIB project came to an end in 2009 \cite{LLN},
the Holifield Radioactive Ion Beam Facility, Oak Ridge,  ceased operations in 2012, and the PAPAP small accelerator awaits
rehabilitation at Demokritos laboratory, Athens. 
}, or in danger of closing (the Tandem of the Orsay ALTO facility).

\begin{acknowledgement}

We are grateful to A. Boudard, P. Descouvemont, N. de S\'er\'eville, F. de Oliveira, S. Goriely, M. Hernanz, C. Iliadis, J. Jos\'e,
R. Longland, V. Tatischeff, J.-P. Uzan and  E. ~Vangioni, for useful discussions and to D. Lunney for his careful reading of the manuscript.  
Finally, we thank Nicolas Alamanos for inviting us to write this review.

\end{acknowledgement}

\end{document}